%% file: SCE_Arxiv_v2.tex
\documentclass[12pt]{article}
\input{header.tex}
\usepackage{amsmath}
\usepackage{authblk}

\usepackage{setspace}
\usepackage{multirow}
\usepackage{algorithm,algpseudocode}
\usepackage{cleveref}
\usepackage{gensymb}
\usepackage{rotating}

\providecommand{\keywords}[1]
{
  \small	
  \textbf{Keywords---} #1
}

\graphicspath{}
\newcommand{\commentOut}[1]{}
\title{Modelling Extremes of Spatial Aggregates of Precipitation using Conditional Methods}

\author[A,B]{Jordan Richards}
\author[A]{Jonathan A. Tawn}
\author[C]{Simon Brown}
\affil[A]{STOR-i Centre for Doctoral Training, Department of Mathematics and Statistics, Lancaster University, LA1 4YR, UK}
\affil[B]{Computer, Electrical and Mathematical Sciences and Engineering Division,
King Abdullah University of Science and Technology (KAUST), Thuwal, 23955-6900, Saudi Arabia}
\affil[C]{Hadley Centre for Climate Prediction and Research, Met Office, FitzRoy Road,
Exeter, EX1 3PB, UK}
\date{}
\begin{document}
  \maketitle
\begin{abstract}
Inference on the extremal behaviour of spatial aggregates of precipitation is important for quantifying river flood risk. There are two classes of previous approach, with one failing to ensure self-consistency in inference across different regions of aggregation and the other imposing highly restrictive assumptions. To overcome these issues, we propose a model for high-resolution precipitation data, from which we can simulate realistic fields and explore the behaviour of spatial aggregates. Recent developments have seen spatial extensions of the \cite{heff2004} model for conditional multivariate extremes, which can handle a wide range of dependence structures. Our contribution is twofold: extensions and improvements of this approach and its model inference for high-dimensional data; and a novel framework for deriving aggregates addressing edge effects and sub-regions without rain. We apply our modelling approach to gridded East-Anglia, UK precipitation data. Return-level curves for spatial aggregates over different regions of various sizes are estimated and shown to fit very well to the data.
\end{abstract}
\hspace{10pt}
  \section{Introduction}
  \label{intro}
  Fluvial flooding is typically not caused by high intensity extreme rainfall at single locations, but by the extremes of precipitation events which are aggregated over spatial catchment areas. Accurate modelling of such events for both the present and the future can help to mitigate the financial impacts associated with floods, for example if river defences are built to withstand a $T-$year event of this kind. Approaches to quantifying the tail behaviour of spatial aggregates exist in the literature; however, these techniques are often simplistic or make unrealistic assumptions about the behaviour of the process for which they are trying to model. We present a novel methodology for making inference on the tail behaviour of spatial aggregates, which we apply in the context of extreme precipitation aggregates.    \par
 We define a spatial process $\{Y(s):s \in \mathcal{S}\}$ for some spatial domain $\mathcal{S}$. Our interest lies in the upper tail behaviour of the spatial aggregate $R_\mathcal{A}$ on regions $\mathcal{A} \subset \mathcal{S} \subset \mathbb{R}^2$,
  \begin{equation}
  \label{aggeq}
  R_\mathcal{A}=\int_{\mathcal{A}}Y(s)\mathrm{d}s,
  \end{equation}
  for different, possibly overlapping, $\mathcal{A}$, and the joint behaviour of $(R_\mathcal{A},R_\mathcal{B})$ for $\mathcal{A},\mathcal{B} \subset \mathcal{S}$.
  Typically, the data we would have available for inference are realisations of $\mathbf{Y}_t=(Y_t(s_1),\dots,Y_t(s_d))$ for $t= 1 ,\dots, n$, which are observations of said process $\{Y(s)\}$ at $d$ sampling locations $\mathbf{s} = (s_1,\dots,s_d) \subset \mathcal{S}$ at $n$ sampling times. Note that $\mathbf{s}$ need not be point locations; they can instead be non-overlapping grid-boxes. Data produced by climate models are often available in this form and observations of $Y(s_i),\; i = 1, \dots, d,$ correspond to spatial aggregates themselves, as they are typically presented as an average over the grid box $s_i$. In these circumstances, the integral in \eqref{aggeq} can be replaced with the equivalent summation, but our methodology is still applicable; see Section \ref{sec-appl}.
  \par
Modelling extreme rainfall events over different spatial aggregation regions is difficult due to spatially-localised and high intensity convection rainfalls. For summer precipitation, convection events are the primary cause of extreme spatial aggregates of rainfall, even over regions much larger than the scale of convection events. This poses significant challenges when using data from any observation network of rain gauges, with such localised events being either recorded at a single rain gauge or missed entirely, with the latter more likely if the network is sparse or the observation window is short. In contrast, radar networks provide spatially complete coverage but are comparatively short in duration and have their own issues, such as calibration and attenuation \citep{harrison2000improving}. The use of climate/numerical models to provide inference on extremes of real-world processes is now a standard approach, see e.g., \cite{cooley2010spatial, northrop2011threshold,Brown2020heatwave} and \cite{ huser2020advances}, which we adopt. However, early generation climate models failed to model convective rainfall due to insufficient spatial resolution. Currently, the only viable approach to capture the physics driving extreme convection events, and any future changes, is to use a new generation of high-resolution climate models that explicitly resolve convection \citep{kendon2014heavier}. In addition to enabling the study of extreme convective rainfall events, these models reproduce well the empirical properties of observational data, benefit from long temporal runs, have no missing data, and provide the opportunity to explore future climatic change scenarios.\par
We are interested in modelling extreme aggregated precipitation data for regions in East-Anglia, UK. We use data from the UKCP18 local model projections \citep{kendon2019ukcp, datacite}. Analysing such high-resolution data poses significant new computational challenges given the large number of sites in our chosen domain, with $d = 934$ exceeding the vast majority of previous spatial extremes studies. The lack of missing data enables us to perform simple and fair comparisons of our approach with existing methods.
 \par 
We assume that both the full marginal behaviour, and dependence, of $\{Y_t(s)\}$ is stationary with respect to time. Marginally, the upper tail behaviour of $Y(s)$ is assumed to be characterised by a generalised Pareto distribution (GPD) with scale and shape parameters, $\upsilon(s)>0$ and $\xi(s)$, respectively, that vary smoothly over $s \in \mathcal{S}$ \citep{Davison1990}. Dependence in $\{Y(s)\}$ is characterised through a marginal transformation to the process $\{X(s):s \in \mathcal{S}\}$, which has standardised margins; see Section~\ref{sec-depmodel}. Letting $F_{Y(s)}(\cdot)$ and $F_{X}(\cdot)$ be the marginal CDFs of $\{Y(s)\}$ and $\{X(s)\}$, respectively, we rewrite \eqref{aggeq} as
    \[
  R_\mathcal{A}=\int_{\mathcal{A}}Y(s)\mathrm{d}s=\int_{A}F^{-1}_{Y(s)}\left\{F_{X}\left[X(s)\right]\right\}\mathrm{d}s.
  \]
We focus on the situation where $\{X(s)\}$ is a stationary process; an assumption that we find holds well for our application (see Section \ref{sec_diag}). If this assumption did not hold, both non-parametric, and parametric, methods exist that account for non-stationarity in extremal dependence \citep{huser2016,castro2020local,richards2021spatial}, and these methods can easily be incorporated into our methodology.   
\par 
As $R_\mathcal{A}$ is essentially a sum of variables, a natural starting point is to consider extreme value theory for sums,
i.e., when $\mathcal{A}$ is a finite collection of indices. Consider the case where $\xi(s)$ varies over $\mathcal{A}$ and let 
$\xi_\mathcal{A}=\max_{s \in \mathcal{A}} \xi(s)$. If $\xi_\mathcal{A}>0$, then Breiman's lemma \citep{breiman1965some} gives that the upper-tail of $R_\mathcal{A}$ is determined solely by the behaviour at sites $\mathcal{A}^*=\{s \in \mathcal{A}, \xi(s)=\xi_\mathcal{A}\}$. Connected results are given 
by \cite{richards2021tail} for $\xi_\mathcal{A} \le 0$. When the value of 
$\xi_\mathcal{A}-\max_{s \in \mathcal{A}\setminus\mathcal{A}^*}\xi(s)$ is small, convergence to the limiting extremal behaviour of $R_\mathcal{A}$ will be slow as the behaviour at the sites not in $\mathcal{A}^*$ is completely ignored. Another perspective is to view $R_\mathcal{A}$ as an infinite sum, and so the central limit theorem may apply when 
$\xi_\mathcal{A}<\frac{1}{2}$ (which ensures finite marginal variances); which if correct would imply that the upper-tail distribution of $R_\mathcal{A}$ should be increasingly more likely to be exponential as $\|\mathcal{A}\|$ increases, i.e., $\xi = 0$ (due to characteristics of maxima of Gaussian random variables), irrespective of whether or not $\xi(s)$ varies over $\mathcal{A}$. Generally this is a false argument, due to Breiman's lemma, when $\xi_\mathcal{A}>0$; however, it is supported in certain circumstances by results in \cite{richards2021tail} when $\xi_\mathcal{A}<0$. We consider regions $\mathcal{A}$  which are small relative to lags at which spatial independence occurs, so we lack sufficient independence for central limit theorem arguments to apply.
\par
  There are three main existing modelling approaches for inference on the upper tails of $R_\mathcal{A}$: univariate methods, spatial approaches that focus on modelling all of the data, and spatial approaches that focus on modelling only the extremes; our approach falls in the latter class, making less restrictive assumptions than previous methods of this type.  \par 
 We first consider the univariate case. Within an extreme value analysis framework, univariate methods for estimating the size of $T-$year events are well studied and cemented in asymptotic theory \citep{Coles2001book}. If we can create a sample of observations of $R_\mathcal{A}$, we can use univariate methods to make inference on its upper tail, i.e., fit a GPD to exceedances of a sample of $R_\mathcal{A}$ above some fixed threshold and then extrapolate to high quantiles. However, creating this sample can be challenging if $\mathbf{s}$ are irregularly spaced point locations. Further complications arise if we have partially missing observations of $\{Y(s)\}$. Even if these issues are overcome, when using univariate methods we lose the information present in the margins of $\{Y(s)\}$ and dependence of $\{X(s)\}$. If the process we are considering is precipitation, this can lead to inference that is not self-consistent and may be physically unrealistic; a trait that can be undesirable to practitioners. To explain this further, observe that, for precipitation, $\{Y(s)\}$ is non-negative everywhere, i.e., $Y(s) \geq 0$ for all $s \in \mathcal{S}$. Trivially it follows that $R_\mathcal{A} \geq R_\mathcal{B}$ for all $\mathcal{B} \subseteq \mathcal{A} \subset \mathcal{S}$ and hence return levels should be similarly ordered. This natural ordering may not follow if we take a simple univariate approach to modelling the upper tail behaviour of the $R_\mathcal{A}$ and $R_\mathcal{B}$ aggregates separately \citep{nadarjah1998}. 
  \par
  In the context of precipitation aggregates, one richly studied approach has developed a class of stationary stochastic processes to model the whole precipitation intensity process, continuous in both time and space. These models typically describe the intensity as the accumulation, at each point in time and space, over a random number of simple shaped individual stochastic rain cells, which cluster in time and space, and move on stochastic trajectories. These models were first developed for a single site by \cite{Rodriguez1987}, then developed spatially and for fine-scale processes, see \cite{benoit2018stochastic}. These models are typically estimated by optimising the fit against a range of characteristics of observed fields. As a result, these models can capture well the broad-scale features of typical precipitation fields. However, for deriving the upper-tail of $R_\mathcal{A}$, the models and their inference have limitations as there is no guarantee that models for the body of a process fit well to the extremes. Yet it is precipitation fields that are extreme somewhere in $\mathcal{A}$ that yield extremes of $R_\mathcal{A}$ unless $\mathcal{A}$ is very large relative to the range of spatial dependence. 
  \par
 From the perspective of spatial extremal methods, $R_\mathcal{A}$ is simply a functional of a spatial field which is extreme in some way. A typical approach to modelling extreme fields is the use of max-stable models, see  \cite{padoan2010likelihood,westra2011detection,reich2012hierarchical}. 
These models are predominately fit to component-wise block maxima, typically annual maxima, at sampling locations.
Typically annual maxima do not occur concurrently for all sampling locations, do not give zeros, and so aggregating over realisations from a max-stable process is not appropriate for inference on aggregates of extreme precipitation events. \cite{coles1993} rectified some of these issues by using a point-process representation of a max-stable field to derive the spatial profile of concurrent events. \cite{coles1996} used this formulation to derive closed form results for the tail behaviour of $R_\mathcal{A}$ where the tail parameters are determined by the marginal GPD parameters of $\{Y(s)\}$ and its dependence structure; \cite{Ferreira2012} formalise these results and provide some non-parametric extensions. Further extensions of this framework by \cite{engelke2018} relate not only the extremal behaviour of $\{Y(s)\}$ and the aggregates $R_\mathcal{A}$, but also the joint behaviour of aggregates over different regions, $\mathcal{A}$. \cite{defondeville2020functional} use functional Pareto processes to model the dependence in $\{Y(s )\}$ and \cite{palacios2020generalized} illustrate non-parametric Pareto process modelling to simulate extreme precipitation fields, re-sampling event profiles from observed, gridded data. All of these modelling approaches rely on the marginal shape parameters of $\{Y(s)\}$ to be spatially homogeneous i.e., $\xi(s) = \xi$ for all $s \in \mathcal{A}$, for each $\mathcal{A}$ of interest.  Furthermore they all have major limitations for applications due to their restrictive choice of dependence structure, as described below.
\par 
The co-occurrence of extremal events is often quantified through the upper tail index $\chi(s_A,s_B)$ \citep{joe1997multivariate} for all $s_A,s_B \in \mathcal{S}$, which can be defined for $\{Y(s)\}$ as ${\chi(s_A,s_B)=\lim_{q \uparrow 1}\chi_q(s_A,s_B)}$, where
 \begin{equation}
\label{chieq}
 \chi_q(s_A,s_B)=\Pr\{Y(s_B) > F^{-1}_{Y(s_B)}(q)| Y(s_A) > F^{-1}_{Y(s_A)}(q)\}.
 \end{equation}
 Positively associated max-stable, or Pareto, processes are asymptotically dependent \citep{Coles2001book} at all spatial distances, i.e., $\chi(s_A,s_B) > 0$ for all $s_A, s_B \in \mathcal{S}$. These models are then unable to account for cases where we have positive association, but $\chi(s_A,s_B) = 0$ for some $s_A,s_B \in \mathcal{S}$ which holds for all Gaussian processes when $s_A \neq s_B$; we refer to this scenario as asymptotic independence, i.e., the tendency for extreme events to occur increasingly independently as the magnitude of the events gets larger. 
\par
\cite{wadsworth2018spatial} have developed a conditional approach to spatial extremes which allows for both asymptotic dependence and asymptotic independence at different spatial distances. They provide a spatial extension of the multivariate \cite{heff2004} model, which enables the modelling of processes given that the process is extreme at least at one location. Dependence parameters within the \cite{heff2004} model are represented as smooth functions, parametric or splines, of distance between variables at the site of interest and the conditioning site, and the residual process is driven by a latent Gaussian process, see Section \ref{sec-depmodel}. Extensions of this approach are given by  \cite{shooterinpress,simpson2020highdimensional,huser2020advances,simpson2020conditional}. These papers provide alternative parametric norming functions, implement Bayesian inference, spatio-temporal extensions, and involve small numbers of sampling locations ($d < 300$), so that full inference is computationally feasible. One exception is \cite{simpson2020highdimensional} who detail an inference approach using INLA for much larger $d$; however, this imposes restrictions on the dependence structure parameters that are not appropriate in our application. Across the papers, they perform  extremal modelling of air and sea temperature fields and spatial wave heights, each of which is spatially smoother than precipitation and are without the issue of zeros. In particular, their primary descriptor of the extremal dependence comes through a location function, see \eqref{alphaeq}.
\par
Compared to the use of limit models for $R_\mathcal{A}$, a much better sub-asymptotic approximation is likely to come from modelling the margins and copula of $\{Y(s)\}$ and deriving the high quantile behaviour of $R_\mathcal{A}$ through computational methods, such as simulation. This is the approach we take but for $\{Y(s)\}$ as a continuous spatial process. We build on the spatial conditional extremes methods and adapt this approach for modelling extreme precipitation fields, i.e., addressing the issues caused by the zeros in the field. We present a novel and flexible solution to the problem of modelling the extremes of spatial aggregated variables and provide improvements to the modelling and inference of spatial extreme fields generally. Using this spatial model we construct a sub-asymptotic tail model for $R_\mathcal{A}$. Our approach overcomes the problems of the existing univariate method with ordering $R_\mathcal{A}$ over different $\mathcal{A}$ and dealing with partially missing data, and is more flexible than approaches that restrict the spatial dependence in $\{Y(s)\}$ to be either asymptotic dependence or asymptotic independence everywhere or impose that $\xi(s)$ is constant over $s \in \mathcal{A}$.
We find that when using Monte-Carlo methods to approximate $\Pr\{R_\mathcal{A} > r\}$, that the position and size of $\mathcal{A}$ within $\mathcal{S}$ are important considerations in the inference for $R_\mathcal{A}$, which require new methods to detect and account for edge effects.
Due to the high value of $d$ in our application, there are serious problems with evaluation of likelihood and confidence interval construction for quantiles of $R_\mathcal{A}$, so we adapt recently proposed stratified sampling schemes for model fitting. We also propose new parametric forms for the spatial dependence functions. We observe high variability in the fitted model for $\{Y(s)\}$ as we move further away from the centre of an event, which corresponds to the observed roughness in events that generate extreme precipitation, which unlike applications for other physical processes requires specific care with modelling the scale function, see \eqref{betaeq}.  
We find strong indication that our model for $\{Y(s)\}$ fits well and that we are able to handle comfortably the structure of observed extremes fields, including their zeroes, and estimate very well the distribution of $R_\mathcal{A}$ for our data.
\par
The layout of this paper is as follows: Section \ref{sec-model} describes our model for the process $\{Y(s)\}$. We describe methods for model inference and simulation of events in Section \ref{sec-infer}, which includes our censoring technique for handling zero values. In Section \ref{sec-appl} we discuss the marginal and dependence model fits for the precipitation data, and inference on the tail behaviour of spatial aggregates of these data. We compare the results from our approach with those using GPD fitted to the sample aggregates and using a spatial asymptotically dependent model in Section \ref{sec_diag}. We end with further discussion and model extensions in Section~\ref{sec-dicuss}.
\section{Modelling the extremes of the spatial process}
\label{sec-model}
\subsection{Marginal Model}
\label{sec-margmodel}
The site-wise marginals of $\{Y(s)\}$ can be modelled using a fitted GPD above some high threshold and the empirical distribution below \citep{Coles2001book}. We extend this approach by incorporating a third component, which we denote $p(s)$, that describes the probability that there is no rain at site $s$.  The marginal distribution function of $Y(s)$ for each $s \in \mathcal{S}$ is
\begin{equation}
\label{MargTransform}
F_{Y(s)}(y)=
\begin{cases}
p(s)&\text{if}\;\;y = 0\\
\frac{1-\lambda(s)-p(s)}{F_{Y_+(s)}(q(s))}F_{Y_+(s)}(y)+p(s)&\text{if}\;\;0 < y \leq q(s)\\
1-\lambda(s)\left[1+\frac{\xi(s)(y-q(s))}{\upsilon(s)}\right]^{-1/\xi(s)}_+ &\text{if}\;\;y > q(s),\\
\end{cases}
\end{equation}
where $\upsilon(s) > 0$ and $F_{Y_+(s)}(y)$ denotes the distribution function of strictly positive values of $Y(s)$ and $p(s) \geq 0,\;\lambda(s) > 0$ and $ p(s) + \lambda(s) < 1$; this ensures that the marginal distribution is continuous across components. We expect spatial smoothness over $F_{Y(s)}$ and so define full spatial models for the three components of \eqref{MargTransform} which also enable us to make inference about $F_{Y(s)}$ for all $s \in \mathcal{S}$, i.e., including where $Y(s)$ is not observed.
\par
We first consider the distribution of $Y(s)$ above $q(s)$. Following the approach of \cite{youngman}, we fit a generalised additive GPD model (GAM) to exceedances $Y(s)-q(s)$. This allows us to represent the GPD scale and shape parameters, $\upsilon(s)$ and $\xi(s)$, respectively, through a basis of smooth splines. We set $\lambda(s)= \lambda$ for all $s \in \mathcal{S}$, allowing us to estimate $q(s)$ for $s \in \mathcal{S}$. We use a non-parametric approach, and simply fit a thin-plate spline to point-wise estimates of $q(s)$ for the associated $\lambda$, as the parametric method of \cite{youngman} fails as we have point masses below $q(s)$, caused by rounding of data. \par
We estimate $p(s)~=\Pr\{Y(s)=0\}$ for $s \in \mathcal{S}$ as a spatially smooth surface by using a logistic GAM \citep{wood2006generalized}; that is, we fit ${ \text{logit}\{\mathbb{E}[p(s)]\}= g(s)}$ where $g(\cdot)$ is a smoothing spline. The degree of smoothness in the surface $p(s)$ is determined by the choice of spline used for $g$. We estimate $F_{Y_+(s)}$ using the empirical distribution of strictly positive values of $Y(s)$, which we denote $\tilde{F}_{Y_+(s)}(\cdot)$. We use the site-wise empirical distribution for fitting and recognise that, should we require simulation of $Y(s)$ for $s \in \mathcal{S}\setminus\mathbf{s}$, we can use additive quantile regression \citep{qgam} to compute the empirical distributions at unobserved locations. This can be performed, if necessary, using only a local neighbourhood of sampling locations, as this method is too computationally expensive for large $d$ and so is not viable without dimension reduction. \par
We use \eqref{MargTransform} to perform a site-wise transformation of the margins of the data to a standardised distribution, albeit with discrete mass in each lower-tail. For modelling dependence within a process $\{X(s):s \in \mathcal{S}\}$ using the \cite{wadsworth2018spatial} conditional extremes framework, we require its margins to have standard exponential upper tails, i.e., $\Pr\{X(s) > x\} \sim C\exp(-x)$ for some $C > 0$, as $x \rightarrow \infty$ and for all $s\in \mathcal{S}$. We follow \cite{wadsworth2018spatial} and use Laplace margins. We handle discrete components in the margins of $\{X(s)\}$ using censoring techniques, see Section~\ref{censorsec}.
\subsection{Dependence Modelling}
\label{sec-depmodel}
\cite{wadsworth2018spatial} model the underlying extremal dependence in our standardised process $\{X(s)\}$, given that it is extreme for some $s \in \mathcal{S}$, by first conditioning on the process being above some high threshold $u$ at a specified site $s_O \in \mathcal{S}$. We introduce the function $h(s_A,s_B)=\|s_A-s_B\|$ for $s_A, s_B \in \mathcal{S}$, where $\|\cdot\|$ is some distance metric (we use the anisotropic measure \eqref{anisoeq}). Under the assumption that there exists normalising functions $\{a: (\mathbb{R},\mathbb{R}_+) \rightarrow \mathbb{R}\}$, with $a(x,0) = x$,  and $\{b: (\mathbb{R},\mathbb{R}_+) \rightarrow (0,\infty)\}$, such that as $ u\rightarrow \infty$, they assume that for each $s_O \in \mathcal{S}$ 
\begin{equation}
\label{depmodeleq}
\left(\left\{\frac{X(s)-a\{X(s_O),h(s,s_O)\}}{b\{X(s_O),h(s,s_O)\}}:s \in \mathcal{S}\right\},X(s_O)-u\right)\Bigg|\bigg(X(s_O)>u\bigg)\xrightarrow{d} \Bigg(\bigg\{Z(s|s_O):s \in \mathcal{S}\bigg\},E\Bigg),
\end{equation}
where $E$ is a standard exponential variable and process $\{Z(s|s_O)\}$ which is non-degenerate for all $s \in \mathcal{S}$ where $s \neq s_O$. That is, there is convergence in distribution of the normalised process to $\{Z(s|s_O) : s \in \mathcal{S}\}$, termed the residual process, which is independent of $E,$ and $Z(s_O|s_O) = 0$ almost surely. Characterisations of the normalising functions, $a$ and $b$ and the residual process $Z(s|s_O)$ are given in Sections \ref{normfuncs} and \ref{residprocess}, respectively. \par
To make inference on the upper tail of $R_\mathcal{A}$ for any $\mathcal{A}\subset \mathcal{S}$ we require the process $\{X(s)\}$ given an extreme value somewhere in the domain $\mathcal{S}$, i.e., 
\begin{equation}
\label{eqn:JT}
\bigg\{X(s):s \in \mathcal{S}\bigg\}\bigg|\left(\max_{s \in \mathcal{S}}X(s) > u\right)
\end{equation}
for large $u$.  Limit~\eqref{depmodeleq} conditions only on observing an exceedance at a specific site $s_O \in \mathcal{S}$, so cannot be immediately used. However, this limit provides a core building block for what is required when combined with a limiting model for ${\{ X(s_O)>u : s_O \in \mathcal{S} \}|(\max_{s \in \mathcal{S}}X(s) > u)}$ as $u \rightarrow \infty$. For a stationary process, this limiting model will be invariant to $s_O$ for all $s_O \in \mathcal{S}$ which are sufficiently far from the boundaries of $\mathcal{S}$. \cite{wadsworth2018spatial} show how simulation from process~\eqref{eqn:JT} can be achieved through an importance sampling method; the outline of this is given in Section \ref{sim}.\par
For the process $\{X(s)\}$ to exhibit long-range independence over $\mathbb{R}^2$, we need conditions on $a$ and $b$ and $Z(s|s_O)$ so that independence is achieved as $h=h(s,s_O)\rightarrow \infty$ for any $s_O \in \mathcal{S}$ and suitably distanced $s \in \mathcal{S}$. This requires for any fixed $x > 0$, that $a(x,h) \rightarrow 0$ and $b(x,h)\rightarrow 1$ as $h \rightarrow \infty$. Further, the residual process $Z(s|s_O)$ must have identical margins to $X(s)$ as $h\rightarrow \infty$; in particular, we require standard Laplace margins for  $Z(s|s_O)$ as $h \rightarrow \infty$.
\subsubsection{Normalising functions}
\label{normfuncs}
For inference, we assume parametric forms for the location and scaling functions $a$ and $b$, respectively, in {limit \eqref{depmodeleq}}. \cite{wadsworth2018spatial} provide a discussion of these normalising functions and provide some suggestions for their possible parametric forms. We considered the range of parametrics forms discussed by \cite{wadsworth2018spatial,shooterinpress}, but for brevity we report only the models that provided the best fit. We let
\begin{equation}
\label{alphaeq}
a(x,h)=x\alpha(h), \;\;\;\text{with}\;\;\;\alpha(h)=\begin{cases}1, &h \leq \Delta,\\
\exp(-\{(h - \Delta)/\kappa_{\alpha_1}\}^{\kappa_{\alpha_2}}), &h > \Delta,
\end{cases}
\end{equation}
where $\Delta \geq 0$ and $\kappa_{\alpha_1},\kappa_{\alpha_2} > 0$ which allows $\{X(s)\}$ to be asymptotically dependent up to distance $\Delta$ from $s_O$, and asymptotically independent thereafter. We also take 
\begin{equation}
\label{betaeq}
b(x,h)=x^{\beta(h)}, \;\;\;\text{with}\;\;\;\beta(h)=
\kappa_{\beta_3}\exp(-\{h/\kappa_{\beta_1}\}^{\kappa_{\beta_2}})
\end{equation}
for $\kappa_{\beta_1},\kappa_{\beta_2} > 0$ and $\kappa_{\beta_3} \in [0,1]$, and so $b(0,x) = x^{\kappa_{\beta_3}}$. Long-range independence holds for $\{X(s)\}$, as $a(x,h) \rightarrow 0$ and $b(x,h) \rightarrow 1$ as $h \rightarrow \infty$ for fixed $x > 0$, whatever the parameters.
\subsubsection{Residual process $\{Z(s|s_O)\}$}
\label{residprocess}
We follow \cite{shooterinpress} by imposing that the residual process $\{Z(s|s_O)\}$ has delta-Laplace margins; a random variable follows a delta-Laplace distribution, i.e., $DL(\mu,\sigma,\delta)$, with location, scale and shape parameters $\mu \in \mathbb{R}$, $\sigma >0$ and $\delta >0$, respectively, if its density is
\begin{equation}
\label{deltalapfunc}
f(z)=\frac{\delta}{2k\sigma\Gamma\left(\frac{1}{\delta}\right)}\exp\left\{-\left\lvert\frac{z-\mu}{k\sigma}\right\rvert^\delta\right\},\;\;\;(z \in \mathbb{R})
\end{equation}
with $\Gamma(\cdot)$ as the standard gamma function and $k^2=\Gamma(1/\delta)/\Gamma(3/\delta)$. The scaling by $k$ is used to improve identifiability between $\sigma$ and $\delta$, as the random variable has expectation $\mu$ and variance $\sigma^2$ regardless of the value of $\delta$. Use of the delta-Laplace distribution introduces flexibility in the marginal choice for $Z(s|s_O)$, as for $\delta = 1$ or $2$, we have the Laplace or Gaussian distributions respectively. As with the normalising functions, we parametrise the delta-Laplace parameters as smooth functions of distance from the conditioning site $s_O$. That is, $Z(s|s_O)\sim \mbox{DL}( \mu\{h(s,s_O)\},\sigma\{h(s,s_O)\}, \delta\{h(s,s_O)\})$, with
\begin{align}
\label{sigmudeltaeqs}
\mu(h)&=\kappa_{\mu_1}h^{\kappa_{\mu_2}}\exp\{-h/\kappa_{\mu_3}\},&(\kappa_{\mu_2} >0, \kappa_{\mu_3} >0),\\
\sigma(h)&=\sqrt{2}\left(1-\exp\{-(h/\kappa_{\sigma_1})^{\kappa_{\sigma_2}}\}\right),&(\kappa_{\sigma_1}>0, \kappa_{\sigma_2} >0),\nonumber\\
\delta(h)&=\max\{1,1+(\kappa_{\delta_1}h^{\kappa_{\delta_2}}-\kappa_{\delta_4})\exp\{-h/\kappa_{\delta_3}\}\},\;\;\;&(\kappa_{\delta_1} \geq 0, \kappa_{\delta_2} >0, \kappa_{\delta_3} >0, \kappa_{\delta_4}<1),\nonumber
\end{align}
for $h \geq 0$. These functions satisfy the constraint that $\mu(0)=\sigma(0) = 0$, which ensures that $Z(s_O|s_O) = 0$ holds and provides a flexible modelling choice for $\delta$. Use of the $max$ operator in $\delta$ ensures that the tails of $Z(s|s_O)$ are not heavier than $X(s)$, i.e., $\delta\{h(s,s_O)\} \geq 1$ for all $s \in \mathcal{S}$. Furthermore, long-range independence in $X(s)$ is achieved as $\mu(h) \rightarrow 0$, $\sigma^2(h) \rightarrow 2$, and $\delta(h) \rightarrow 1$ as $h \rightarrow \infty$, where the variance of a standard Laplace random variable is 2.\par
Following the approach of \cite{shooterinpress}, dependence in $\{Z(s|s_O)\}$ is induced by first considering the process $\{W(s|s_O)\}=\{W(s)|(W(s_O)=0)\}$ for all $s \in \mathcal{S}$, where $\{W(s)\}$ is a standard stationary Gaussian process with correlation function $\rho(h)$. We set ${\{Z(s|s_O)\}=\{F^{-1}_{Z(s|s_O)}\{\Phi[W(s|s_O)]\}\}}$ for all $s \in \mathcal{S}$, where $\Phi(\cdot)$ and $F_{Z(s|s_O)}$ are the CDFs of a standard Gaussian distribution and $Z(s|s_O)$, respectively. The corresponding density function to $F_{Z(s|s_O)}$ is $f_{Z(s|s_O)}$, defined by \eqref{deltalapfunc} and \eqref{sigmudeltaeqs}. \par
To illustrate the dependence in $\{Z(s|s_O): s \in \mathcal{S}\}$, we consider the joint distribution of $Z(s|s_O)$  in a finite-dimensional setting, which we achieve by using a Gaussian copula model. Consider any $s_O \in (s_1,\dots,s_d)$, and without loss of generality, rewrite the sampling locations as $s_O,(s_1,\dots,s_{d-1})$, i.e., here we illustrate with $s_O = s_d$. The joint distribution of $\{Z(s_1|s_O),\dots,Z(s_{d-1}|s_O)\}$ for $s_O$ is, for $\mathbf{z}=(z_1,\dots,z_{d-1})$,
\begin{equation}
\label{gausscopeq}
F_{s_O}(\mathbf{z})=\Phi_{d-1}\left\{\Phi^{-1}(F_{Z(s_1|s_O)}(z_1)),\dots,\Phi^{-1}(F_{Z(s_{d-1}|s_O)}(z_{d-1}));\mathbf{0},\Sigma\right\},
\end{equation}
where $\Phi_{d-1}(\cdot; \mathbf{0},\Sigma)$ is the CDF of a $(d-1)-$dimensional Gaussian distribution with mean $\mathbf{0}$. The correlation matrix $\Sigma$ must account for the conditioning $W(s)|(W(s_O) = 0)$. To create $\Sigma$, we initialise a stationary correlation matrix $\Sigma^*$ using correlation function $\rho(\cdot)$ evaluated for all pairwise distances, and we condition on observing $W(s_O)=0$. That is, the correlation matrix $\Sigma$ has $(i ,j)$-th element
\begin{equation}
\label{cormat}
\Sigma_{ij}=\frac{\Sigma^*_{ij}-\Sigma^*_{i0}\Sigma^*_{j0}}{(1-\Sigma^{*2}_{i0})^{1/2}(1-\Sigma^{*2}_{j0})^{1/2}}.
\end{equation}
Note the elements of $\Sigma$ are normalised such that the diagonal elements are equal to one. In our application, $\rho(\cdot)$ is taken to be the Mat\'ern correlation function
\[
\rho(h)=\frac{2^{1-\kappa_{\rho_2}}}{\Gamma(\kappa_{\rho_2})}\left(\frac{2h\sqrt{\kappa_{\rho_2}}}{\kappa_{\rho_1}}\right)^{\kappa_{\rho_2}}K_{\kappa_{\rho_2}}\left(\frac{2h\sqrt{\kappa_{\rho_2}}}{\kappa_{\rho_1}}\right),\;\;\;\;(\kappa_{\rho_1} > 0, \kappa_{\rho_2} >0),
\]
where $K_{\kappa_{\rho_2}}(\cdot)$ is the modified Bessel function of the second kind of order $\kappa_{\rho_2}$. \par 
To account for spatial anisotropy in the extremal dependence structure of $\{X(s)\}$ we use the transformation of coordinates
\begin{equation}
\label{anisoeq}
s^*=\begin{pmatrix}
1 & 0\\ 
0 & 1/L \end{pmatrix}\begin{pmatrix}
\cos\theta & -\sin\theta \\
\sin\theta & \cos\theta \end{pmatrix}s,
\end{equation}
where $\theta \in [-\pi/2, 0]$ controls rotation and $L > 0$ controls the coordinate stretching effect; with $L=1$ recovering the isotropic model. We define our distance metric $\|s_A-s_B\| = \|s^*_A-s^*_B\|_*$, where $\|\cdot\|_*$ denotes great-circle, or spherical, distance.
\subsubsection{Extensions for censored precipitation data}
\label{censorsec}
The non-zero probability of zeroes for precipitation $Y(s)$ causes the $X(s)$, for all $s \in \mathcal{S}$, to have non-zero mass at a finite lower endpoint. Consequently, the Gaussian copula and delta-Laplace marginal model described in \eqref{gausscopeq} are not appropriate for the transformed precipitation data in these lower tail regions. To circumvent this issue, we apply censoring at all points where ${Y_t(s) = 0}$ for all $s \in \mathcal{S}$ and $t = 1,\dots,n$; an approach adopted by \cite{benoit2018stochastic}, albeit for small $d$. A spatially-varying censoring threshold $c(s)$ is attained by transforming $p(s)$ in Section \ref{sec-margmodel} to the Laplace scale, using $c(s) = F_L^{-1}\{p(s)\}$, where $F_L(\cdot)$ is the standard Laplace CDF. We then assert that ${Y_t(s) = 0 \Leftrightarrow X_t(s) \leq c(s)}$, which in turn implies that $Z_t(s|s_O) \leq c_t^{(s_O)}(s)$ where $c_t^{(s_O)}(s)$ is dependent on the value observed at the conditioning site and given in \eqref{residequation}. \par
For inference, the number of censored components varies at each time point; the number, locations and censoring values can all vary, with a maximum value of $d-1$ locations with censoring. If the number of censored components is large, it is clear from \eqref{gausscopeq} that evaluation of a censored distribution function will be computationally expensive. We take a pseudo-likelihood approach to inference, which we detail in Section \ref{modelfit}; this requires only a bivariate density. We also detail its multivariate analogue; although this is not used for inference, it has a variety of uses, i.e., infilling of extreme events with missing observations or inference at sites ${s \in \mathcal{S}\setminus \mathbf{s}}$. These features are detailed in Appendix~\ref{censorsec-append} of the Supplementary Materials \citep{papersup}.
\section{Inference and Simulation}
\label{sec-infer}
\subsection{Model Fitting}
\label{modelfit}
Our censored triplewise likelihood approach for model fitting is based on the pseudo-likelihood approach of \cite{padoan2010likelihood}; their pairwise approach provides unbiased estimation of model dependence parameters. Recall that some observations are right-censored at different sampling locations with varying rate of occurrence over time.
 To define a single likelihood contribution at time $t$, we begin by considering a single conditioning site amongst the observed sites $s_i \in (s_1,\dots,s_d)$ such that $y_t(s_i) > F^{-1}_{Y(s_i)}\{F_L(u)\}$ and hence $x_t(s_i) > u$. We then define the set of all such times by $\mathcal{T}^{(s_i)}~=\{t= ~1,\dots, n: x_t(s_i)\geq u\}$. For the observed sites, we define $h_{i,j} = h(s_i,s_j)$ for $i,j = 1,\dots,d$ with $ i \neq j $. Then for each site $s_j\;,j=1,\dots,d, j \neq i$, we define the residual for time $t \in \mathcal{T}^{(s_i)}$ for conditioning site $s_i$ as
\begin{equation}
\label{residequation}
z_t^{(s_i)}(s_j)=\begin{cases}
[x_t(s_j)-a\{x_t(s_i),h_{i,j}\}]/b\{x_t(s_i),h_{i,j},\} \;\;&\text{if}\;\;x_t
(s_j) > c(s_j),\\
c^{(s_i)}_{t}(s_j), \;\;&\text{otherwise,}
\end{cases}
\end{equation}
where $a$ and $b$ are described in Section \ref{normfuncs}, and $
{c^{(s_i)}_{t}(s_j)=\frac{c(s_j)-a\{x_t(s_i),h_{i,j}\}}{b\{x_t(s_i),h_{i,j}\}}}$ is the censored residual for site $s_j$ with conditioning site $s_i$ and $t \in \mathcal{T}^{(s_i)}$. The full pseudo-likelihood is given, for residuals $\mathbf{z}_t^{(s_i)}$ with conditioning site $s_i$ and parameter vector $\boldsymbol{\psi}$, by
\begin{align}
\label{fulllik}
L_{CL}(\boldsymbol{\psi})&=\prod^d_{i=1}\prod_{t \in \mathcal{T}^{(s_i)}}L^{s_i}_{CL}(\boldsymbol{\psi};\mathbf{z}^{(s_i)}_t)=\prod^d_{i=1}\prod_{t \in \mathcal{T}^{(s_i)}}\prod_{\forall j<k ; j \& k \neq i}\frac{g_{s_i}(z^{(s_i)}_t(s_j),z^{(s_i)}_t(s_k),c^{(s_i)}_{t}(s_j),c^{(s_i)}_{t}(s_k))}{J(z^{(s_i)}_t(s_j),z^{(s_i)}_t(s_k),c^{(s_i)}_{t}(s_j),c^{(s_i)}_{t}(s_k))},
\end{align}
where $L_{CL}^{s_i}$ is the censored likelihood contribution for $s_i$ and the bivariate density $g$ is defined in Appendix~\ref{censorsec-append} of the Supplementary Materials \citep{papersup}. The Jacobian term is
\[
J\left(z^{(s_i)}_t(s_j),z^{(s_i)}_t(s_k),c^{(s_i)}_{t}(s_j),c^{(s_i)}_{t}(s_k)\right)=b\{x_t(s_i),h_{i,j}\}^{\mathrm{1}\{z_t(s_j)>c^{(s_i)}_{t}(s_j)\}}b\{x_t(s_i),h_{i,k}\}^{\mathrm{1}\{z_t(s_k)>c^{(s_i)}_{t}(s_k)\}},
\]
where $\mathrm{1}\{\cdot\}$ is the indicator function.
The parameter vector $\boldsymbol{\psi}$ contains all parameters of the normalising functions $a$ and $b$, the residual parameter functions $\mu,\sigma$ and $\delta$, the correlation $\rho$ and the anisotropy. Estimation of $\boldsymbol{\psi}$ can be achieved by maximising \eqref{fulllik}. 
\subsection{Stratified sampling regime}
\label{sec-strat}
Clearly, maximising \eqref{fulllik} is computationally infeasible if $d$ is large, as evaluation of \eqref{fulllik} requires $(d-1)(d-2)\sum^d_{i=1}|\mathcal{T}^{(s_i)}|/2$ evaluations of $g_{(s_i)}$, which can require double integrals and grows as $O(d^3n)$. We detail a stratified sampling regime to create a pseudo-likelihood that circumvents the computational issue. To construct a sub-sample of data that can be used to estimate the parameters $\boldsymbol{\psi}$ via pseudo-likelihood estimation, we first need to consider what these parameters represent. These parameters control characteristics of the dependence functions described in Sections~\ref{normfuncs} and \ref{residprocess}, which are functions of either distance to the conditioning site $h_{i,j}$ or pairwise distances $h_{j,k}$ for $s_i,s_j,s_k \in \mathbf{s}$. Thus, we construct our sub-sample by drawing triples of sites $(s_i,s_j,s_k) \in \mathbf{s}$;
 there are $d(d-1)(d-2)/2$ possible triples of sites, and our sub-sample must adequately represent the distribution of the distances in the full data. However, not all distances can be represented in the sub-sample. If we pick triples randomly, then we are more likely to pick sites with larger pairwise distances. This has two disadvantages: pairs with larger distances are not informative about the dependence parameters, as at these distances the process may exhibit near-independence, and we are also unlikely to learn about the dependence for small distances, as less pairs with smaller pairwise distances are sampled. To ensure this is not the case, each triple $(s_i,s_j,s_k)$ is chosen so that the distances $h_{i,j}, h_{i,k}$ and $h_{j,k}$ do not exceed a specified threshold. This is a natural extension of the approach of \cite{Huser2013} who suggest using only pairs of locations that are within some low distance $h_{max} > 0$ of each other - we instead impose this constraint on triples. We investigate the reliability of our approach in Appendix~\ref{append-infersimstudy} of the Supplementary Materials \citep{papersup}.\par
To sub-sample $d_s \ll d(d-1)(d-2)/2$ triples of locations for inference, we begin by uniformly sampling a conditioning site $s_i \in (s_1,\dots, s_d)$. A pair of sites are then drawn randomly from the set ${\{s_j,s_k \in \mathbf{s}\setminus s_i, j < k : \max\{h_{i,j},h_{i,k}\} < h_{max}\}}$ without replacement, and the process is repeated. In sampling in this way, only sites within distance $h_{max}$ of the conditioning site are used for inference. That is, we estimate the spatial functions of the dependence parameters for $h_{i,j}< h_{max}$ and $h_{i,k}< h_{max}$, and so $h_{j,k} < 2h_{max}$ only, and then extrapolate to larger distances. There is a trade-off involved in choosing the value of $h_{max}$: if too low, then extrapolations to larger distances are likely to be poor; if too high, fit at small distances is compromised. In Section~\ref{sec-appl-dep}, we describe a heuristic technique for choosing $h_{max}$ and we find that this works well in practice. The number of triples $d_s$ is chosen to be as large as possible, whilst pseudo-likelihood estimation remains computationally feasible. 
\subsection{Simulation of an event}
\label{sim}
We now detail a technique that will allow us to draw realisations of $\{Y(s): s \in \mathcal{S}\}$. First, we note that the model in Section \ref{sec-depmodel} does not describe the dependence in all of $\{Y(s)\}$; instead, it describes
\begin{equation}
\label{simexp}
\bigg\{Y(s):s \in \mathcal{S}\bigg\}\bigg|\left( \max\limits_{s \in\mathcal{S}}\left\{ F^{-1}_L(F_{Y(s)}\{Y(s)\})\right\} > v\right) \equiv \bigg\{F^{-1}_{Y(s)}(F_{L}\{X(s)\}):s \in \mathcal{S}\bigg\}\bigg|\left( \max\limits_{s \in\mathcal{S}}\left\{X(s)\right\} > v\right),
\end{equation}
for $v \geq u$ with $u$ used for fitting in Section \ref{modelfit}. Thus, to create a realisation of $\{Y(s): s \in \mathcal{S}\}$, we draw realisations of \eqref{simexp} with probability
\begin{equation}
\label{probmax}
\Pr\left\{ \max\limits_{s \in\mathcal{S}}\left\{ F^{-1}_L(F_{Y(s)}\{Y(s)\})\right\} > v\right\},
\end{equation}
and otherwise draw realisations of 
\begin{equation}
\label{empiricalobs}
\bigg\{Y(s):s \in \mathcal{S}\bigg\}\bigg|\left( \max\limits_{s \in\mathcal{S}}\left\{ F^{-1}_L(F_{Y(s)}\{Y(s)\})\right\} < v\right).
\end{equation}
As we do not expect realisations of \eqref{empiricalobs} to contribute to the tail behaviour of $R_\mathcal{A}$, we simply draw realisations of \eqref{empiricalobs} from the observed data. We estimate \eqref{probmax} empirically; although this could be inferred using the parametric model of Section \ref{sec-model}. If $\mathcal{S}$ does not correspond to the set of sampling locations, then we would have to approximate \eqref{empiricalobs} though some form of infilling, i.e., using the quantile regression technique \citep{qgam} discussed in Section \ref{sec-margmodel}.\par
We now describe a simulation technique that will allow us to draw realisations of \eqref{simexp}. That is, the field $\{Y(s):s \in \mathcal{S}\}$ given that an extreme value above a threshold is observed anywhere in the domain. This threshold varies with $s$ and corresponds to the relative quantile $v$ on the Laplace scale. \cite{wadsworth2018spatial} detail
the procedure for achieving this. There are three steps: drawing conditioning sites $s_O \in \mathcal{S}$, simulating the fields ${\{Y(s): s \in \mathcal{S}\}|( F^{-1}_L(F_{Y(s_O)}\{Y(s_O)\}) > v)}$ using the fitted model described in Section \ref{sec-model}, and then using importance sampling to approximate \eqref{simexp}, see Algorithm~\ref{sim-algo}, Step \ref{import-op}. The first step requires random sampling of conditioning sites $s_O$ for some $s_O \in \mathcal{S}$; we do this uniformly, which provides a good first approximation of the occurrence of these sites in $\mathcal{S}$ and then improve on this via the importance sampling regime described below. \par
To simulate $N$ realisations from process \eqref{simexp}, we follow \cite{wadsworth2018spatial} and draw an initial $N' > N$ realisations of the process $\{X(s): s \in \mathcal{S}\}$ on the Laplace scale. Then, using importance sampling, we sub-sample $N$ realisations from ${\{X(s): s \in \mathcal{S}\}|\max_{s \in \mathcal{S}}X(s) > v}$, and transform the margins of the sample to the original scale, $\{Y(s)\}$. The sub-sampling regime adds extra weight to realisations for which the conditioning site is near the boundary of the domain. This is to alleviate the edge effect caused by not using conditioning sites outside of the boundaries of $\mathcal{S}$. A discussion of a related issue is given in Section \ref{sec-agg-infer}. We found that setting $N' \approx 5N$ was sufficient for our application, although this may be dependent on the size of $\mathcal{S}$ and value of $N$. 
\begin{algorithm}[h]
\caption{Simulating \eqref{simexp}}
\label{sim-algo}
\begin{enumerate}
\item For $i = 1,\dots, N'$ with $N' > N$: \begin{enumerate}
\item Draw a conditioning location $s^{(i)}_O$ from $\mathcal{S}$ with uniform probability density $1/|\mathcal{S}|$.
\item Simulate $E^{(i)} \sim Exp(1)$ and set $x_i(s^{(i)}_O) = v + E^{(i)}$.
\item Simulate a field $\{z_i(s|s_O^{(i)}): s \in \mathcal{S}\}$ from the residual process model defined in Section \ref{residprocess}.
\item Set $\{x_i(s): s \in \mathcal{S}\}=a\{x_i(s^{(i)}_O),h(s,s^{(i)}_O)\}+b\{x_i(s^{(i)}_O),h(s,s^{(i)}_O)\}\times \{z_i(s|s_O^{(i)}): s \in \mathcal{S}\}$.
\end{enumerate}
\item Assign each simulated field  $\{x_i(s): s \in \mathcal{S}\}$ an importance weight of
\[
\left\{\int_\mathcal{S}\mathrm{1}\{x_i(s)>v\}\mathrm{d}s\right\}^{-1},
\]
for $i=1,\dots,N^{'}$, and sub-sample $N$ realisations from the collection with probabilities proportional to these weights. \label{import-op}
\item Transform each $\{x_i(s): s \in \mathcal{S}\}$ to $\{y_i(s): s \in \mathcal{S}\}$ using the marginal transformation \eqref{MargTransform}. If $x_i(s) \leq c(s)$, set $y_i(s)=0$, where for some $s^{'} \in \mathcal{S}$, $y_i(s^{'})$ is above its $F_L(v)$-th quantile.
\end{enumerate}
\end{algorithm}
\subsection{Inference on Spatial Aggregates}
\label{sec-agg-infer}
 Using the sample of realisations of $\{Y(s): s \in \mathcal{S}\}$ generated in Section \ref{sim}, we make inference about the tail behaviour of $R_\mathcal{A}$ in \eqref{aggeq}
or the corresponding sum; here we focus on the latter, but a discussion of the integral is given in Section \ref{sec-dicuss}. The possible size of the aggregation region $\mathcal{A}$ in relation to the region $\mathcal{S}$ is of particular interest. Trivially, we require $\mathcal{A} \subseteq \mathcal{S}$. However, we cannot have $\mathcal{A}=\mathcal{S}$, as if we did, the simulation algorithm will never generate an event for which the conditioning site lies outside of the boundaries of $\mathcal{S}$, but we still observe an extreme event somewhere inside $\mathcal{A}$. To avoid such edge-effects, we require the boundaries of $\mathcal{A}$ to be far enough inside the interior of $\mathcal{S}$, such that the distribution \eqref{simexp} does not change if the size of $\mathcal{S}$ increases. Informally, we require a buffer zone between the boundaries of $\mathcal{A}$ and $\mathcal{S}$ which is large enough, such that any event with conditioning site outside of $\mathcal{S}$ has negligible effect on the distribution within $\mathcal{A}$. We select the width $\tau$ of this buffer zone by using the measure $\chi_q(s_A,s_B)$ given in \eqref{chieq} and stationarity. 
We choose $\tau$ such that for any $s \in \mathcal{A}$ and $s_O \in \mathbb{R}^2\setminus \mathcal{S}$, such that for $h(s,s_O)>\tau$, we have $\chi_q(s_O,s)<\gamma$ for small $\gamma > 0$ and for all large $q$; that is, we have a small probability, less than $\gamma$, of observing a large event at $s$ given that there is an extreme event at any site outside of $\mathcal{S}$. This measure can be evaluated empirically or by simulating from the fitted model; we take the latter approach in Section \ref{sec_diag}.
\section{Application}
\label{sec-appl}
\subsection{Data}
We consider data consisting of average hourly precipitation rate (mm/hour) taken from the UK climate projections 2018 (UKCP18) \citep{kendon2019ukcp}. Data are from a convection permitting model which produces values over hourly intervals between the years 1980 and 2000, using the observed atmospheric conditions. The sampling locations are $(5km)^2$ grid boxes corresponding to the British National Grid from Ordnance Survey (OSGB). The spatial domain $\mathcal{S}$ of interest is East-Anglia, UK (see Figure~\ref{GAMfig}) and only data sampled over land have been included, leaving $934$ sampling locations. Each observation corresponds to the average over the assigned spatio-temporal grid-box. The data represent the average in each grid-box, and so a natural quantity of interest is $\bar{R}_\mathcal{A}:=R_\mathcal{A}/|\mathcal{A}|$, rather than $R_\mathcal{A}$, but we present results on $R_\mathcal{A}$ as this variable must satisfy the ordering constraints discussed in Section~\ref{intro}. To remove any seasonal effect observed in the data, we use summer, i.e., July-August, observations only, leaving $43200$ fields\footnote{Note that the UKCP18 data uses a 360 day calendar, and so each month is composed of 30 days.}. We chose to take summer precipitation events as these typically exhibit higher intensity than winter events \citep{sharkeywinter}.
 We treat the centre of each grid box as a sampling location, and as the grid-boxes are non-overlapping and contiguous, we can approximate the integral $R_\mathcal{A}$ in \eqref{aggeq} using a sum. We use the great-circle distance as our distance metric described in Section~\ref{residprocess}.
\subsection{Marginal Analysis}
\label{sec_appl-marg}
\begin{figure}[h]
\centering
\begin{minipage}{0.32\linewidth}
\centering
\includegraphics[width=\linewidth]{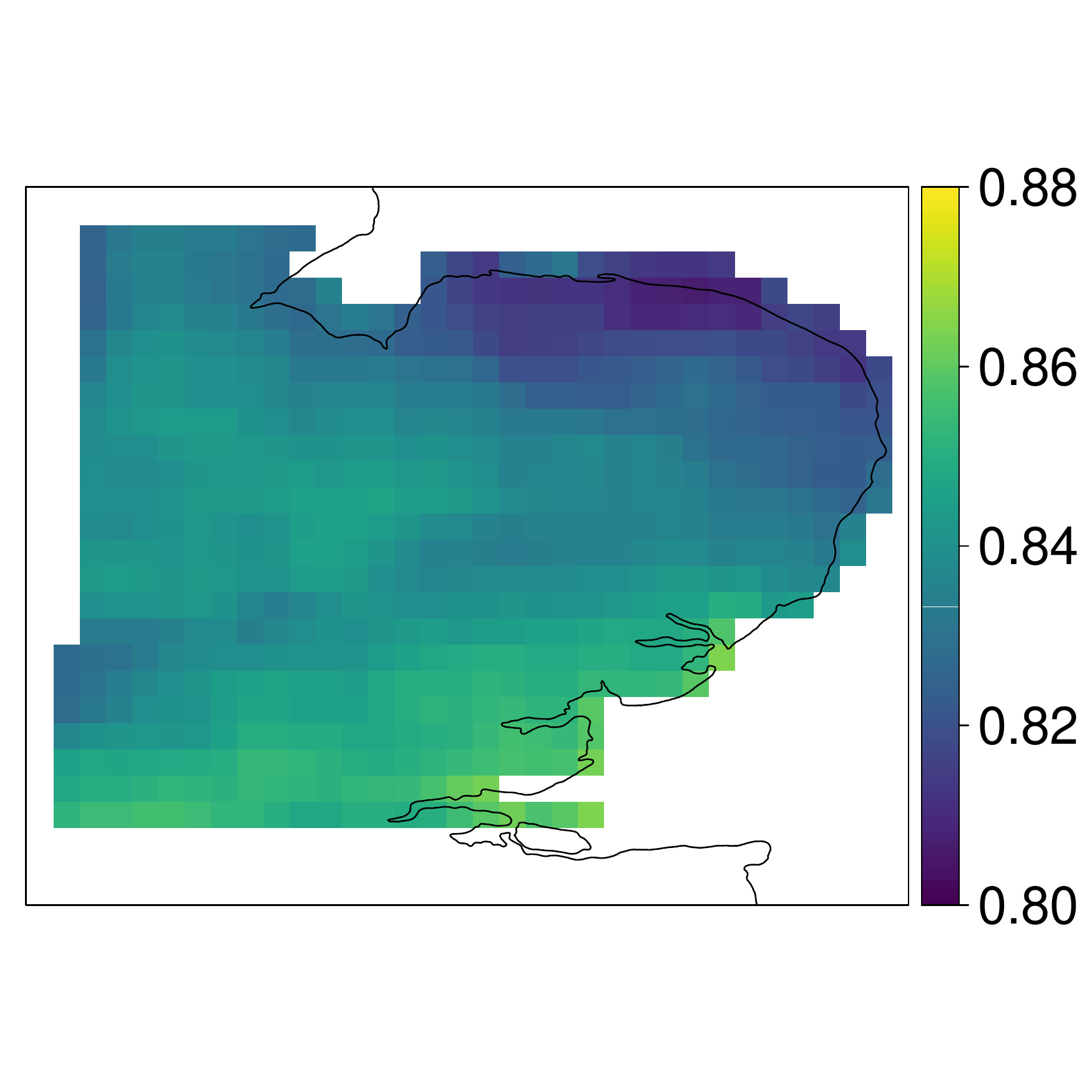} 
\end{minipage}
\begin{minipage}{0.32\linewidth}
\centering
\includegraphics[width=\linewidth]{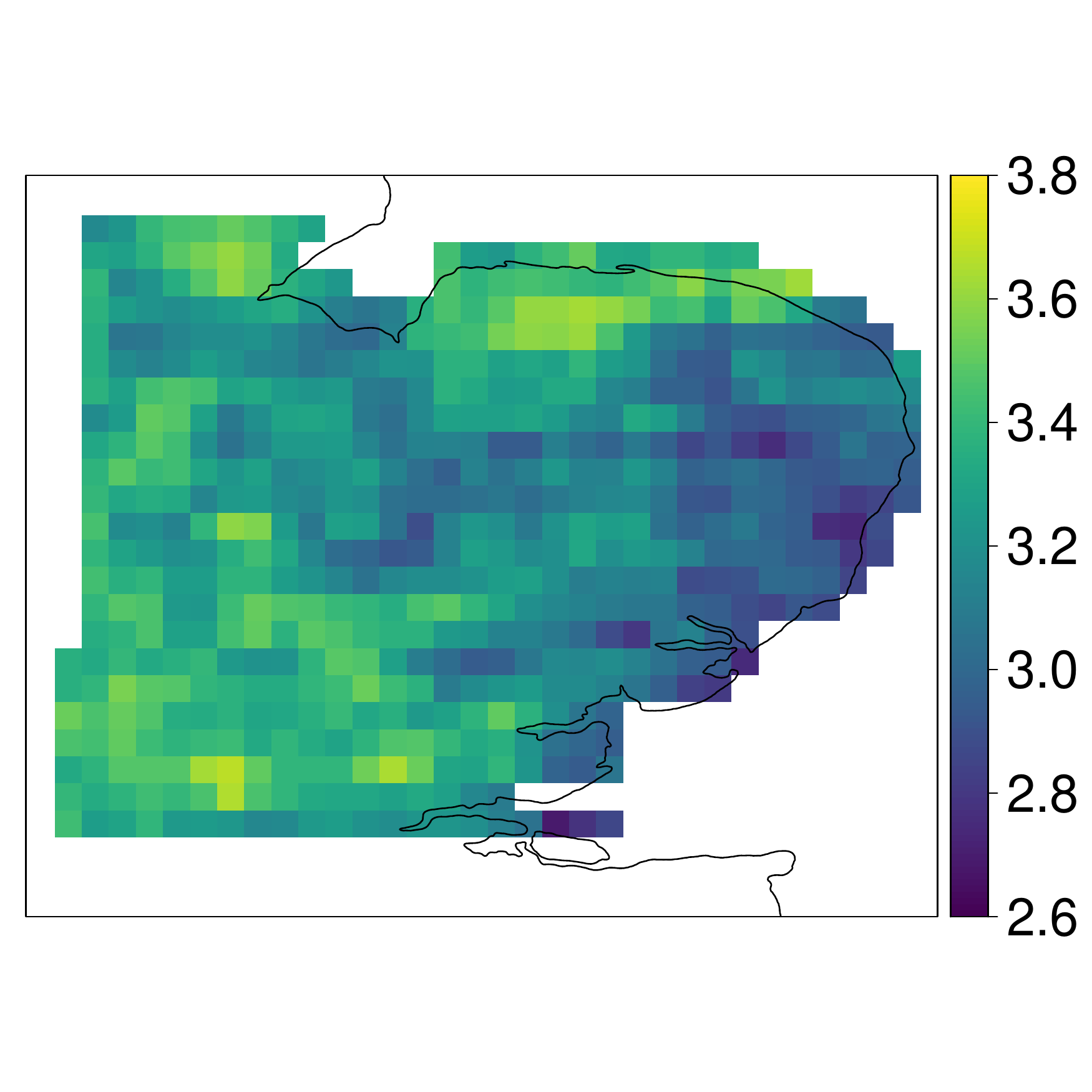} 
\end{minipage}
\begin{minipage}{0.32\linewidth}
\centering
\includegraphics[width=\linewidth]{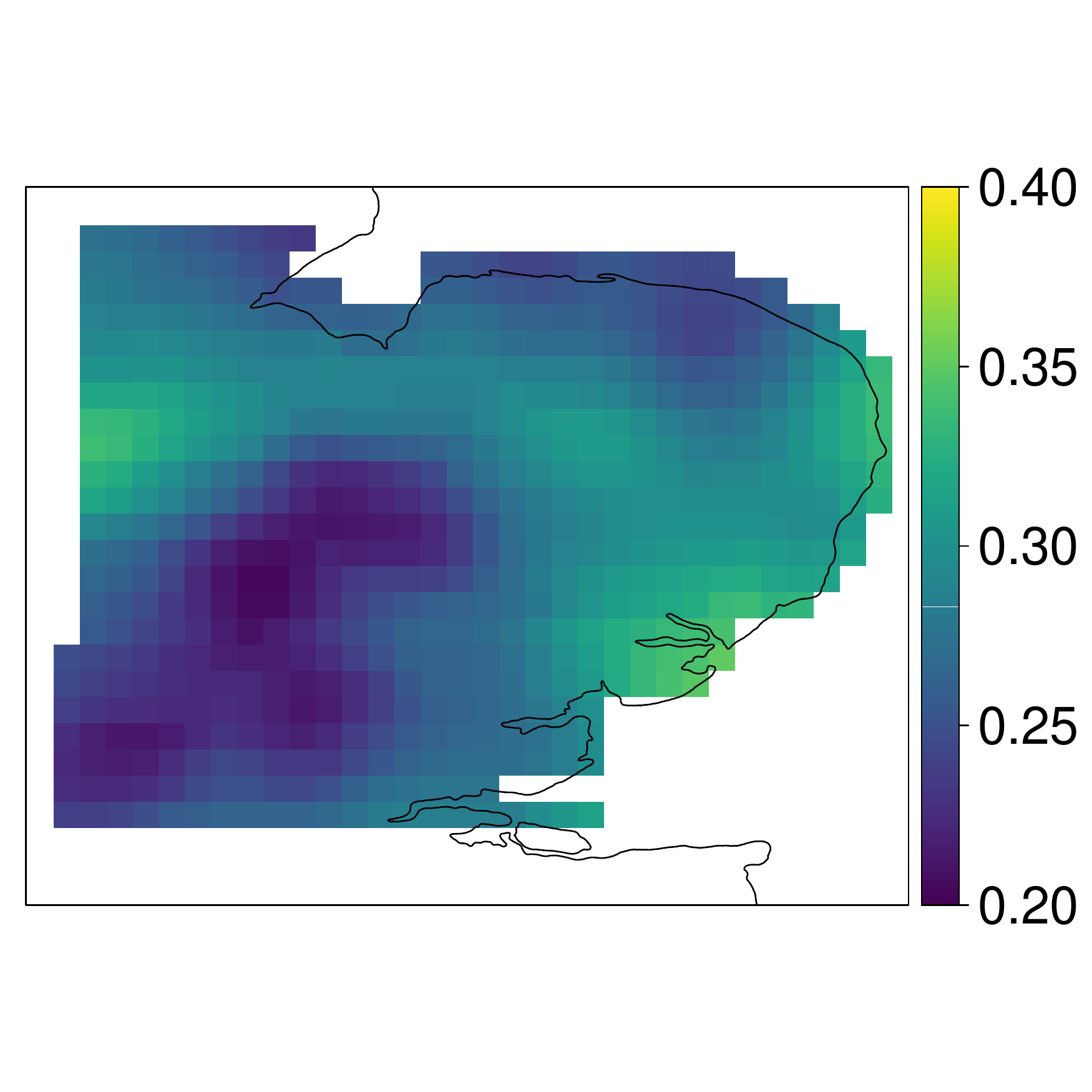} 
\end{minipage}
\caption{Spatially smoothed marginal distribution parameter estimates for East Anglia. Left: $\hat{p}(s)$, centre: $\hat{q}(s)$, right: $\hat{\xi}(s)$. $\hat{\upsilon}(s)
$ is illustrated in Figure~\ref{GAMscale} of the Supplementary Materials \citep{papersup}. }
\label{GAMfig}
\end{figure}

Initial analysis shows that the data consists of $8.7\%$ hours with zero precipitation, but much of the data with non-zero values exhibits noise around zero produced by the climate model. Thus, the data less than $1\times 10^{-5}$mm/hour were set to zero\footnote{This level of precipitation would be recorded as zero by a rain gauge.}, increasing the average number of dry hours to $83.7\%$.  Figure \ref{GAMfig} gives a spatial map of the estimated probability of zero precipitation $p(s)$ within a given hour; this is estimated using the logistic regression GAM detailed in Section \ref{sec-margmodel}. We observe some spatial variation in $p(s)$, with slightly lower estimates being found along the north-east coast. \par
We fit the spatial marginal model detailed in Section \ref{sec-margmodel}. We take $\lambda(s) = 0.005$ for all $s \in \mathcal{S}$ in \eqref{MargTransform} and the corresponding GPD threshold $q(s)$, estimated using a thin-plate spline, is illustrated in Figure~\ref{GAMfig} with $q(s)$ varying roughly over $\mathcal{S}$; larger values are found along the east coast. The GPD GAM model with spatially smooth estimate parameters is then fit to site-wise exceedances above $\hat{q}(s)$ at each site; a spatial map of the shape parameters is given in Figure~\ref{GAMfig}. We take the approach of \cite{youngman} and use as many knots as is computationally feasible in the thin-plate splines, which is 300. This creates a potentially overly rough spline which may overfit the data and not capture true physical smoothness; however, our primary interest is in the dependence structure when studying aggregates as this is the novel element of our model, so we chose this approach to ensure that the empirical marginal distributions are as well modelled as possible. We observe $\hat{\xi}(s) > 0$ for all $s \in \mathcal{S}$, and so the marginal upper tails are unbounded at each site.  Q-Q plots of the marginal fits at five randomly sampled locations are presented in Figure~\ref{Gamdiag} of the Supplementary Materials \citep{papersup}, all showing good fits.  To evaluate the fit over all locations, we use a pooled Q-Q plot, transforming all data onto standard exponential margins using the fitted model, see Figure~\ref{Gamdiag}. Again the fit is remarkably good, although confidence intervals are not provided due to the spatial dependence in the pooled data.
\subsection{Dependence Model}
\label{sec-appl-dep}
All dependence models are fitted by taking the exceedance threshold $u$ in \eqref{depmodeleq} to be the standard Laplace $98\%$ quantile. This leaves 864 fields for fitting the extremal dependence model given an observed extreme at a single conditioning site. A lower threshold $u$ was considered; however, we found that this leads to poorer model fits as the data exhibits a partial mixing of dependence structures. We believe this is due to the presence of multiple data generating processes in the climate model. Precipitation is typically generated by either high intensity events with localised spatial profiles, i.e., convective cells, or low intensity events with much large spatial profiles, i.e., frontal storms \citep{Thomassen2020}. In the absence of covariates to distinguish between these events in the data, we use a higher exceedance threshold to remove any frontal events; this is discussed further in Section~\ref{sec-dicuss}. The empirical estimate (and $95\%$ confidence interval) for the probability in \eqref{probmax} is $0.273\; (0.257, 0.290)$; this corresponds to the proportion of all observed fields used for fitting when we pool over all $934$ conditioning sites. Confidence intervals for \eqref{probmax} were created using a stationary bootstrap \citep{politis1994stationary} with $1000$ samples and expected block size $m_K$, here taken to be 48 hours. To compute a single bootstrap sample, we repeat the following until a sample of length greater than or equal to $n$ is obtained; draw a starting time $t^*\in\{1,\dots,n\}$ uniformly at random and a block size $K$ from a geometric distribution with expectation $m_K$, then add the block of observations $\{y_t(s):s \in \mathcal{S}, t\in \{t^*,\dots,t^*+K-1\}\}$ to the bootstrap sample. However, in cases where $t^*$ is generated with $t^*+K-1 > n$, we instead add $\{y_t(s):s\in\mathcal{S}, t\in \{1,\dots,t^*+K-n-1\}\cup\{t^*,\dots,n\}\}$. We then truncate the sample to have length $n$.\par
\begin{figure}
\centering
\begin{minipage}{0.32\linewidth}
\centering
\includegraphics[width=\linewidth]{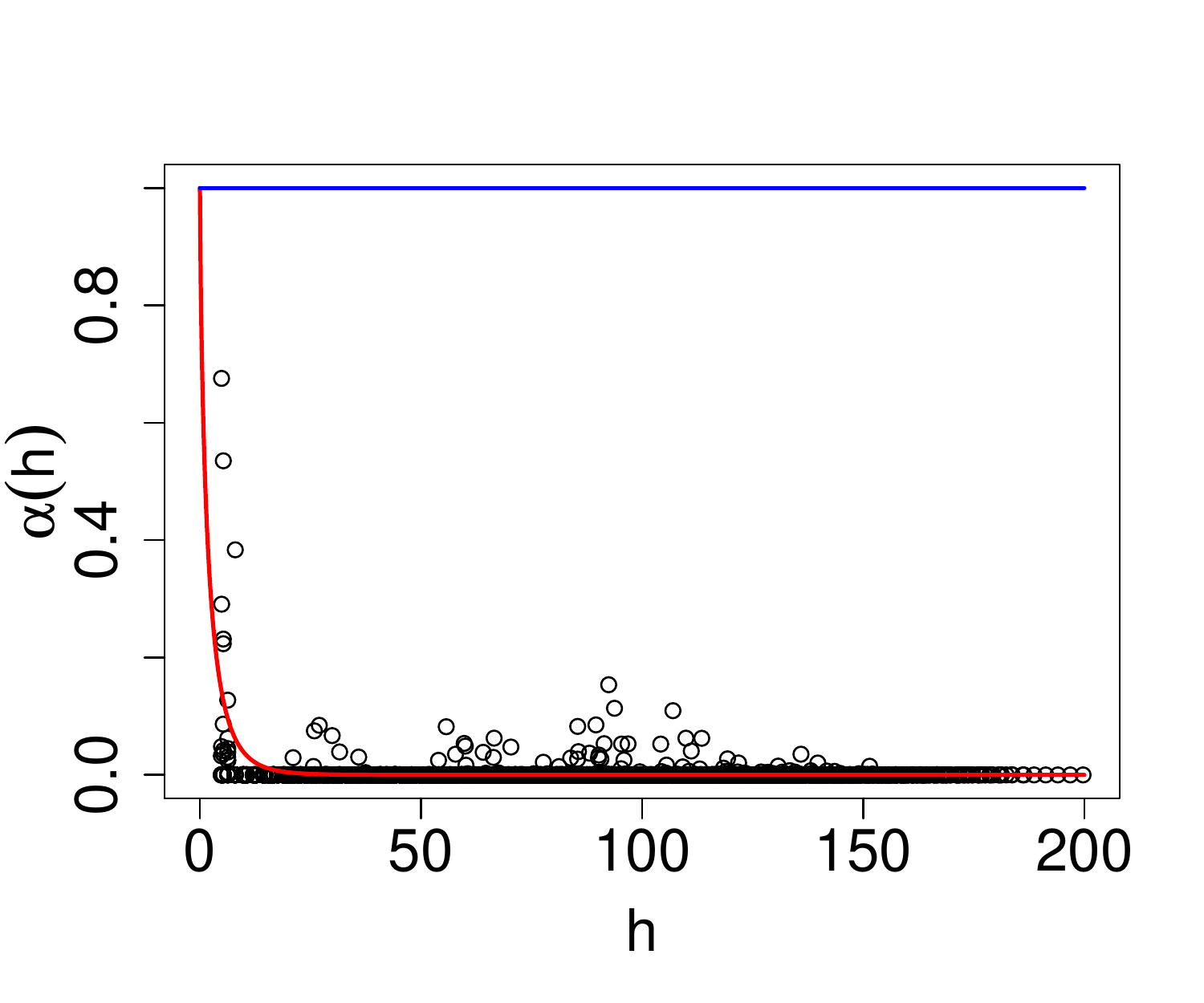} 
\end{minipage}
\begin{minipage}{0.32\linewidth}
\centering
\includegraphics[width=\linewidth]{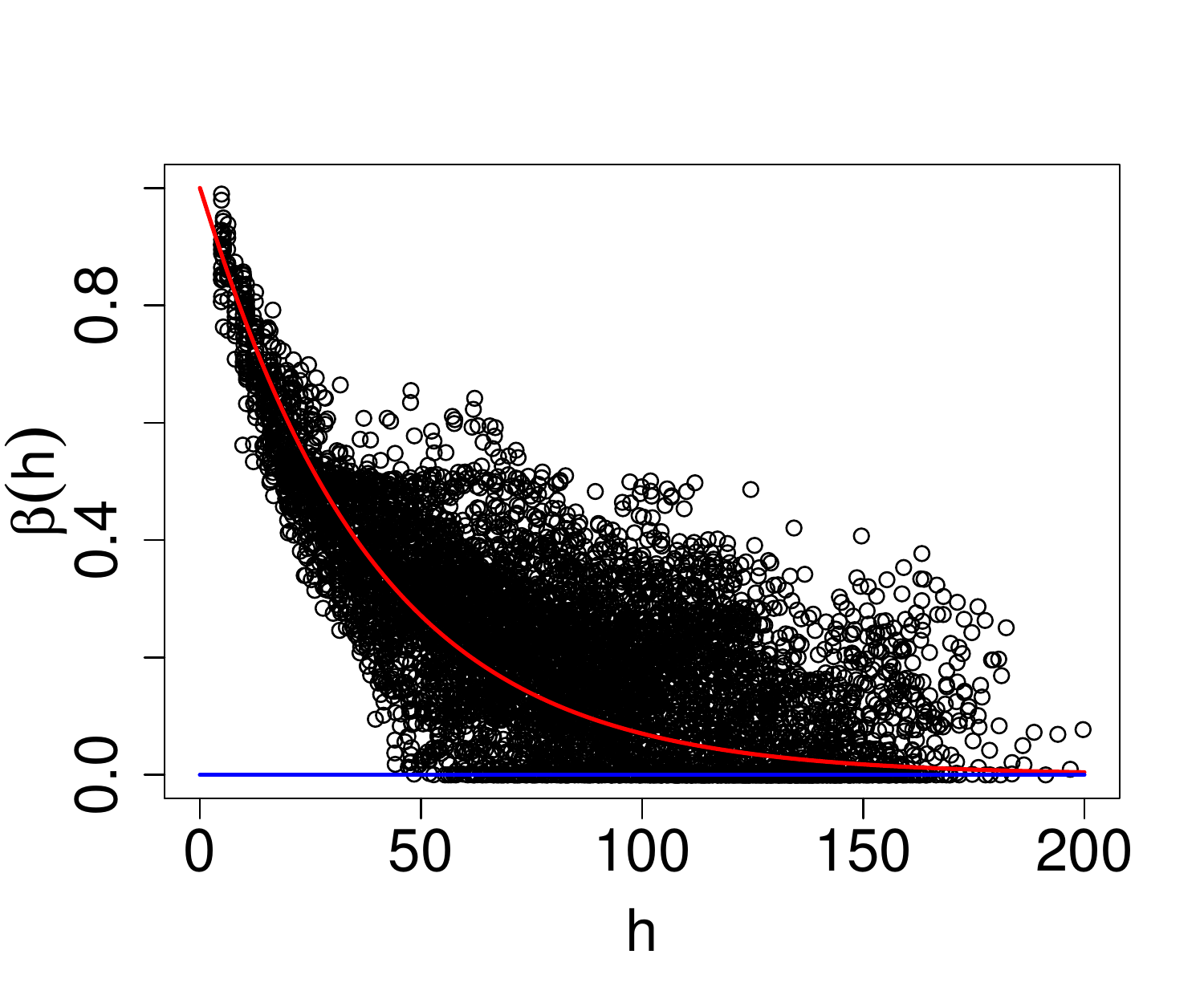} 
\end{minipage}
\begin{minipage}{0.32\linewidth}
\centering
\includegraphics[width=\linewidth]{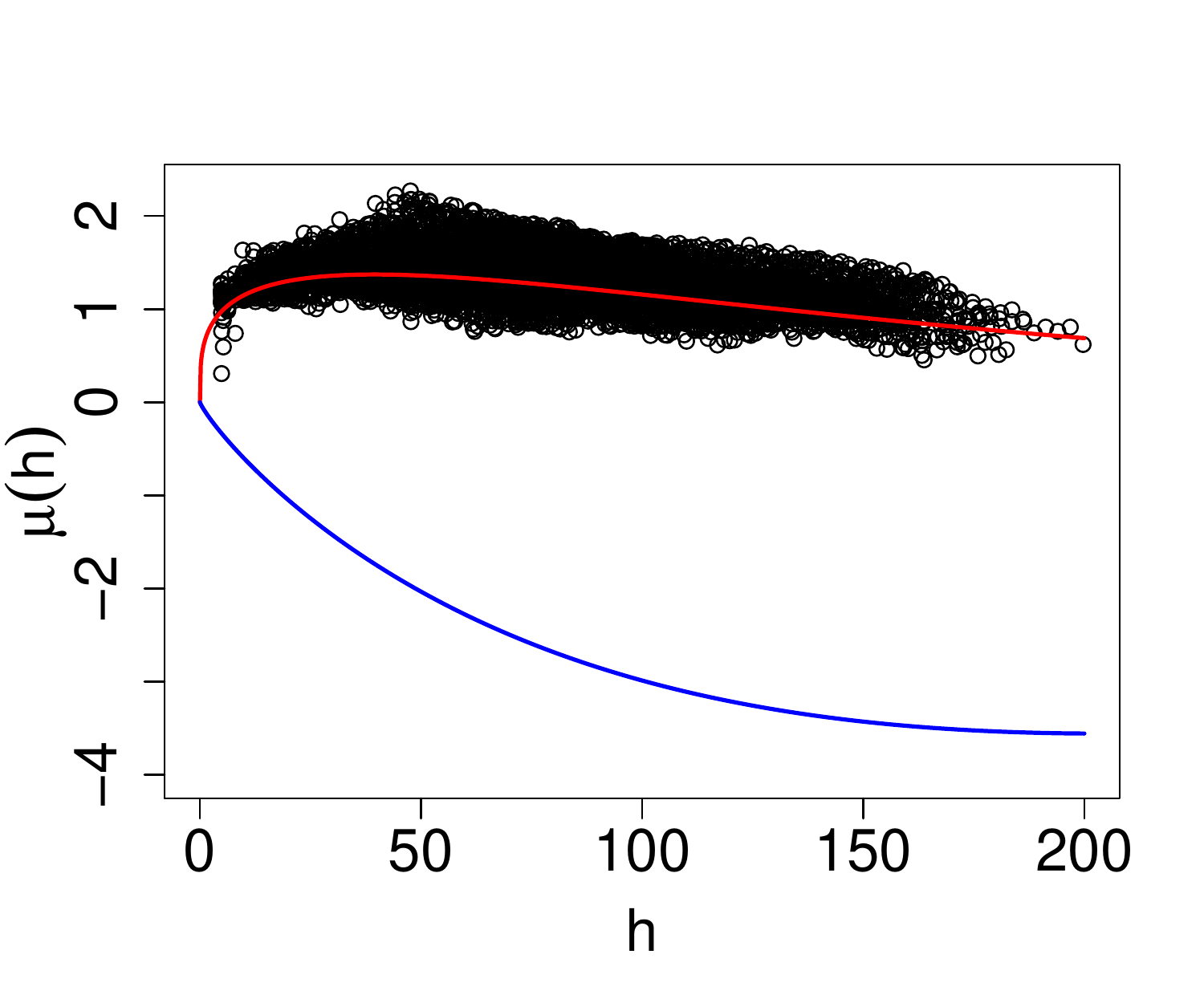} 
\end{minipage}
\begin{minipage}{0.32\linewidth}
\centering
\includegraphics[width=\linewidth]{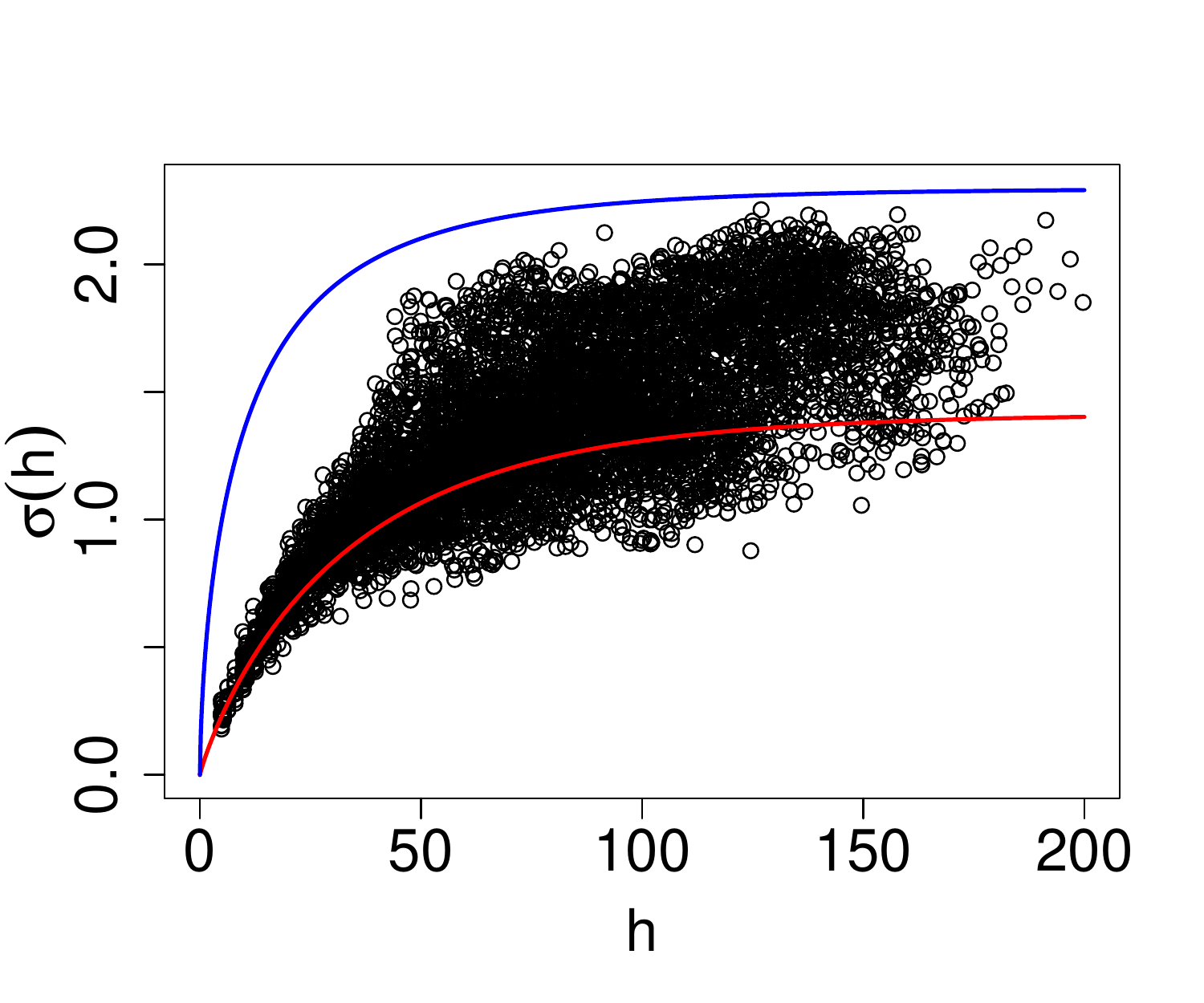} 
\end{minipage}
\begin{minipage}{0.32\linewidth}
\centering
\includegraphics[width=\linewidth]{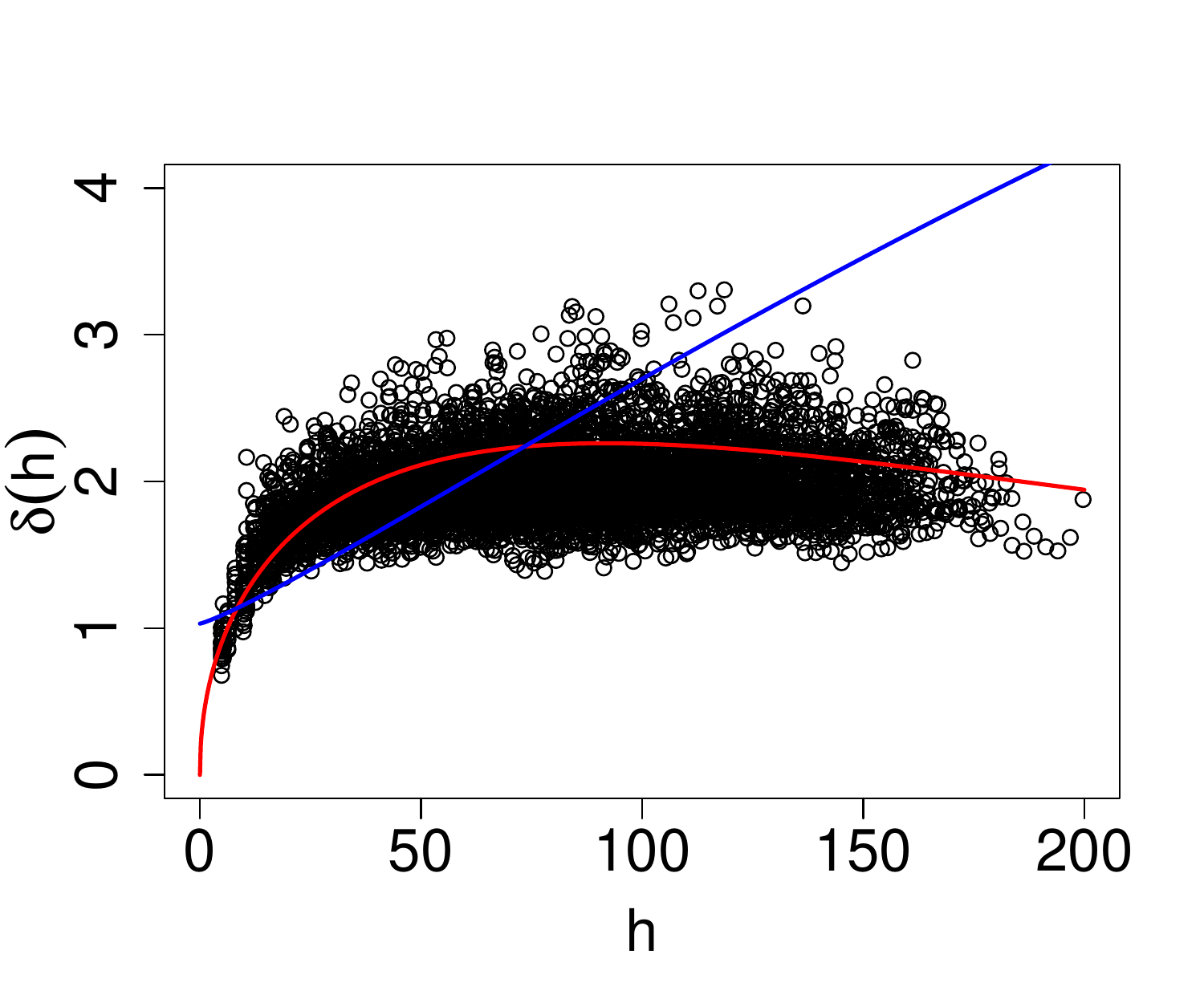} 
\end{minipage}
\begin{minipage}{0.32\linewidth}
\centering
\includegraphics[width=\linewidth]{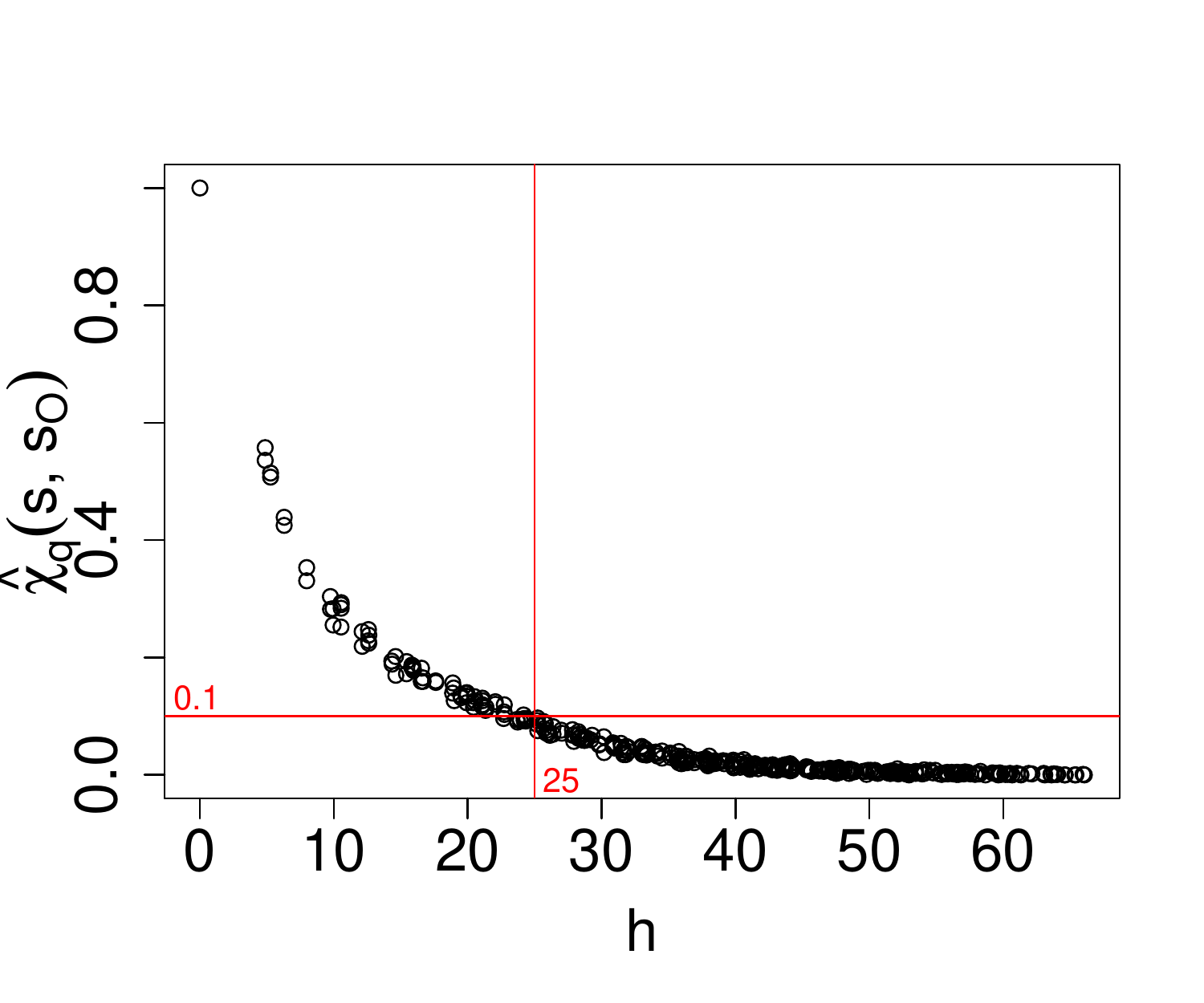} 
\end{minipage}
\caption{Estimates of parameters that determine the extremal dependence structure plotted against inter-site distance $h$, which is calculated under the anisotropy transformation for the full spatial model. Estimates from the free fits described in Section \ref{sec-appl-dep} are given by the black points, parametric spatial functions are given in red (asymptotically independent model) and blue (asymptotically dependent model). Bottom right: estimates from model for $\chi_q(s,s_O)$ in \eqref{chieq} with $q=1/(24\times 90)$. Distances ($km$) are given in the spatial anisotropy setting.}
\label{depestimates}
\end{figure}

We proceed with an initial analysis by fitting a simple version of the model of Section~\ref{sec-depmodel} to these data. We fit the model with two caveats: we make the temporary assumption that the residual process $\{Z(s|s_O)\}$ is independent at all distances; and evaluate a sequence of ``free" pairwise parameter estimates \citep{wadsworth2018spatial} for the normalising functions and those functions that describe the marginal characteristics of $\{Z(s|s_O)\}$. That is, we fit individual parameters, i.e., $\alpha^{(s_O)}_{s_i}$ etc. for $i = 1,\dots, d$ with $ s_i \neq s_O$, rather than a spatial function $\alpha\{h(s,s_0)\}$, and we do this for seven different conditioning sites $s_O$ sampled randomly over $\mathcal{S}$. This approach can be used to assess the stationarity of $\{X(s)\}$; if we observe clear disagreement in the parameter estimates for the different conditioning sites, then the assumption of stationarity of $\{X(s)\}$ is unlikely to be appropriate. We find no evidence for non-stationarity in the parameter estimates presented in Figure \ref{depestimates}; while we observe some volatility in the free parameter estimates, the general patterns appear to be the same regardless of the choice of conditioning site. We use the spatial structure in the free estimates to motivate our choice for the forms of the parameter functions detailed in Sections~\ref{normfuncs}~and~\ref{residprocess}. \par
Using the sampling method described in Section \ref{sec-strat}, the full spatial fit uses $d_s= 5000$ triples of sites with each sampling location being used as a conditioning site at least once. As the estimates of $\alpha$ and $\beta$ in Figure \ref{depestimates} decay quickly with increasing spatial distance, i.e., for any distance greater than $25km$, $\alpha \approx 0$ and $\beta < 0.5$; this illustrates that the underlying process $Y(s)$ exhibits fairly localised strong extremal dependence.  This suggests that we should focus on modelling extremal dependence locally, as this will be the driving factor of the aggregate behaviour. A distance of $25km$ in the anisotropic setting corresponds a distance of at most $28km$ in the original setting, and so we set $h_{max}= 28km$. Although $5000$ triples of sites represents a very small proportion of all possible triples, we observe a good model fit from Figure \ref{depestimates}, which shows that, even at distances greater than $28km$, the fitted parametric functions for the dependence parameters correspond well to the sequences of free estimates. We fitted a variant of \eqref{sigmudeltaeqs} for $\delta$, without the $\max$ operator constraint, and found the best fit was achieved with this $\delta$ function. That gave values for $\delta\{h(s,s_O)\}<1$ at the two closest sites to $s_O$. We found that this had negligible effect on the analysis, but clearly form \eqref{sigmudeltaeqs} would be more ideal.  \par
Figure \ref{depestimates} can be used to make further inference about the underlying dependence structure of the precipitation process. For example, we find that $\Delta$ in \eqref{alphaeq} can be taken to be zero without restricting the quality of the fit and similarly we can set $\kappa_{\delta_4}=\kappa_{\beta_3}=1$. The estimate $\Delta = 0$ suggests that the process is asymptotically independent at even the closest distances as asymptotic dependence requires both $\alpha(h) = 1$ and $\beta(h) = 0$ for all $h$, which the estimates in Figure \ref{depestimates} suggest is not the case; a fit imposing asymptotic dependence is discussed later. Furthermore, we found that incorporating spatial anisotropy into the dependence model improved the overall fit; stronger extremal dependence was found along an approximate $-10^{\circ}$ bearing, reducing by at most $7\%$ over different directions. Parameter estimates (and standard errors) are provided in Table \ref{parEst} in the Supplementary Materials \citep{papersup}. Although not illustrated in Figure \ref{depestimates}, $\rho$ decays quickly with distance, with $\rho(100)\approx 0.2$.\par
To further support the choice of $h_{max}$, we estimate $\chi_q(s,s_O)$ in \eqref{chieq} for $s_O\in \mathcal{S}$ in the centre of $\mathcal{S}$, taking $q$ corresponding to a one-year return level probability, and look to see how far away $s$ must be for $\chi_q(s,s_O)$ to be less than $\gamma$ for small $\gamma > 0$ (see Section~\ref{sec-agg-infer}). We estimate $\chi_q(s,s_O)$ by simulating $5\times 10^4$ replications, using Algorithm~\ref{sim-algo}, from the fitted model and this is illustrated in Figure~\ref{depestimates}, bottom-right panel; for $\gamma = 0.1$, we find that a distance of $h_{max}$ is sufficient and so we set $\tau=h_{max}$ in Section~\ref{sec-agg-infer}, discussed further in Section~\ref{sec_diag}.\par
\begin{figure}[h]
\centering
\begin{minipage}{0.32\linewidth}
\centering
\includegraphics[width=\linewidth]{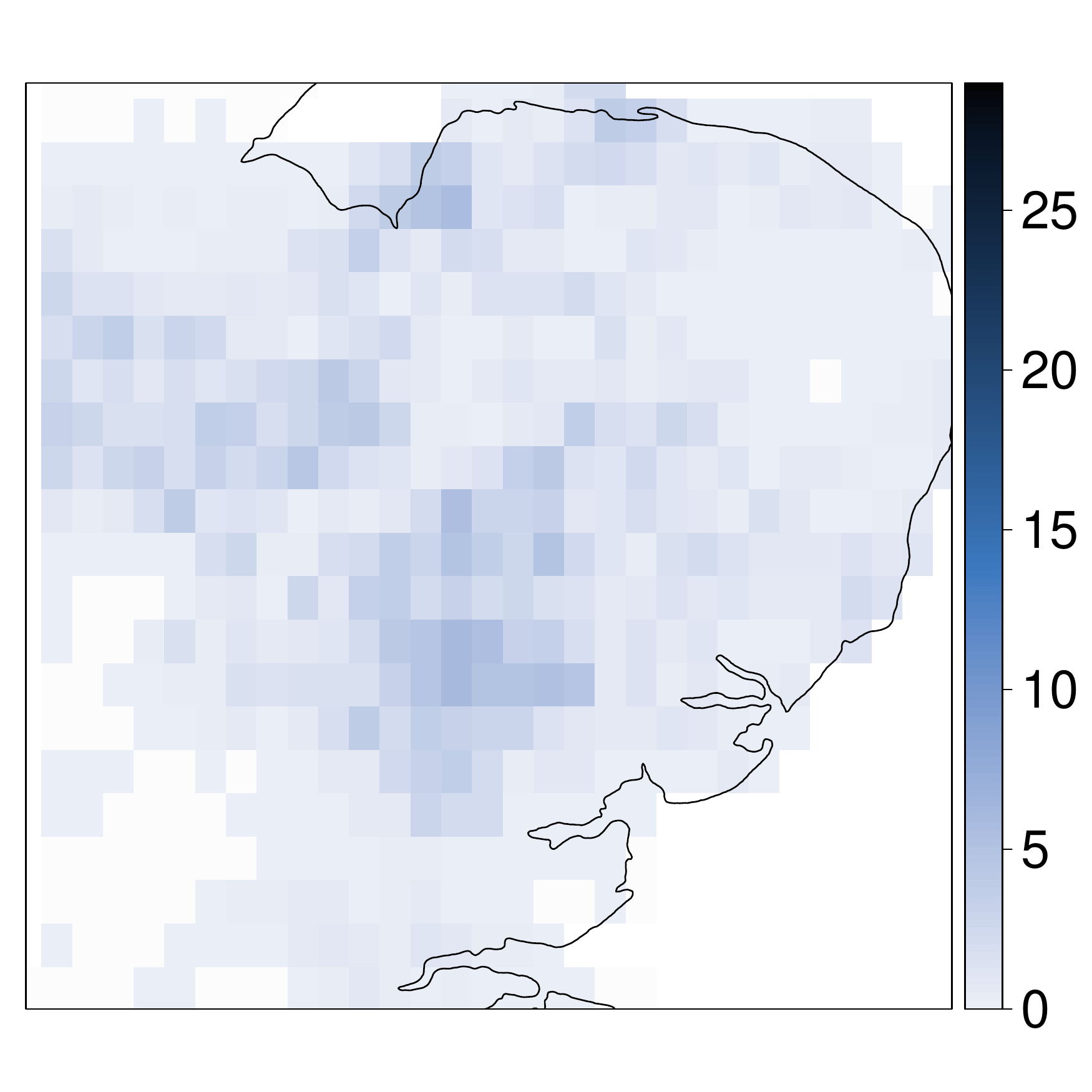} 
\end{minipage}
\begin{minipage}{0.32\linewidth}
\centering
\includegraphics[width=\linewidth]{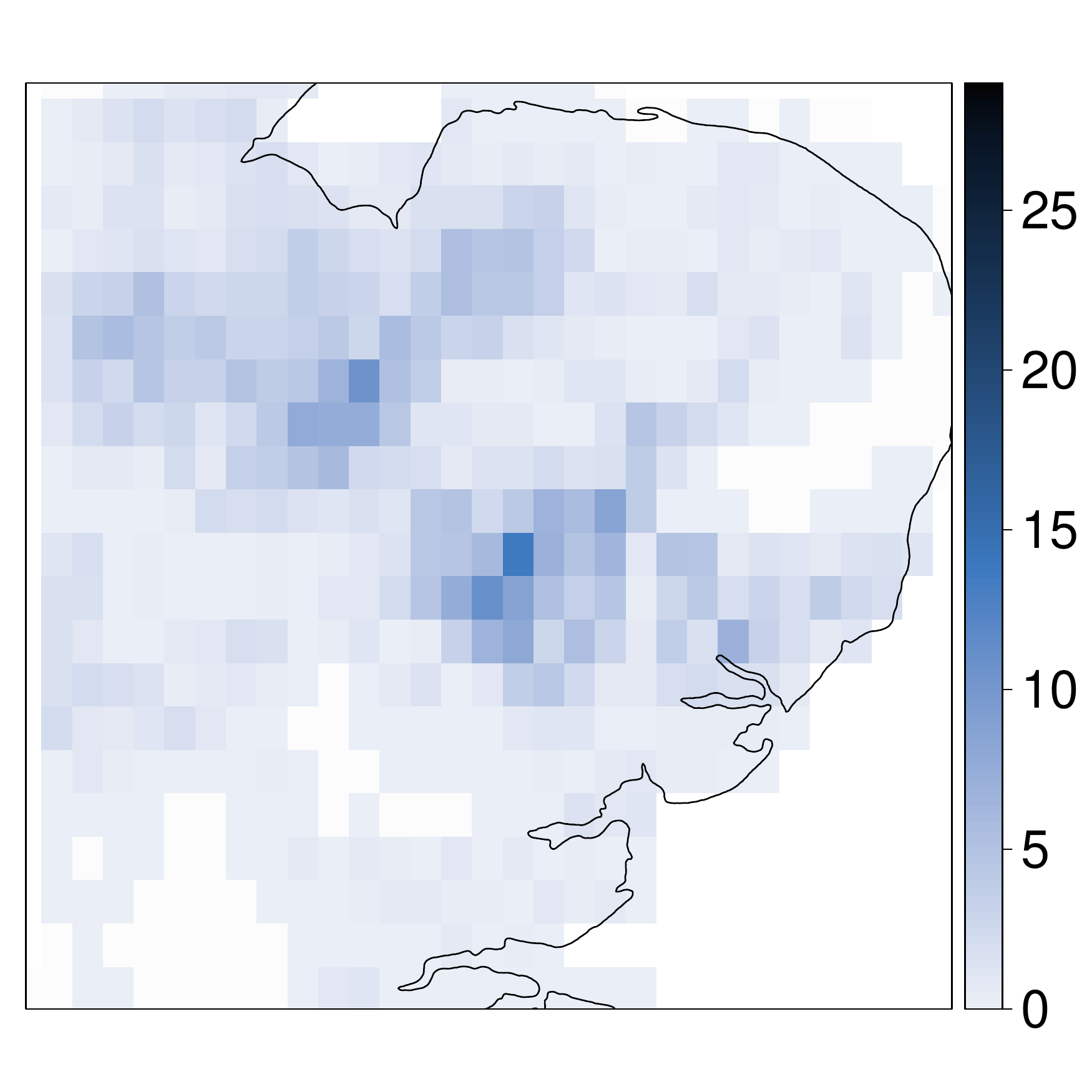} 
\end{minipage}
\begin{minipage}{0.32\linewidth}
\centering
\includegraphics[width=\linewidth]{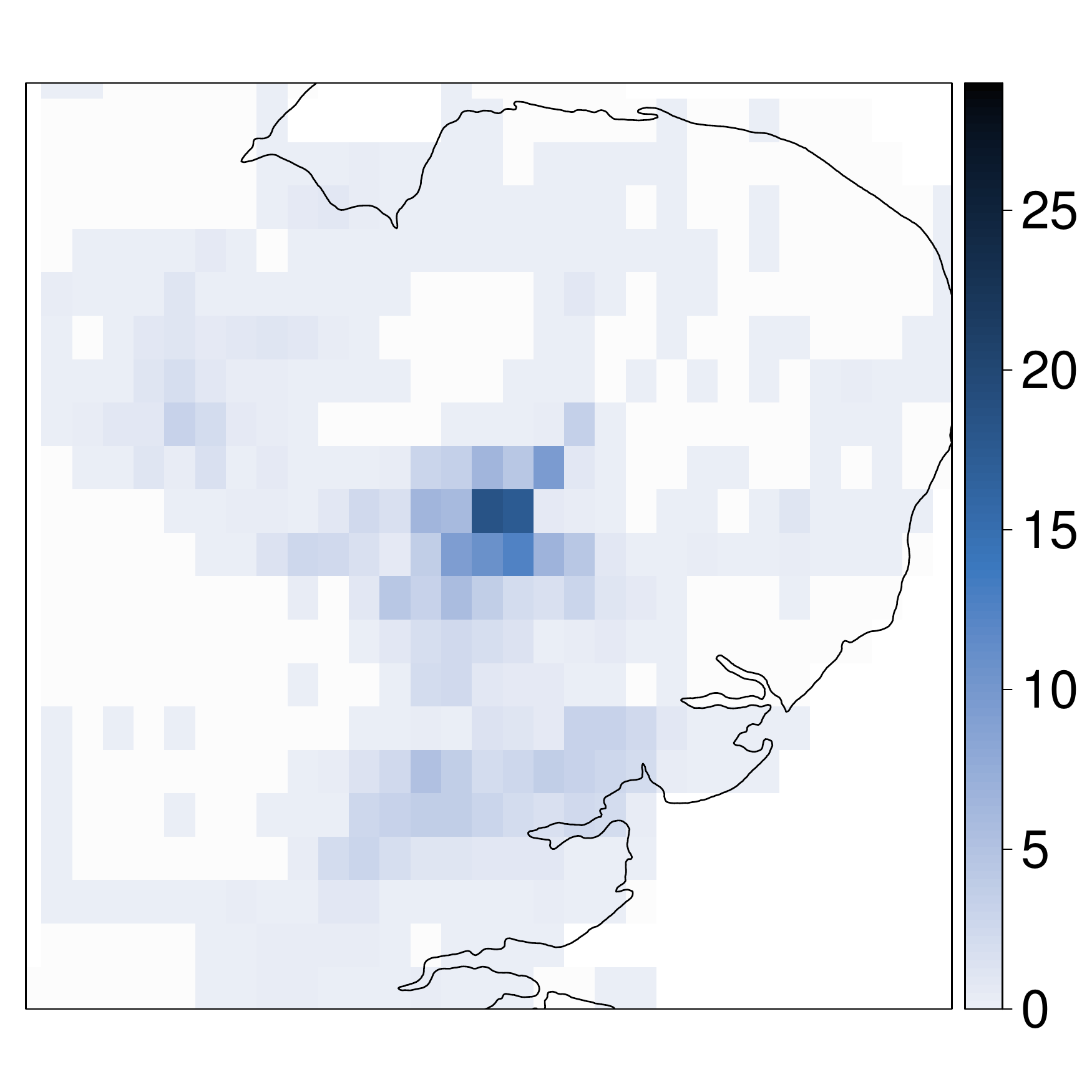} 
\end{minipage}
\begin{minipage}{0.32\linewidth}
\centering
\includegraphics[width=\linewidth]{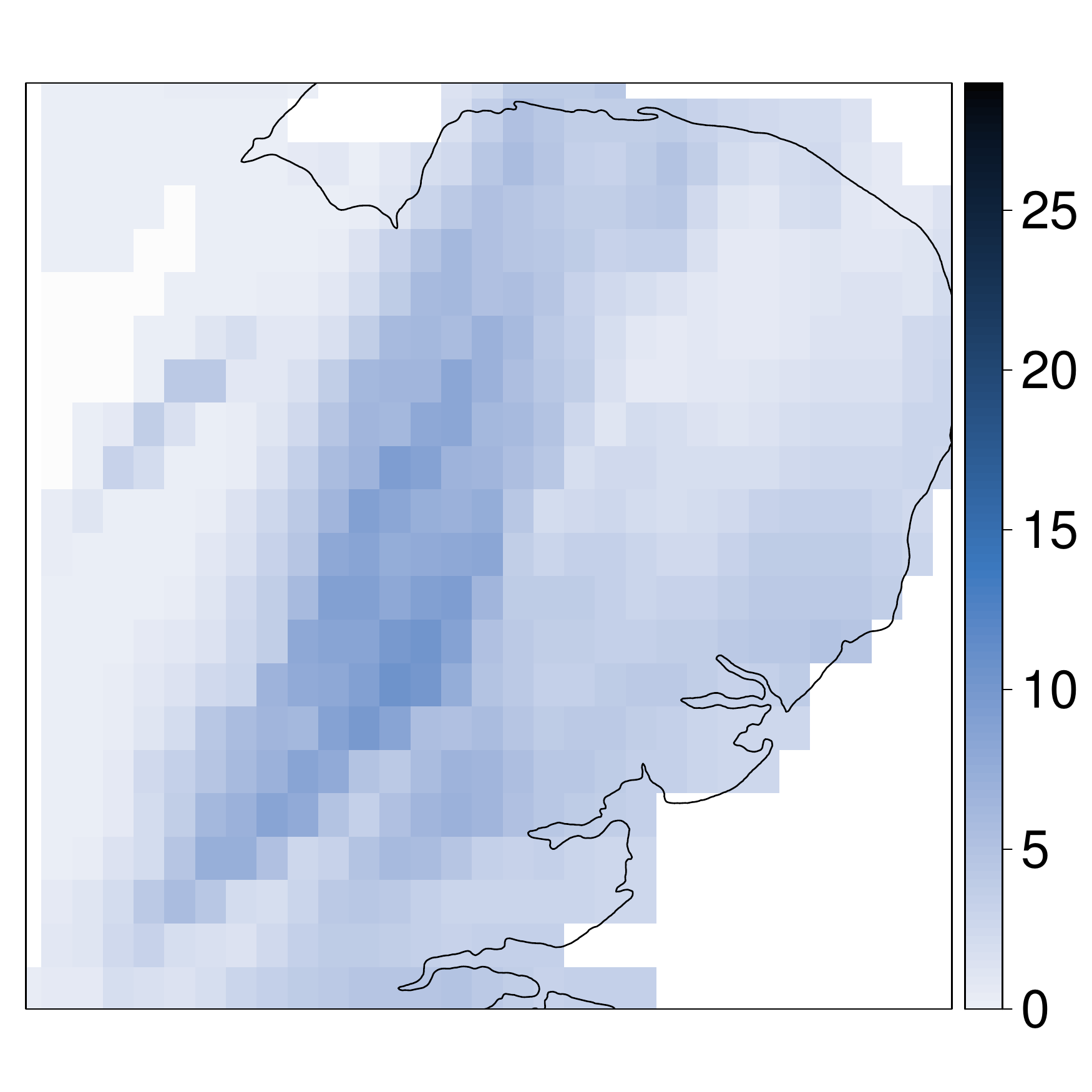} 
\end{minipage}
\begin{minipage}{0.32\linewidth}
\centering
\includegraphics[width=\linewidth]{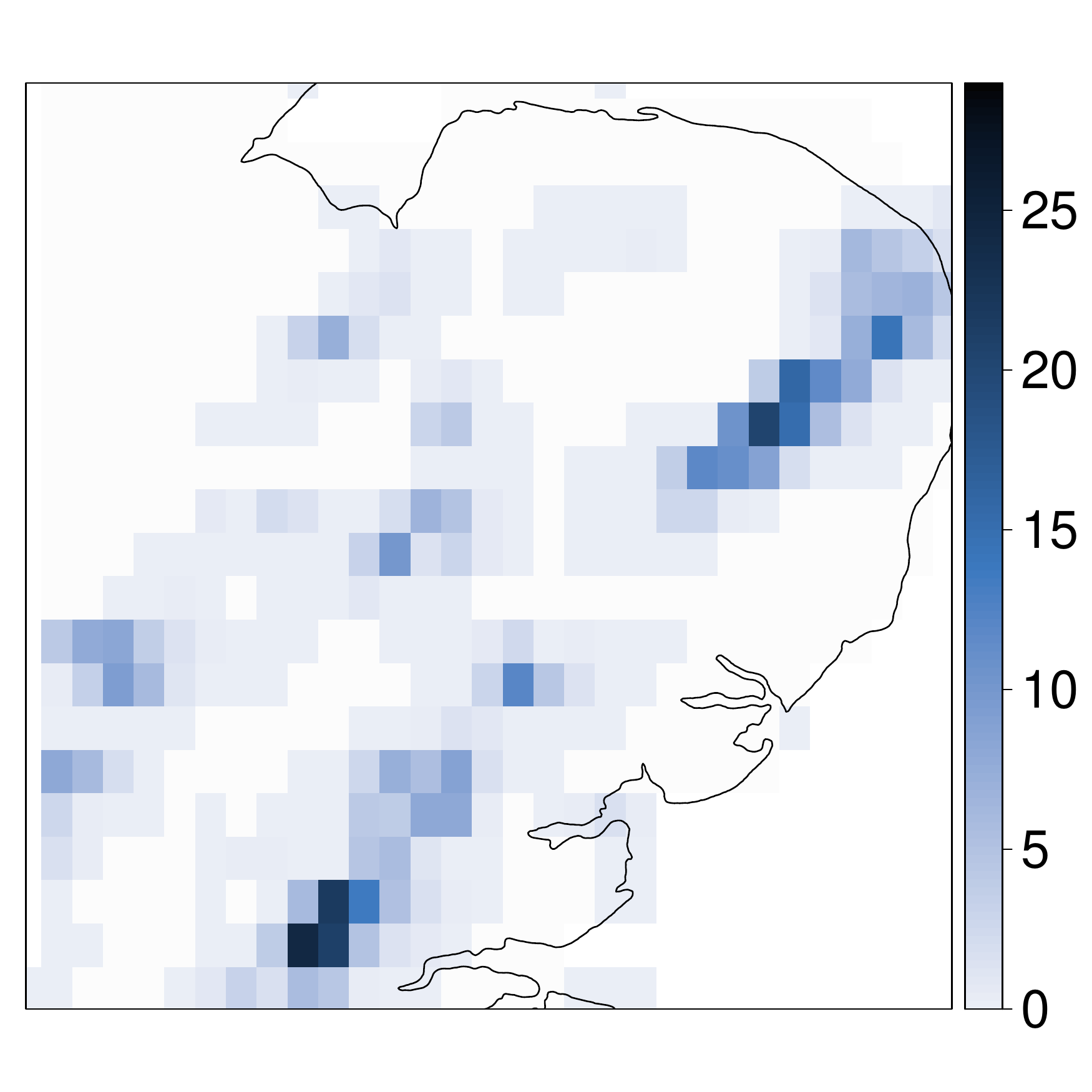} 
\end{minipage}
\begin{minipage}{0.32\linewidth}
\centering

\includegraphics[width=\linewidth]{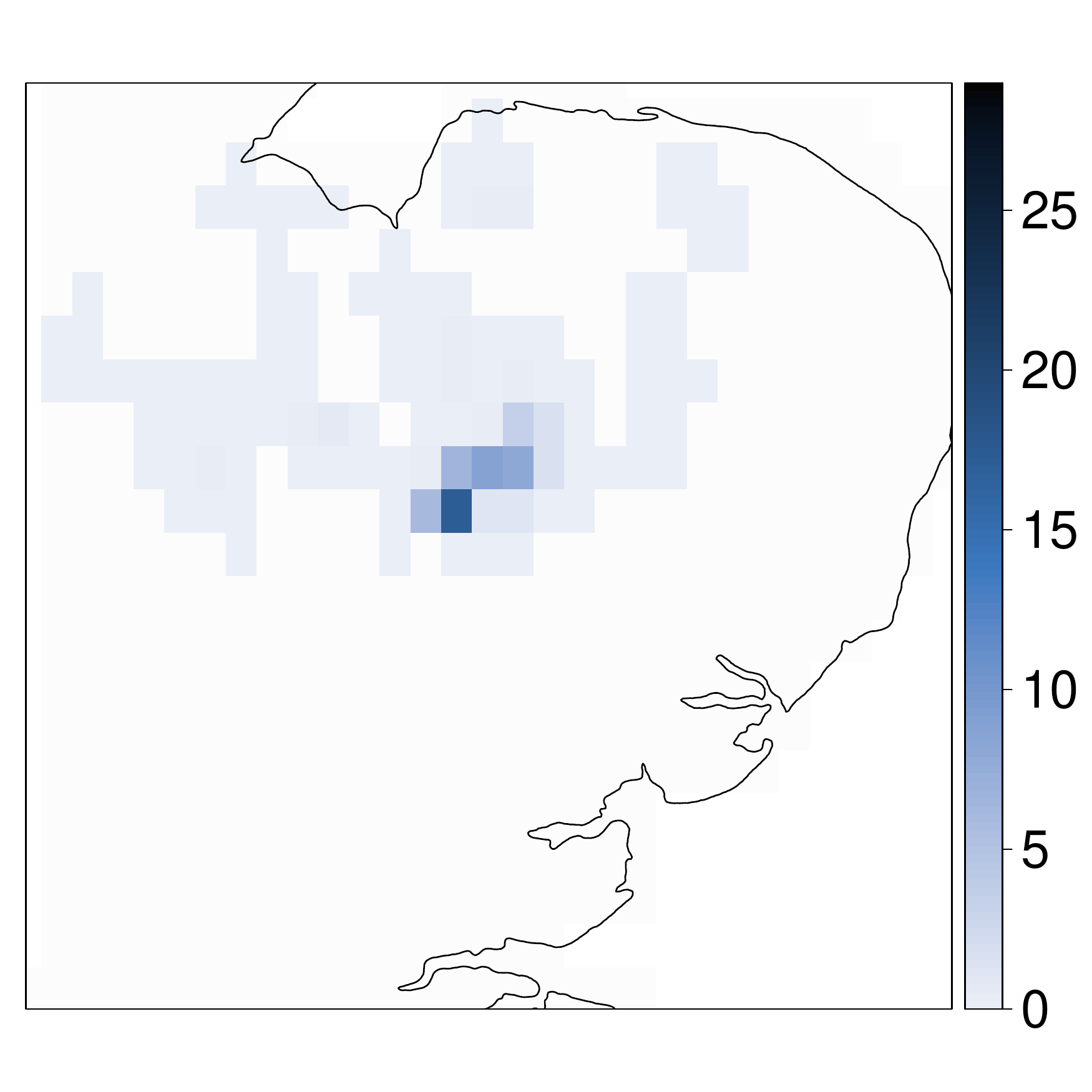} 
\end{minipage}
\caption{Extreme precipitation fields (mm/hr). Top-row: realisations from the fitted model described in \eqref{simexp}. Bottom-row: observed fields from the data. }
\label{realisations}
\end{figure}

 
Figure \ref{realisations} illustrates six extreme fields: three realisations from the model defined in \eqref{simexp} and three observations from the data. Fields are chosen such that the site in the centre of $\mathcal{S}$ exceeds its $99.9\%$-quantile but the maximum over the entire field does not exceed $30$mm/hr; this is to make it easier to compare the spatial structure in the fields. Realisations from the model appear to replicate the roughness in observed events well. Furthermore, in Figure~\ref{realisations} we observe that replications from the model are able to exhibit some of the different physical properties of extreme precipitation. For example, the top-left panel displays a spatially flat event whilst the other two illustrate localised extreme events; multiple sub-events in the top-middle and a single sub-event in the top-right. Further diagnostics for fitted extremal dependence models have been proposed in the literature, e.g., by \cite{blanchet2011spatial}, \cite{huser2020advances}; however, our focus lies on inference for aggregates and so we present diagnostics for these variables only.
\par
As $\alpha$ quickly goes to zero with distance, all extremal dependence is instead exhibited through $\beta$. This is atypical of fits of this model for other applications \citep{wadsworth2018spatial,shooterinpress,simpson2020conditional}, where the $\alpha$ function drives the extremal behaviour of their modelled processes, e.g., temperature and sea wave heights. Having $\beta$ controlling extremal dependence would suggest that the process that generates the extreme precipitation we are modelling is somewhat rough; this concurs with the observed fields containing an extreme value shown in Figure \ref{realisations}, and consistent with the spatial nature of strong convective rainfall.  To illustrate this, we note that, for small $h=h(s,s_O)$, we have $\alpha(h) \approx 0$ and $\beta(h) \approx 1$, then $\mathbb{E}[X(s)|X(s_O)=x] \approx \mu(h)x$ and $\var(X(s)|X(s_O)=x) \approx x^2\sigma^2(h)$, and so the largest events at $s_O$ are the most variable. This has not been observed in other applications as the extremal dependence in these processes is typically quite smooth with $\var(X(s)|X(s_O)=x) \approx \sigma^2(h)$ as $\beta(h)\approx 0$ when $h$ is small. Even at the largest $h$ value possible in $\mathcal{S}$, the process $\{X(s)\}$ does not exhibit independence; although $\alpha$ and $\beta$ tend to zero the residual process does not attain standard Laplace margins with $\delta(h)=1$.\par
Existing literature for approaches that rely on modelling the underlying process to make inference the extremal behaviour of spatial aggregates of precipitation typically use models that only allow for asymptotic dependence \citep{coles1993,coles1996,buishand2008}. We fit such a model to illustrate that imposing asymptotic dependence may lead to poor inference for the tails of spatial aggregates. We term this the ``AD~model" and the model described above as the {``AI model"}. To specify the AD model, we fix $\alpha(h)=1$ and $\beta(h)=0$ for all $h$ and we change $\sigma(h)$ in \eqref{sigmudeltaeqs} to ${\sigma(h)=\kappa_{\sigma_3}\left(1-\exp\{-(h/\kappa_{\sigma_1})^{\kappa_{\sigma_2}}\}\right)}$ with $\kappa_{\sigma_1}, \kappa_{\sigma_2}, \kappa_{\sigma_3}>0,$ as we no longer require that $\sigma(h)\rightarrow \sqrt{2}$ as $(h) \rightarrow \infty$. The corresponding $\mu(\cdot)$ and $\delta(\cdot)$ functional forms remain the same and the spatial anisotropy setting described in \eqref{anisoeq} is still used.  To fully capture the behaviour of $\mu(\cdot)$, we found we had to take $h_{max} = 75km$. The estimated spatial functions for the AD model are illustrated in Figure \ref{depestimates}. With $\alpha$ and $\beta$ fixed, we observe that the other parameters are forced to compensate for this misspecification. For example, we observe a strictly negative $\mu$ function; this is to compensate for fixing the $\alpha$ value too large for the data. Given this, we re-estimated the free parameters with $\alpha=1$ and $\beta=0$ fixed and observed good agreement between the spatial functions and these new estimates. However, this does not imply that the model as a whole fits well, this is emphasised in Section~\ref{sec_diag} where spatial aggregates of simulated fields $\{Y(s)\}$ are studied. 
 \subsection{Diagnostics and tails of spatial aggregates}
 \label{sec_diag}
\begin{figure}[h!]
\begin{minipage}{0.43\textwidth}
\centering
\includegraphics[width=\textwidth]{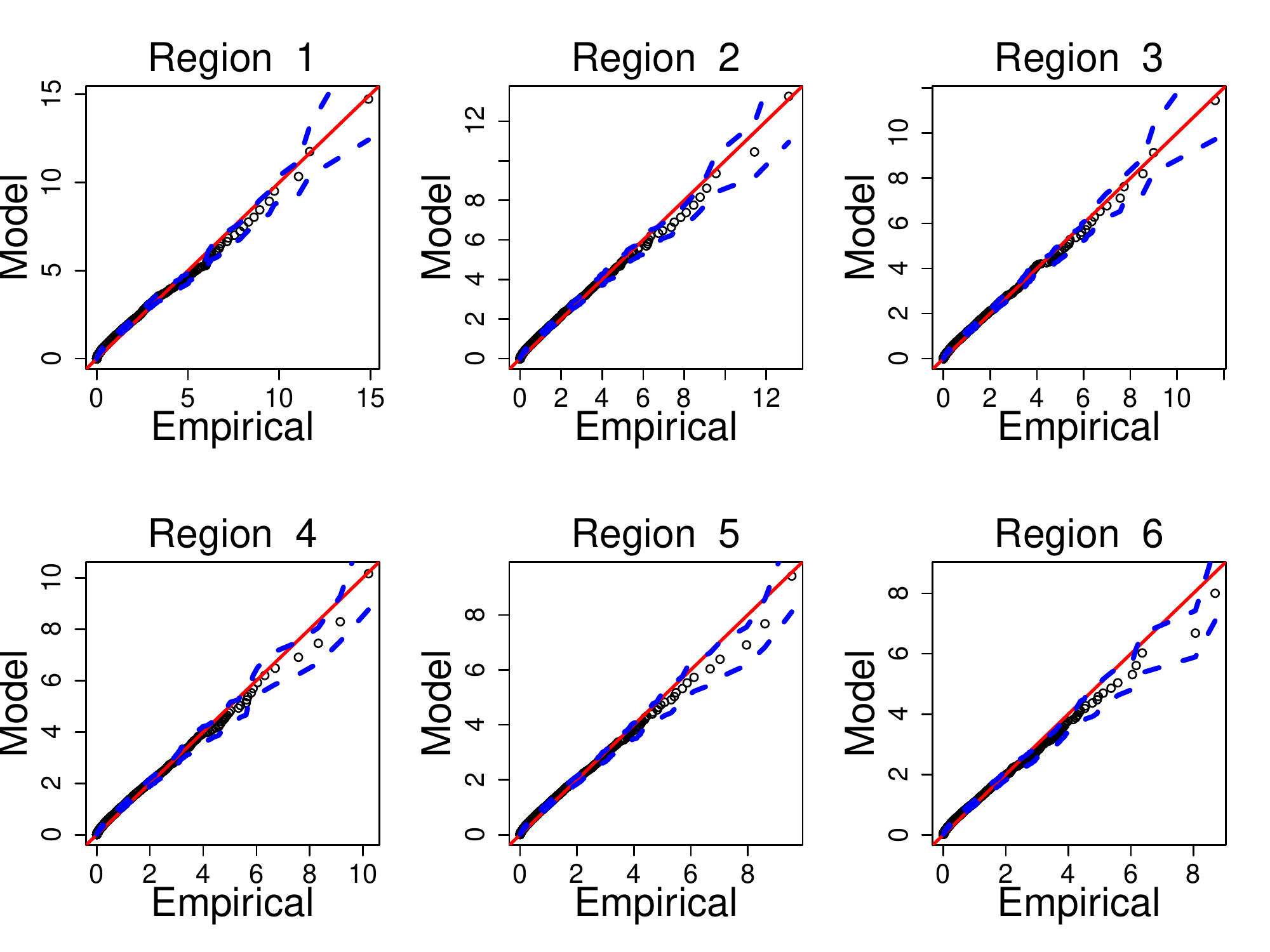} 
\end{minipage}
\begin{minipage}{0.38\textwidth}
\centering
\includegraphics[width=\textwidth]{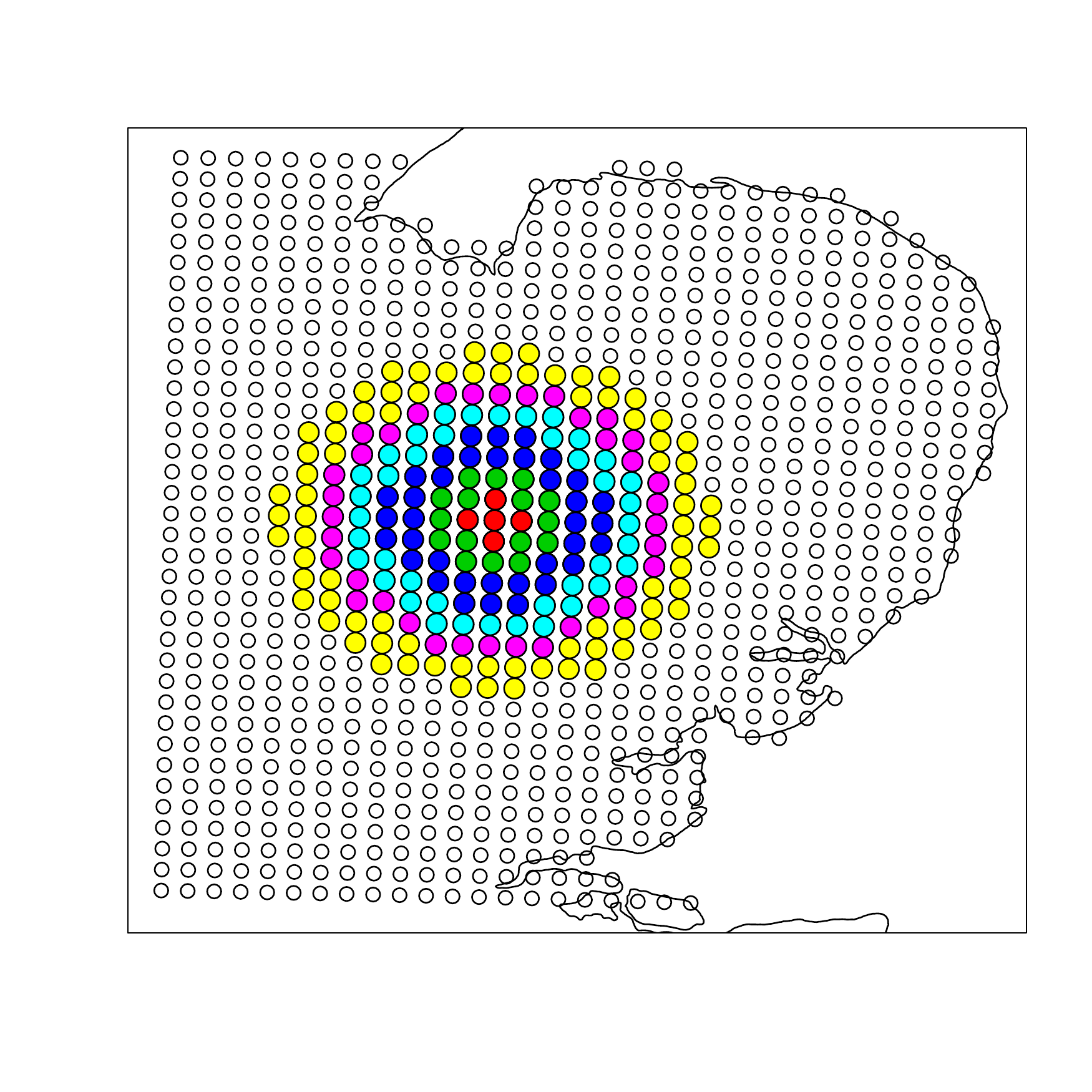} 
\end{minipage}
\begin{minipage}{0.17\textwidth}
\centering
\includegraphics[width=\textwidth]{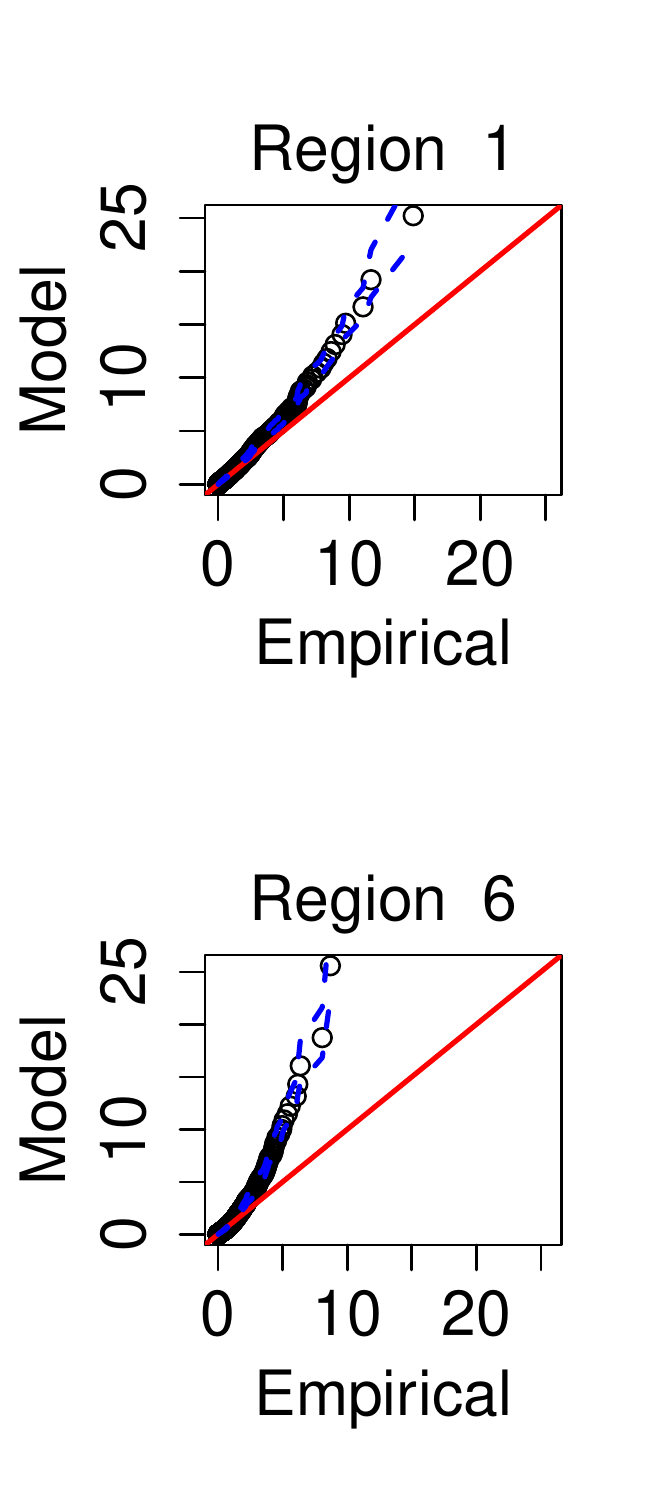} 
\end{minipage}
\caption{Q-Q plots for model, and empirical, $\bar{R}_{\mathcal{A}}$ of regions of increasing size. Left: AI model, right: AD model. Probabilities range from $0.7$ to a value corresponding to the 20 year return level. $95\%$ confidence intervals are given by the blue dashed lines. Q-Q plots for all six regions for the AD model are given in Figure~\ref{agg_ad} of the Supplementary Materials \citep{papersup}. Centre: aggregate regions $\mathcal{A}$ with corresponding areas $(125,525,1425,2425,3350,5425)-km^2$. Regions 1-6 are coloured red, green, blue, cyan, purple, yellow; regions include both the coloured and interior points. }
\label{agg_diag2}
\end{figure}

Q-Q plots, presented in Figure~\ref{agg_diag2}, assess how well the tails of the simulated distributions compare against the tails of the empirical distribution of the spatial averages. Confidence intervals given for the simulated quantiles are derived by applying to the observations the stationary bootstrap with expected block size of $48$ hours. We evaluate a sample of $50$ parameter estimates by applying the procedure described in Section~\ref{sec-strat} to stationary bootstrap samples of the data; note that we keep $h_{max}=28km$, but use different, randomly selected triples of sites for each sample. Only $50$ estimates were derived due to the computational resources required for fitting.  For each parameter set estimate, we draw ${5 \times 10^5}$ realisations of $\{Y(s):s \in \mathcal{S}\}$ using the regime described in Section~\ref{sim}. Then $\bar{R}_{\mathcal{A}}$ is calculated for each sample and for each region $\mathcal{A}$; per the discussion in Section \ref{sec-appl-dep}, each $\mathcal{A}$ is at least $\tau = 28km$ away from the boundaries of $\mathcal{S}$. Figure~\ref{agg_diag2} illustrates generally very good fits for the tails of $\bar{R}_{\mathcal{A}}$ with nested regions $\mathcal{A}$. The AI model appears to slightly underestimate the true magnitude of the largest aggregates for the largest regions; while this may suggest that the model is not capturing dependence at further distances from the conditioning site, Figure \ref{depestimates} suggests that the model fits well even at the furthest distances. This leads us to suspect that there is a mixture of events present in the data when we consider large spatial regions for aggregation, and that the model is not flexible enough to capture these mixtures, see Section \ref{sec-dicuss}. To illustrate the benefits of using our approach, Figure \ref{agg_diag2} also illustrates the same diagnostics for the AD model described in Section~\ref{sec-appl-dep} for the smallest and largest regions. The AD model provides much poorer fits than the AI models, as it always overestimates the quantiles; this suggests that the AD model overestimates the dependence within the original process even for the smallest aggregation regions. A similar plot for non-overlapping regions is illustrated in Figure \ref{agg_diag} of the Supplementary Materials \citep{papersup}; here we observe some underestimation in the largest estimated return levels for the regions closest to the east coast, which is a possible indication of non-stationarity along this coast. \par
\begin{figure}[h]
\centering
\begin{minipage}{0.49\textwidth}
\centering
\includegraphics[width=\textwidth]{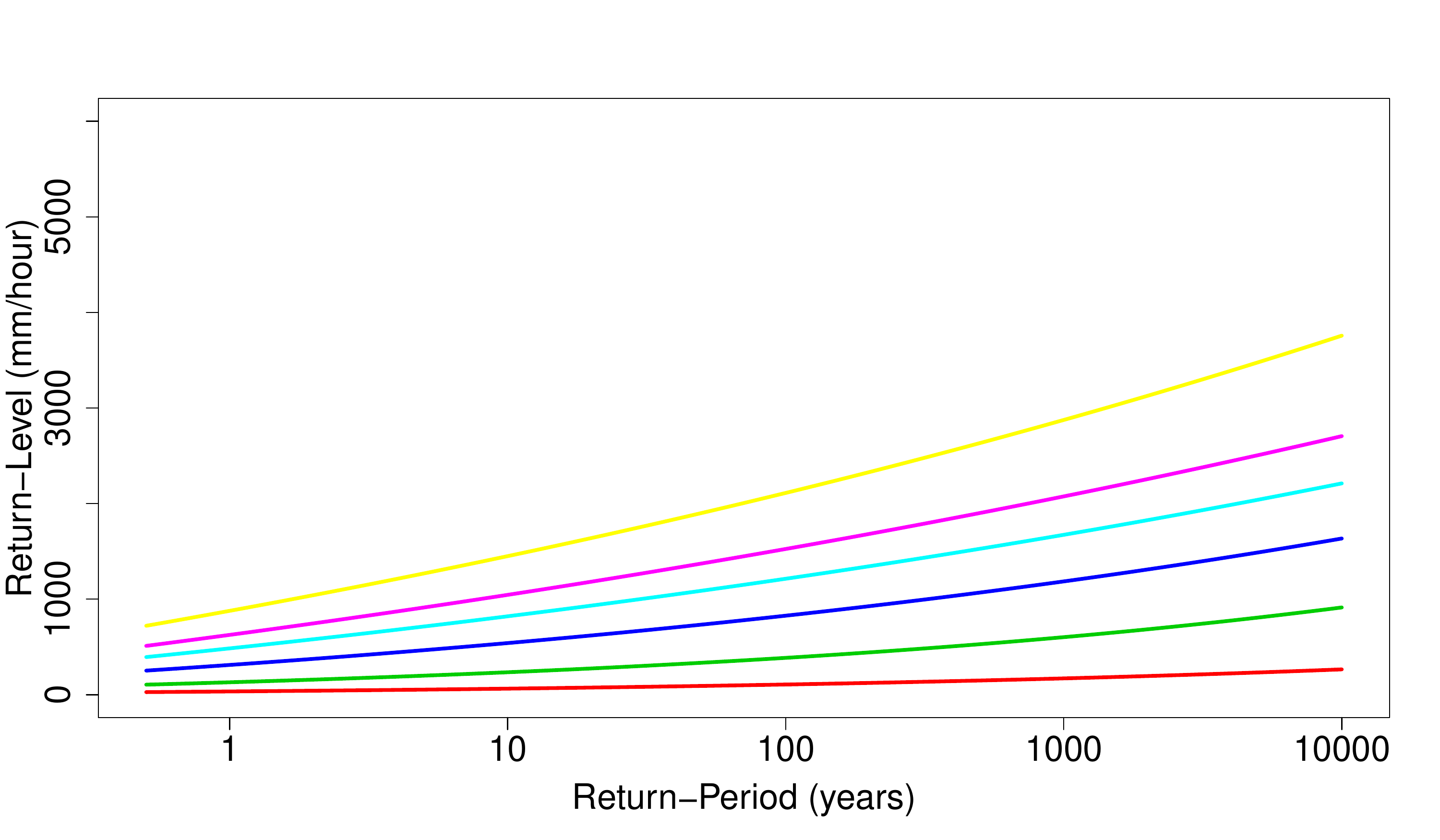} 
\end{minipage}
\begin{minipage}{0.49\textwidth}
\centering
\includegraphics[width=\textwidth]{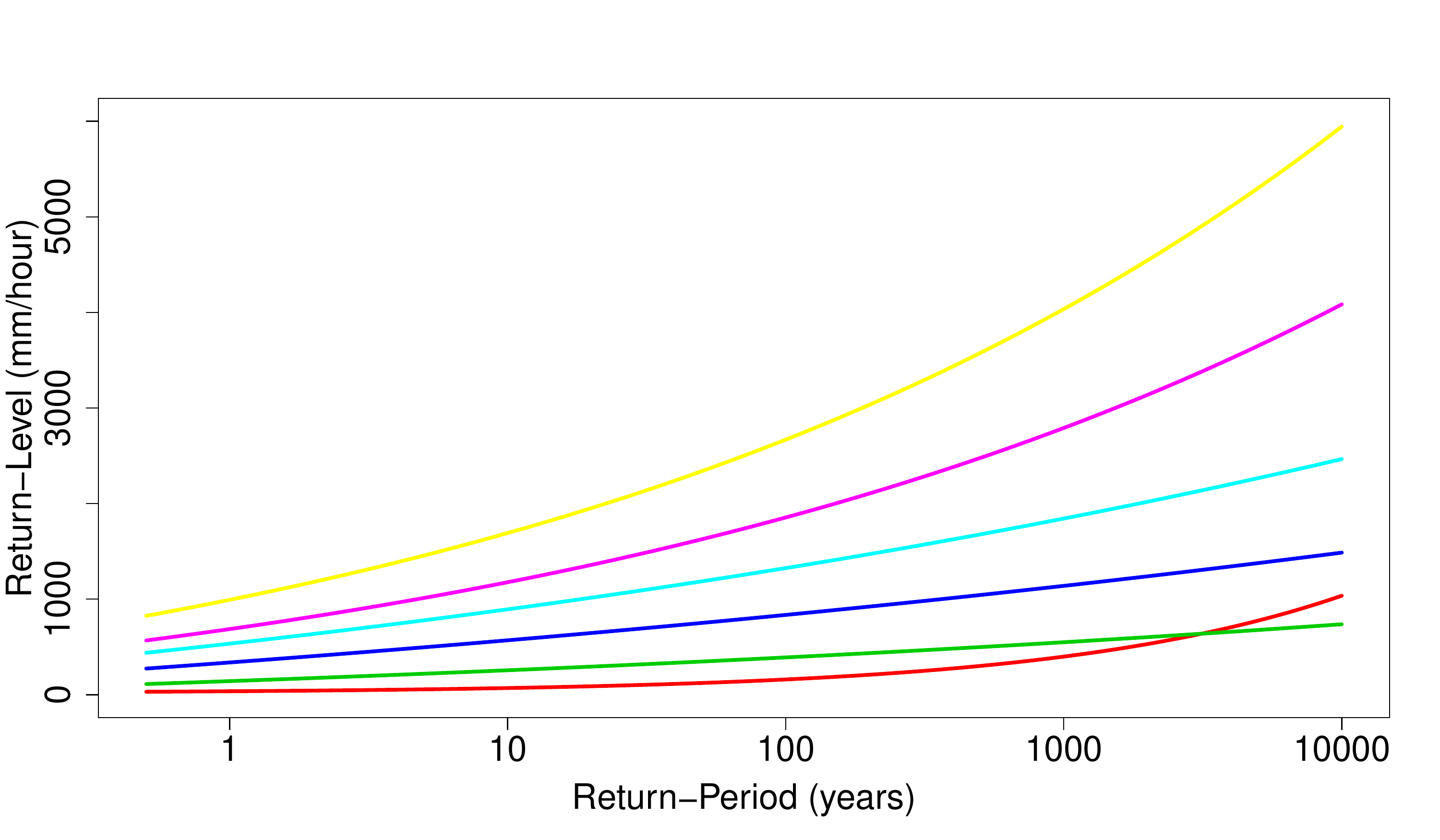} 
\end{minipage}
\begin{minipage}{0.49\textwidth}
\centering
\includegraphics[width=\textwidth]{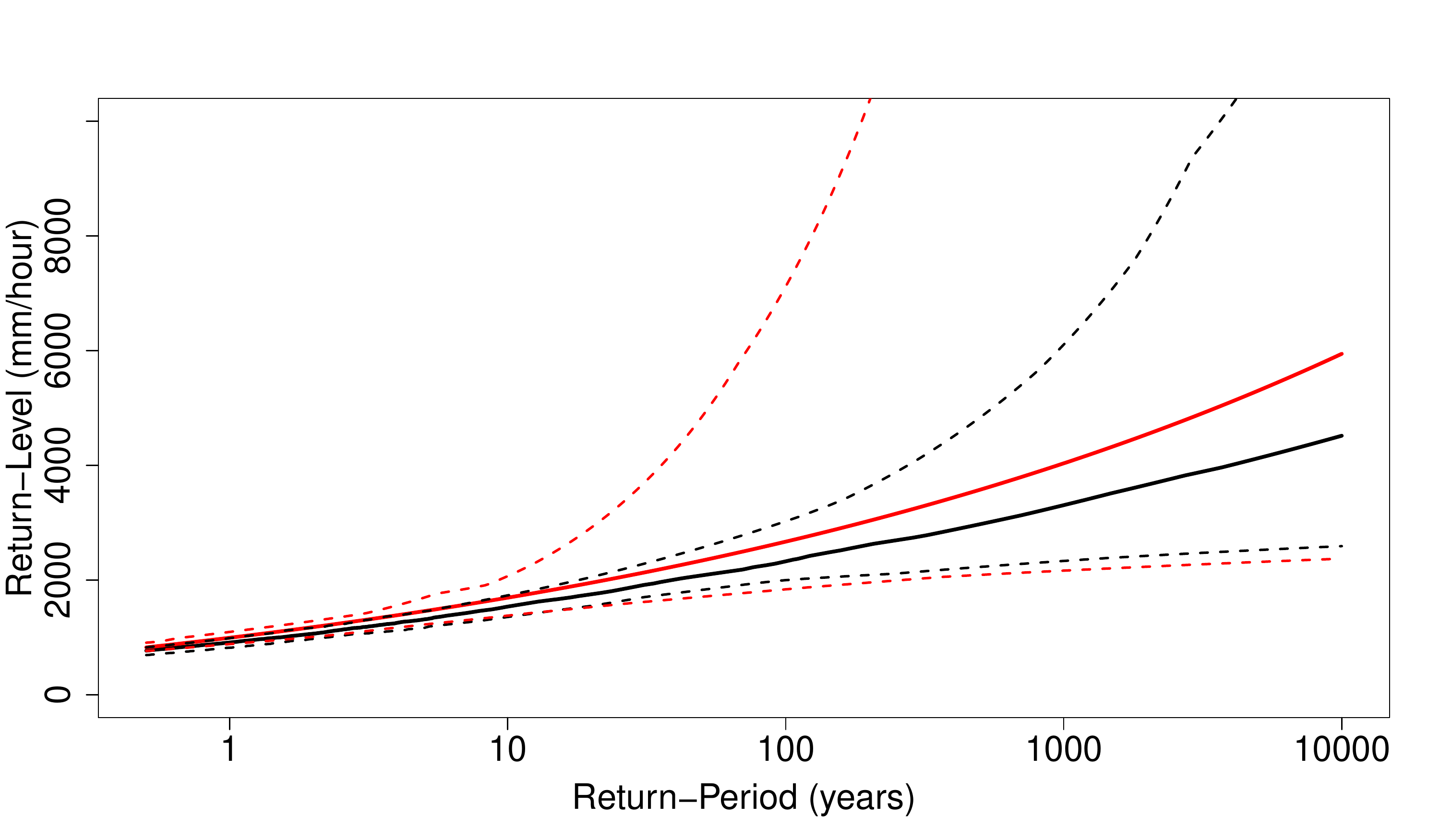} 
\end{minipage}
\begin{minipage}{0.49\textwidth}
\centering
\includegraphics[width=\textwidth]{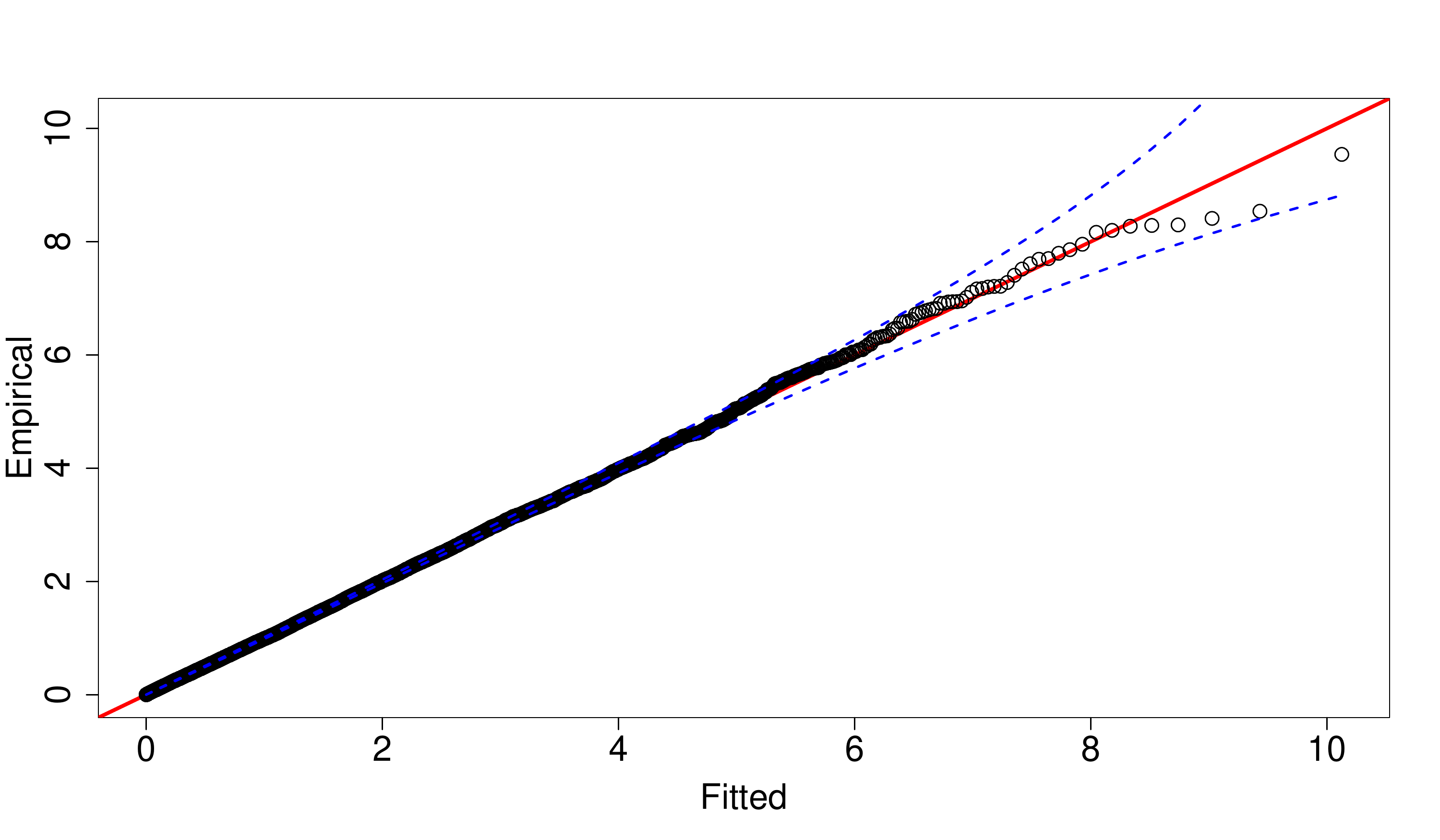} 
\end{minipage}
\caption{Top: Estimated return level curves of $R_{\mathcal{A}}$ using the model (left) and observations (right).  Colours correspond to the regions illustrated in Figure \ref{agg_diag2}. Bottom-left: return level estimates for Region 6 in Figure \ref{agg_diag2} using methods (i) and (ii) in black and red, respectively. $95\%$ confidence intervals for the methods are given by the coloured dashed lines. Bottom-right: Q-Q plot for pooled GPD fit for approach (i), over all $50$ bootstrap samples, on standard Exponential margins. $95\%$ tolerance bounds are given by the dashed lines. }
\label{return_levels}
\end{figure}

As discussed in Section \ref{intro}, obtaining physically consistent return level estimates of spatial aggregates is essential. We compare two methods for achieving this: (i) performing a long-run simulation from our model, deriving empirical estimates of return-levels from these replicates; (ii) fitting a GPD to the observed aggregate tails and extrapolating to the desired return-level. Ideally, we want to use only the former approach as this mitigates the potential issues with using method (ii) discussed in Section \ref{intro}; however, for computational efficiency, we perform a shorter run for method (i) with $5 \times 10^5$ realisations and use a fitted GPD to extrapolate to the largest return-levels. Figure~\ref{return_levels} presents estimates of return level curves for $R_\mathcal{A}$ over the nested regions, illustrated in Figure~\ref{agg_diag2}, using methods (i) and (ii), top-left and top-right panels, respectively. For each region $\mathcal{A}$, a GPD is fitted to exceedances of the respective sample $R_{\mathcal{A}}$ above the $99.9\%$ quantile and return level curves are estimated from these fits.  In the top-right panel of Figure \ref{return_levels}, we observe intersection in the return level curves estimated for the two smallest regions using method (ii). This problem does not arise using the computationally efficient version of approach (i), e.g., in the top-left panel of Figure \ref{return_levels}, where we have $5\times 10^5 \times 20/43200 \approx 231$, and $20$, years of data for inference, respectively. Furthermore, the confidence intervals produced by method (i) are tighter, as more data are used for extrapolation; we illustrate this in the bottom-left panel of Figure \ref{return_levels}, where we overlay return level estimates $R_\mathcal{A}$ using both methods, for a single region $\mathcal{A}$. Confidence intervals are derived for both methods by fitting a GPD to 50 bootstrap samples of $R_\mathcal{A}$: in method (i), these are the samples as described at the top of this section; for (ii), we perform  a simple bootstrap of the observed data, assuming temporal independence. A higher exceedance threshold for approach (i) was considered, but we found that the difference in estimates was negligible; to support the use of the $99.9\%$ quantile, we illustrate a pooled {Q-Q} plot in Figure~\ref{return_levels}, transforming exceedances from all $50$ bootstrap samples onto standard Exponential margins using their respective GPD fits and observe an excellent overall fit.\par
\begin{figure}[h]
\centering
\begin{minipage}{0.32\textwidth}
\centering
\includegraphics[width=\textwidth]{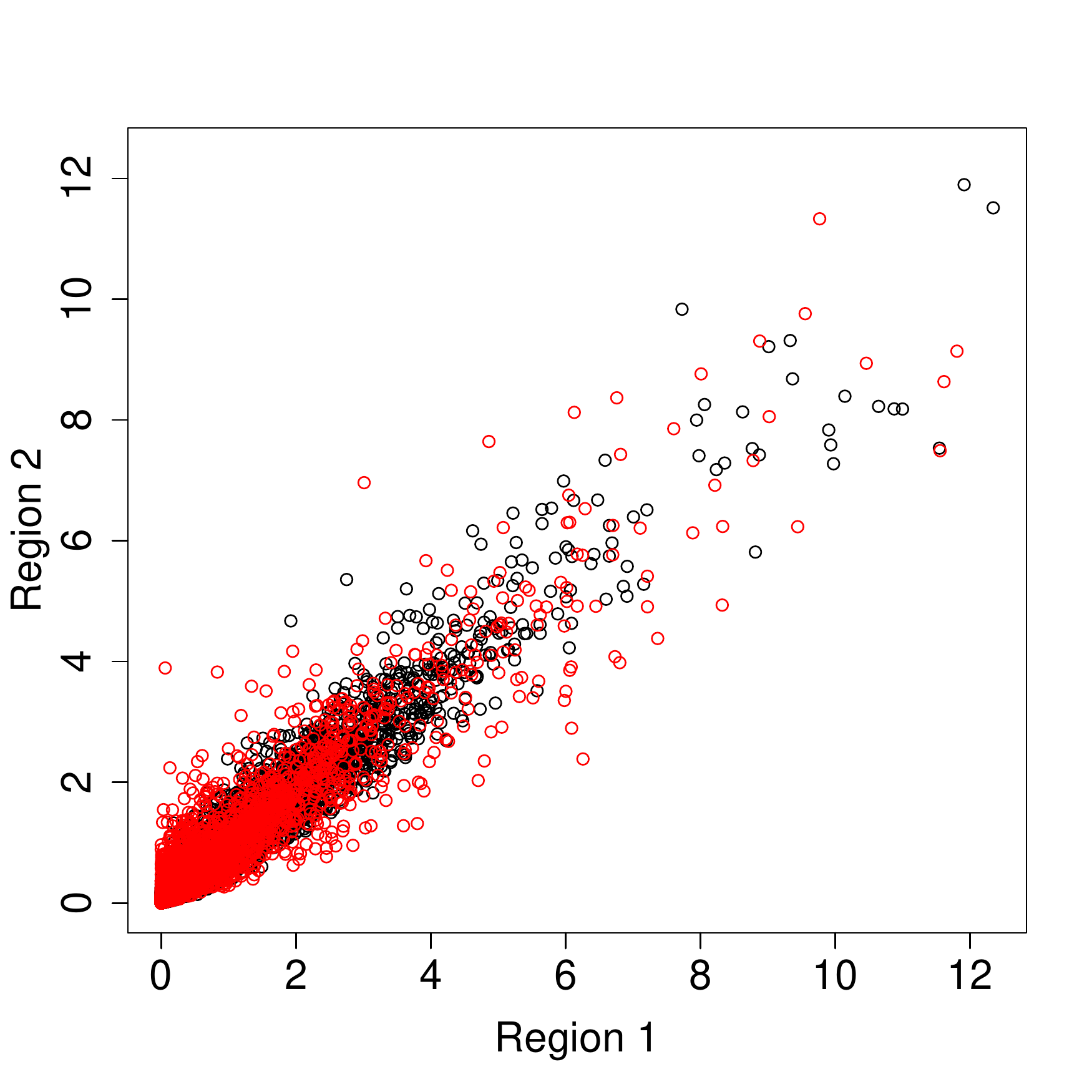} 
\end{minipage}
\begin{minipage}{0.32\textwidth}
\centering
\includegraphics[width=\textwidth]{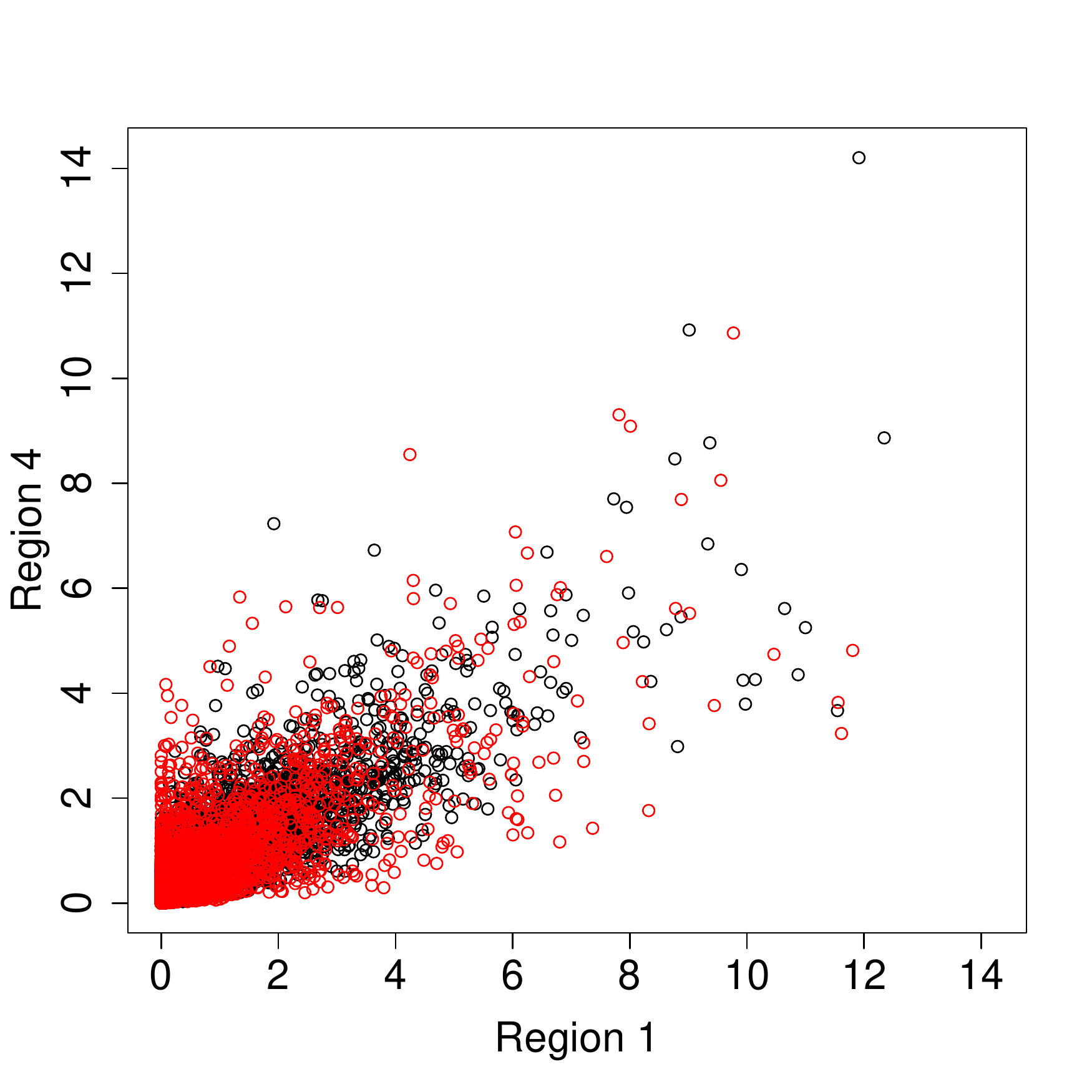} 
\end{minipage}
\begin{minipage}{0.32\textwidth}
\centering
\includegraphics[width=\textwidth]{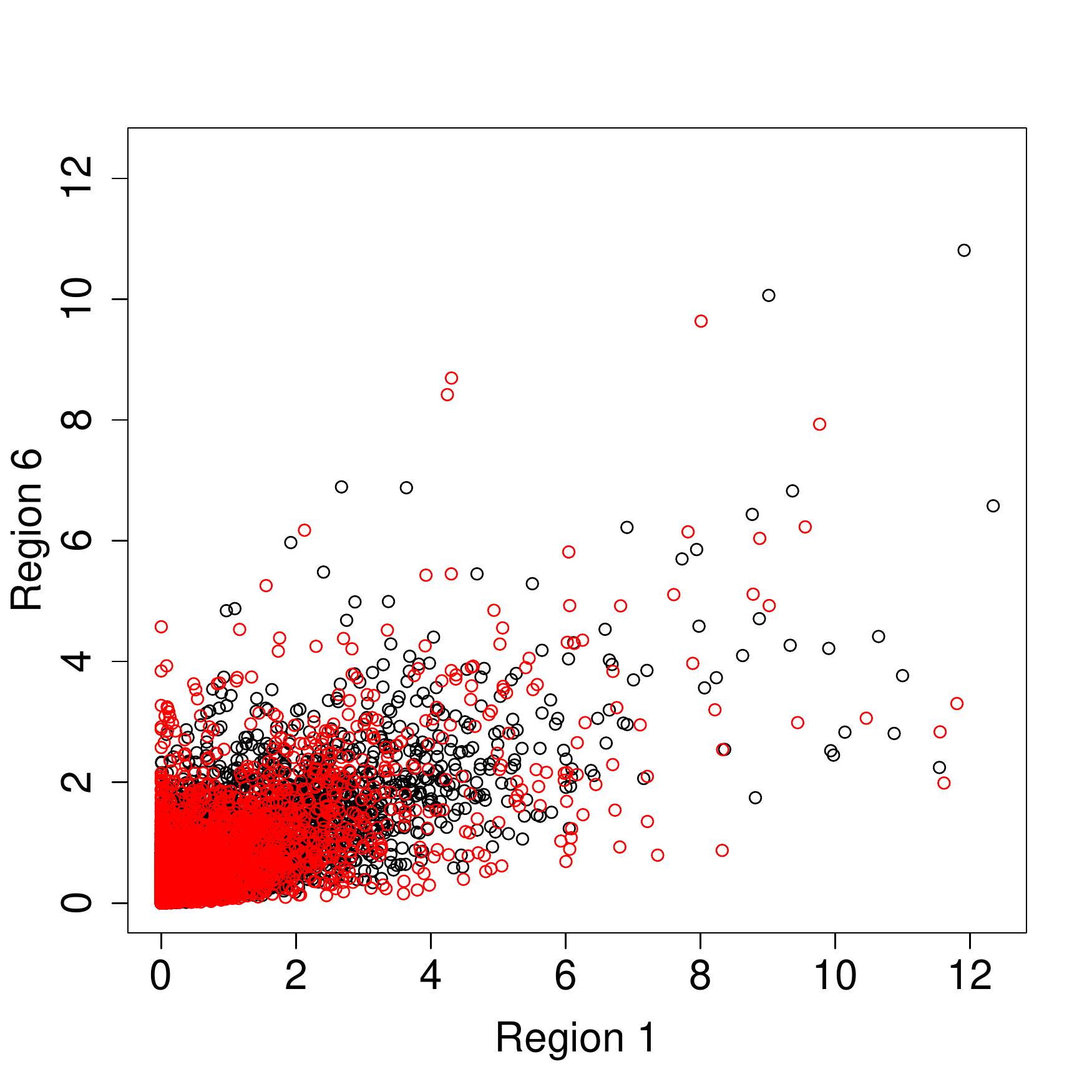} 
\end{minipage}
\caption{Plots of $2\times10^4$ realisations of pairwise $(\bar{R}_{\mathcal{A}},\bar{R}_{\mathcal{B}})$ for nested regions $\mathcal{A},\mathcal{B}$, illustrated in Figure \ref{agg_diag2}. Black points are model estimates, red points are from the data. The regions $\mathcal{A},\mathcal{B}$ are labelled on the respective panels.}
\label{figjoint}
\end{figure}

A further point of interest for practitioners is inference on the joint behaviour of $(\bar{R}_{\mathcal{A}},\bar{R}_{\mathcal{B}})$ for different regions $\mathcal{A},\mathcal{B} \in \mathcal{S}$. We investigate this joint behaviour for the different aggregate regions given in Figure~\ref{agg_diag2}. Figure~\ref{figjoint} illustrates realisations of pairwise $(\bar{R}_{\mathcal{A}},\bar{R}_{\mathcal{B}})$ for both model and empirical estimates with nested regions $\mathcal{A},\mathcal{B}$, showing that the model captures the joint distributions well; a similar figure is given for non-overlapping regions in Figure~\ref{figjointns} of the Supplementary Materials \citep{papersup}, in which we observe that extreme events do not typically occur together. This suggests that the extremal behaviour of the aggregates is driven by spatially-localised events. Further evidence for this can be found in Figure~\ref{figjoint} for aggregates over nested regions; if the extremes of $R_\mathcal{A}$ were driven by events with a spatial profile that encompassed $\mathcal{S}$, then we would not observe weakening extremal dependence between aggregates over the smallest, and increasingly larger, regions. Instead, we would observe no change in the strength of extremal dependence.
 \section{Discussion}
 \label{sec-dicuss}
 We have presented extensions of the \cite{heff2004} and \cite{wadsworth2018spatial} models for modelling the extremal dependence for precipitation data. As illustrated in Section \ref{sec-depmodel}, this model provides flexibility over existing models for extreme precipitation as it can capture asymptotic independence.  Simulating from this model is simple, and replications can be used to make reliable inference about the tail behaviour of spatial aggregates of the underlying process once issues linked to edge effects are addressed. This approach circumvents an issue that is common with independent inference on the tails of spatial aggregates over different regions, namely that they run the risk of making inference that is inconsistent with the physical properties of the process. \par
 A particular drawback of our approach is that inference using the full likelihood is computationally infeasible. To overcome this issue, we proposed methods for model fitting and assessing parameter uncertainty that are based on a pseudo-likelihood approach which requires specification of a hyper-parameter $h_{max}$. We found that these methods worked well in our application, as we were able to choose a suitable $h_{max}$ quite low for which the model fits well in a reasonable time-frame, see Figure~\ref{depestimates}. This is because our data exhibits fairly localised extreme events; in applications where this is not the case, a larger $h_{max}$ will be required which could potentially lead to more samples being required for fitting.\par
Data used in Section \ref{sec-appl} are from a climate model, which means that sampling locations are comprised of non-overlapping grid-boxes, rather than point locations. In our application, we take $R_\mathcal{A}$ to be the corresponding summations, rather than the integrals defined in \eqref{aggeq}; however, this is not to say that our approach cannot be used if we require inference on the tail of an integral. We have detailed a fully spatial model for both the dependence and marginal behaviour of $\{Y(s)\}$, and so it is possible to create a sample of $\{Y(s): s \in \mathcal{S}\}$ where $\mathcal{S}$ is not necessarily the sampling locations. We can then approximate $R_\mathcal{A}$ by specifying $\mathcal{S}$ as a fine-grid and taking the sum of $\{Y(s): s \in \mathcal{A}\}$.\par
When considering spatial aggregates over the largest regions $\mathcal{A}$, we find that our approach slightly underestimates the largest events, see Figure \ref{agg_diag2}. Whilst this may be caused by boundary effects, it could also be caused by a complexity of the data generating process that is not captured by the model. As the size of $\mathcal{A}$ increases, it becomes less likely that the tail behaviour of $R_\mathcal{A}$ is driven by a single type of extreme event. There are two possible areas of complexity that are missed for regions that are sufficiently large: (i) multiple occurrences of localised high-intensity convective events \citep{Schroeer2018}, whereas we modelled single occurrences in Section~\ref{sec-appl}; (ii) events consisting of a mixture of localised high-intensity convective and widespread low-intensity non-convective,  events. Our data appears to exhibit these; recall in Section \ref{sec-appl-dep} we remarked that we considered a lower threshold $u$ in \eqref{depmodeleq} for modelling, but we found that this was not feasible as the data exhibits mixtures of dependence. A higher threshold had to be specified to remove observed fields that exhibited long-range spatial dependence to improve model fitting. Further improvements can be made to inference on the tails of $R_\mathcal{A}$ by modelling frontal events. To illustrate this, consider that we model $R_\mathcal{A}|( \max_{s \in \mathcal{S}} X(s) > v)$, i.e., $R_\mathcal{A}$ given an extreme event somewhere in $\mathcal{S}$, and undo said conditioning using the data. We do not model $R_\mathcal{A}$ given that there is no extreme event anywhere in $\mathcal{S}$, i.e., such as caused by a frontal event. As the size of $\mathcal{A}$ grows, we will increasingly find that these events will drive the extremal behaviour of $R_\mathcal{A}$; this could be further explanation behind the underestimation in Figure~\ref{agg_diag2}, and so should be incorporated into the model. We also considered another measure of the extremal dependence in Appendix~\ref{append-depmodel} of the Supplementary Materials \citep{papersup}, which suggests improvement may be possible using mixture modelling. A possible approach to this problem is to incorporate covariates on precipitation field type into the model.
 
\section*{Acknowledgements} Jordan Richards and Jonathan Tawn gratefully acknowledge funding through the STOR-i Doctoral Training Centre and Engineering and Physical Sciences Research Council (grant EP/L015692/1). Simon Brown was supported by the Met Office Hadley Centre Climate Programme funded by BEIS and Defra. The authors are grateful to Robert Shooter of the Met Office Hadley Centre, UK, and to Jennifer Wadsworth and Emma Simpson of  Lancaster University and University College London, respectively, for helpful discussions. The data can be downloaded from the CEDA data catalogue, see \cite{datacite}. Code that supports our findings can be found in the \href{http:/github.com/Jbrich95/scePrecip}{Jbrich95/scePrecip} repository on GitHub.
\bibliographystyle{imsart-nameyear} 
\bibliography{ref}       

\section*{Appendix}
\setcounter{figure}{0}
\renewcommand{\thefigure}{S\arabic{figure}}
 \renewcommand{\thetable}{S\arabic{table}}%
\begin{appendix}
\section{Delta-Laplace conditional distribution functions}
\label{censorsec-append}

\subsection{Connection to main text}
This appendix supports the material presented in Sections \ref{censorsec} and \ref{modelfit} of the main paper. This appendix describes contributions to the pairwise likelihood function in \eqref{fulllik} given by the conditional distribution associated with the residual distribution described in Section \ref{residprocess} of the main paper.
\subsection{Bivariate conditional distribution} 
We note that if we had $Y(s_i) > 0$ for all $i=1,\dots,d$, then the corresponding joint density of \eqref{gausscopeq} for observed residuals $\mathbf{z}=(z_1,\dots,z_{d-1}) \in ((c_1,\infty)\times\dots\times(c_{d-1},\infty))$ is
\begin{align}
\label{gausscopeq-dens}
f_{s_O}(\mathbf{z})=\phi_{d-1}\left\{\Phi^{-1}(F_{Z(s_1|s_O)}(z_1)),\dots,\Phi^{-1}(F_{Z(s_{d-1}|s_O)}(z_{d-1}));\mathbf{0},\Sigma\right\} \prod_{i=1}^{d-1}\frac{f_{Z(s_{i}|s_O)}(z_i)}{{\phi(\Phi^{-1}(F_{Z(s_{i}|s_O)}))}},
\end{align}
where $c_1,\dots,c_{d-1}$ are the censoring thresholds defined in Section~\ref{modelfit} of the main text, $\phi(\cdot)$ is the PDF of a standard Gaussian distribution and $\phi_{d-1}(\cdot; \mathbf{0},\Sigma)$ is the CDF of a $(d-1)-$dimensional Gaussian distribution with mean $\mathbf{0}$ and correlation matrix $\Sigma$ defined in \eqref{cormat}; components of $\mathbf{z}$ are not censored when in this domain.  \par 
For censoring thresholds $c_1,c_2$, the bivariate density with the conditioning constraints is
\begin{equation}
\label{compLikeq}
g_{s_O}(z_1,z_2;c_1,c_2)=\begin{cases}
f_{s_O}(z_1,z_2) \;\;&\text{if}\;\;z_1 > c_1, z_2 > c_2,\\
f_{s_O}(z_1)F_{s_O,\; 2|1}\{c_2,z_1\}\;\;&\text{if}\;\;z_1 > c_1, z_2 \leq c_2,\\
f_{s_O}(z_2)F_{s_O,\; 1|2}\{c_1,z_2\}\;\;&\text{if}\;\;z_1 \leq c_1, z_2 > c_2,\\
F_{s_O}(c_1,c_2)\;\;&\text{if}\;\;z_1 \leq c_1, z_2 \leq c_2,
\end{cases}
\end{equation}
where the conditional distribution $F_{s_O,\; i|j}(c,z)$ is given by
\begin{equation}
\label{condCDFbiv}
F_{s_O,\; i|j}(c_i,z_j)=\Phi\left\{\Phi^{-1}(F_{Z(s_{i}|s_O)}(c_i));\mu_{i|j},\Sigma_{i|j}\right\},
\end{equation}
where $\mu_{i|j}=\Sigma_{ij}\Sigma^{-1}_{jj}\Phi^{-1}\{F_{Z(s_{j}|s_O)}(z_j)\}$ and $\Sigma_{i|j}=\Sigma_{ij}-\Sigma_{ij}^2\Sigma_{jj}^{-1}$ with $\Sigma$ defined in \eqref{cormat}. Both $f_{s_O}(\cdot,\cdot)$ and $F_{s_O}(\cdot,\cdot)$ are the bivariate analogues of \eqref{gausscopeq} and \eqref{gausscopeq-dens} and $F_{Z(s_{i}|s_O)}$ is the CDF of a delta-Laplace distribution with parameters given in \eqref{sigmudeltaeqs}; these and $\Sigma$ are all identifiable as they are all fully determined by pairwise distances. 
\subsection{Multivariate extension of $\eqref{condCDFbiv}$}
Suppose we have some $s_O \in \mathcal{S}$ and $\boldsymbol{s}_A\in\mathcal{S}\setminus s_O$, which need not be any of the original sampling locations, and is indexed such that $\mathbf{s}_A=(s_1, \dots, s_{n_A})$. We partition the indices of set $\boldsymbol{s}_A$ into two sets; indices for censored sites  and non-censored sites, $A_c$ and $A_{nc}$, respectively and without loss of generality we write $A_c = (s_1,\dots,s_{n_{c}})$ and $A_{nc} = (s_{n_{c}+1},\dots,s_{n_{A}})$. If sites in $A_c$ are censored with thresholds $c_1, \dots, c_{n_{c}}$, then the conditional distribution function of
\[
\bigg(Z(s_1)|s_O),\dots,Z(s_{n_c})|s_O)\bigg)\bigg|\bigg(Z(s_{n_{c}+1}|s_O)=z_{n_c+1},\dots,Z(s_{n_A}|s_O)=z_{n_A}\bigg)
\]
is $F_{s_O,\; A_c|A_{nc}}(c_1,\dots,c_{n_c}, z_{n_c+1},\dots,z_{n_A})$ equals to
\begin{equation}
\label{CondCDFmultiv}
\Phi_{n_c}\left\{\Phi^{-1}(F_{Z(s_1|s_O)}(c_1)),\dots,\Phi^{-1}(F_{Z(s_{n_c}|s_O)}(c_{n_c}));\mu_{A_c|A_{nc}},\Sigma_{A_c|A_{nc}}\right\},
\end{equation}
where
\begin{align*}
\mu_{A_c|A_{nc}}&=\Sigma_{A_cA_{nc}}\Sigma^{-1}_{A_{nc}A_{nc}}(\Phi^{-1}\{F_{Z(s_{n_c+1}|s_O)}(z_{n_c+1})\}, \dots, \Phi^{-1}\{F_{Z(s_{n_A}|s_O)}(z_{n_A})\})^T\\
\Sigma_{A_c|A_{nc}}&=\Sigma_{A_{c}A_{c}}-\Sigma_{A_{c}A_{nc}}\Sigma_{A_{nc}A_{nc}}^{-1}\Sigma_{A_{nc}A_{c}}.
\end{align*}
The notation $\Sigma_{AB}$ denotes the partition of $\Sigma$ that takes rows indexed by the set $A$ and columns indexed by the set $B$.
\section{Reliability of stratified sampling regime for pseudo-likelihood estimation}
\label{append-infersimstudy}
\subsection{Connection to main text}
This appendix supports the material in Section \ref{sec-strat} of the main paper. Appendix~\ref{ap_simstudy} details a simulation study that we conduct to assess the reliability of parameter estimation using the stratified sub-sampling procedure for inference detailed in Section~\ref{modelfit} of the main paper.
\subsection{Simulation study} 
\label{ap_simstudy}
We proceed by conducting a parametric bootstrap to assess the accuracy of the pseudo-likelihood estimator, denoted here as $\hat{\boldsymbol{\psi}}$, that is described in Section~\ref{sec-strat} of the main paper; in particular, we aim to investigate if the estimator $\hat{\boldsymbol{\psi}}$ is biased. We begin by taking a single conditioning site $s_O$ in the centre of the spatial domain $\mathcal{S}$ (see Section~\ref{sec-appl} of the main paper), and assume that the true underlying distribution of $\{Y(s):s\in\mathcal{S}\}|Y(s_O)>u$ is characterised by the model described in Section~\ref{sec-depmodel} of the main paper, with true parameter values denoted by $\boldsymbol{\psi}_0$; we set $\boldsymbol{\psi}_0$ to be the values provided in Table~\ref{parEst}. For $50$ repetitions, we draw 1000 realisations of 
$\{Y(s):s\in\mathcal{S}\}|Y(s_O)>u$ using $\boldsymbol{\psi}_0$ and to each of these samples we apply our pseudo-likelihood estimation approach with $h_{max}=28km$ and $d_s=1000$; note that the maximum number of triples with a single conditioning site is $4656$, and so to ensure that we use only a sub-sample for parameter estimation, we take $d_s=1000$. Moreover, we follow the justification given in Section~\ref{sec-depmodel} of the main paper and treat $\kappa_{\alpha_3},\kappa_{\beta_3}$ and $\kappa_{\delta_4}$ as fixed. Using this procedure provides $50$ estimates of $\hat{\boldsymbol{\psi}}$ which we use to assess its sampling distribution.\par
\begin{figure}[h]
\centering
\includegraphics[width=0.9\linewidth]{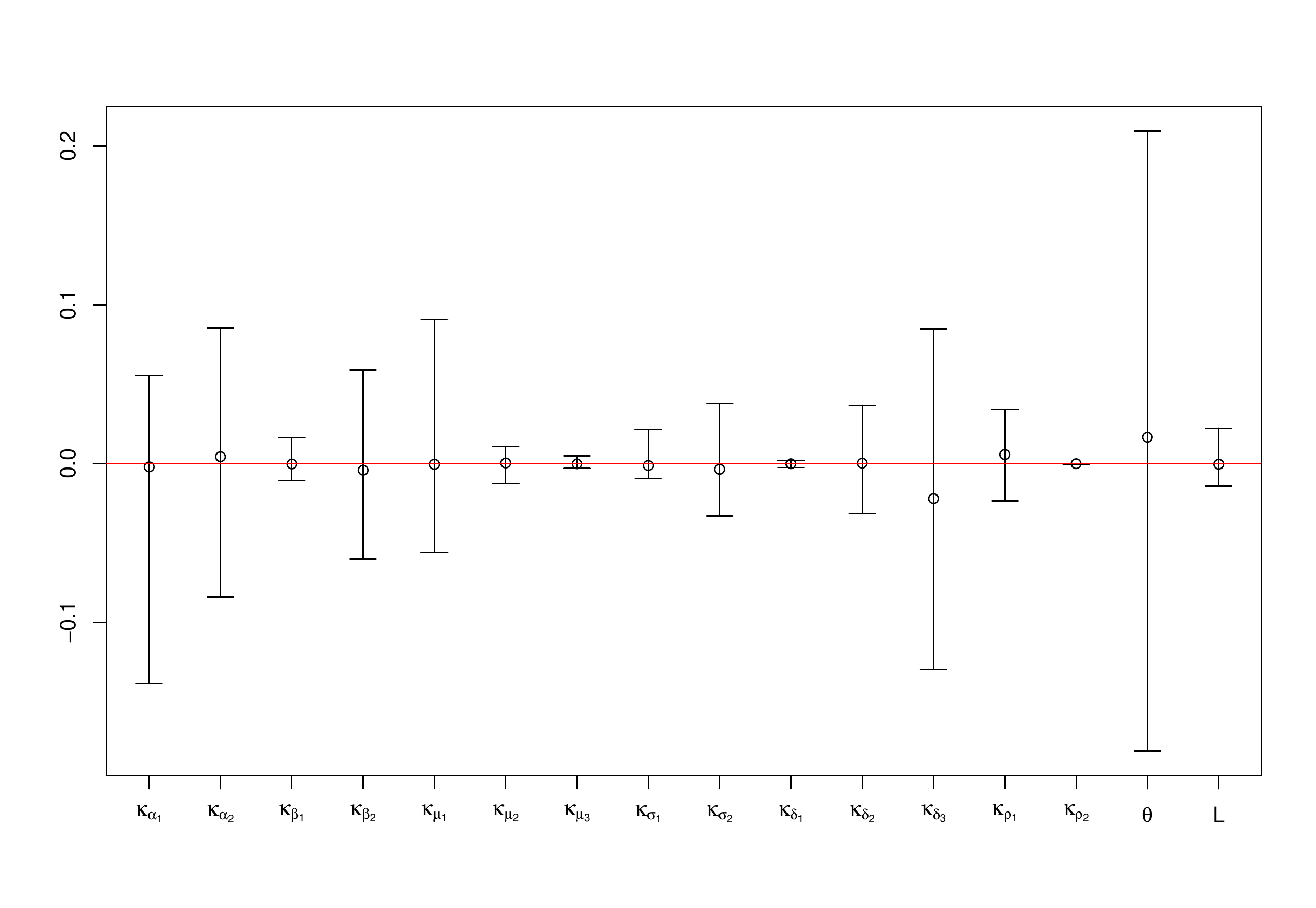} 
\caption{Illustration of the sampling distribution of $(\hat{\boldsymbol{\psi}}-\boldsymbol{\psi}_0)/\boldsymbol{\psi}_0$. The black dots denote estimates of the component-wise medians of the sampling distribution of $(\hat{\boldsymbol{\psi}}-\boldsymbol{\psi}_0)/\boldsymbol{\psi}_0$ with components labelled along the $x$-axis. Estimates of the 95$\%$ confidence intervals of the sampling distribution are denoted by the intervals. The red horizontal line passes through zero. }
\label{siminferCI}
\end{figure}

In Figure~\ref{siminferCI}, we illustrate the sampling distribution of $\hat{\boldsymbol{\psi}}$ by plotting estimates of the componentwise medians and $95\%$ confidence intervals for each component of the relative error, i.e., $(\hat{\boldsymbol{\psi}}-\boldsymbol{\psi}_0)/\boldsymbol{\psi}_0$; recall that $\boldsymbol{\psi}$ is composed of $16$ estimatable parameters. We observe that, for the majority of the parameters, the median of the sampling distribution of $(\hat{\boldsymbol{\psi}}-\boldsymbol{\psi}_0)/\boldsymbol{\psi}_0$ is close to zero, suggesting that the estimator $\hat{\boldsymbol{\psi}}$ provides unbiased estimation in these cases. Whilst small bias may be present for some parameters, e.g., $\theta$ and $\kappa_{\delta_3}$, we note that, due to the strong dependence between components of $\boldsymbol{\psi}$, that $50$ bootstrap repetitions may not be sufficient to accurately estimate the sampling distribution of all components of $\hat{\boldsymbol{\psi}}$; however, as we also find good fits for the aggregate diagnostics in Figure~\ref{agg_diag2} in the main paper, we believe that there is sufficient evidence to suggest that the pseudo-likelihood estimator we use is reliable.

\section{Application dependence model evaluation}
\label{append-depmodel}
\subsection{Connection to main text}
This appendix supports the material in Section \ref{sec-appl-dep} of the main paper. Appendix~\ref{append-deptable} details the parameter estimates for the spatial AI model and Appendix~\ref{chizero-append} details a diagnostic, based on $\chi$, for evaluating the fitted dependence model.
\subsection{Table of AI model parameter estimates}
\label{append-deptable}
Table \ref{parEst} gives the estimates and standard errors for the model parameters discussed in Section \ref{sec-appl-dep} of the main paper. Note that those parameters without standard errors were treated as fixed in the model fit.
\begin{table}[H]
\caption{Parameter estimates (standard errors) to 2 d.p.}
 \centering
 \begin{tabular}{c|c|c|c|c|c|c|c|c} 
\hline
\multicolumn{3}{c|}{$\alpha(h)$}& \multicolumn{3}{|c|}{$\beta(h)$}& \multicolumn{3}{|c}{$\mu(h)$}\\
\hline
   $\kappa_{\alpha_1}$ & $\kappa_{\alpha_2}$ & $\Delta$ & $\kappa_{\beta_1}$ & $\kappa_{\beta_2}$& $\kappa_{\beta_3}$ &  $\kappa_{\mu_1}$ & $\kappa_{\mu_2}$ &$\kappa_{\mu_3}$  \\
 \hline
1.95 (0.21) & 0.73 (0.03) & 0.00& 38.58 (3.10)& 1.02 (0.03) & 1.00 & 0.65 (0.05) &0.28 (0.02) &140.00 (10.20)\\
\hline
\multicolumn{9}{c}{\vspace{0.5em}}\\
  \end{tabular}
  \begin{tabular}{c|c|c|c|c|c} 
  \hline
 \multicolumn{2}{c|}{$\sigma(h)$}&\multicolumn{4}{|c}{$\delta(h)$}\\
 \hline
 $\kappa_{\sigma_1}$ & $\kappa_{\sigma_2}$&$\kappa_{\delta_1}$ & $\kappa_{\delta_2}$ &$\kappa_{\delta_3}$&$\kappa_{\delta_4}$\\
 \hline
 34.22 (2.67) &0.89 (0.03) &0.43 (0.04) &0.46  (0.03) &142.14 (1.72) &1.00 \\
\hline
\multicolumn{6}{c}{\vspace{0.5em}}\\
  \end{tabular}
  \begin{tabular}{c|c|c|c} 
  \hline
\multicolumn{2}{c|}{$\rho(h)$}&\multicolumn{2}{|c}{\eqref{anisoeq}}\\
 \hline
$\kappa_{\rho_1}$ & $\kappa_{\rho_2}$&$\theta$&$L$\\
 \hline
58.71 (3.67)&0.53 (0.01) &-0.18 (0.02)&0.93 (0.01)\\
 \hline
 \end{tabular}
 \label{parEst}
 \end{table}
 \subsection{Probability of no rain diagnostic}
 \label{chizero-append}
\begin{figure}[h]
\centering
\includegraphics[width=0.9\linewidth]{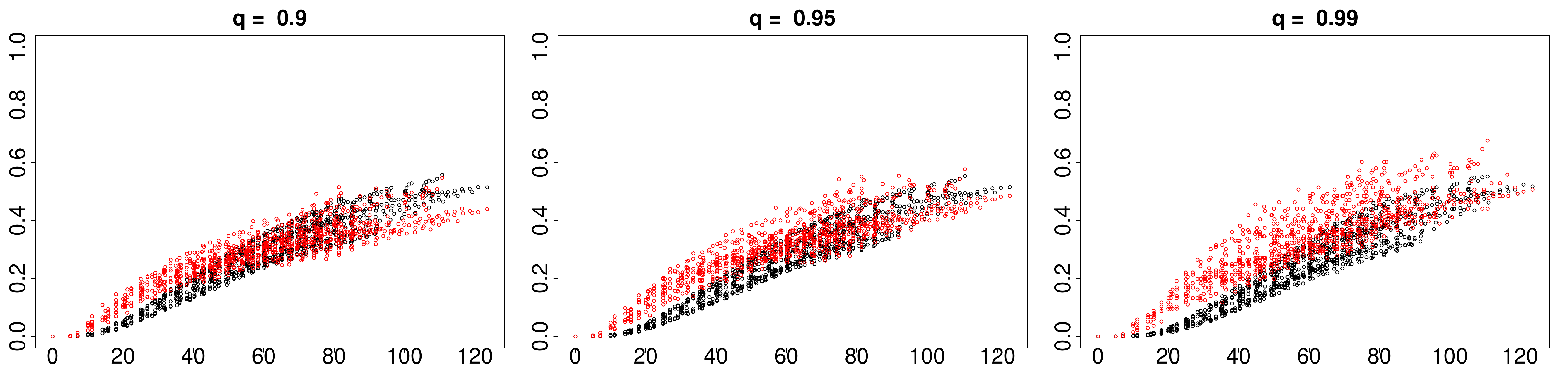} 
\caption{Estimates of $\chi_q^{(0)}(s,s_O)$ against distance $h(s,s_O)$ for $q = (0.9,0.95, 0.99)$. Red points denote empirical estimates, black points denote estimates derived from simulations from the fitted model.}
\label{chizerofig}
\end{figure}

 We further evaluate the extremal behaviour of our fitted model by proposing a novel statistic for extremal dependence of precipitation processes that is based on $\chi$. We define
 \begin{equation}
 \label{chizero}
 \chi^{(0)}_q(s,s_O)=\Pr\left\{\left(Y(s) = 0\right)|\left( Y(s_O) > F^{-1}_{Y(s_O)}(q)\right)\right\},
 \end{equation}
 for $s,s_O \in S$ and $q \in [0,1]$. This is the probability of observing no rain at $s$ given that an extreme is observed at $s_O$. If $\{Y(s)\}$ is truly stationary in both its marginal and dependence structures, we would expect this measure to be a function of distance; results in Section \ref{sec_appl-marg} of the main text suggest the former is not true, but that there is only limited marginal non-stationarity. Figure~\ref{chizerofig} compares estimates of \eqref{chizero} from the model against estimates from the data, for different $q$, with $s_O$ in the centre of $\mathcal{S}$. Model estimates are calculated empirically from $1 \times 10^6$ realisations of $\{Y(s): s \in \mathcal{S}\}$, which are drawn using the scheme described in Section \ref{sim} of the main text. For lower values of $q$, the model captures the empirical values for the probability reasonably well. However, as $q$ increases, model estimates of \eqref{chizero} decrease whilst empirical estimates remain broadly the same. This could suggest that our model for $\{Y(s)\}$ is unable to capture some of the dependence behaviour in the data, i.e., that caused by a mixture of processes.
 \section{Supplementary Figures}
 \label{supfigs}
 \begin{figure}[h]
\centering
\includegraphics[width=0.6\linewidth]{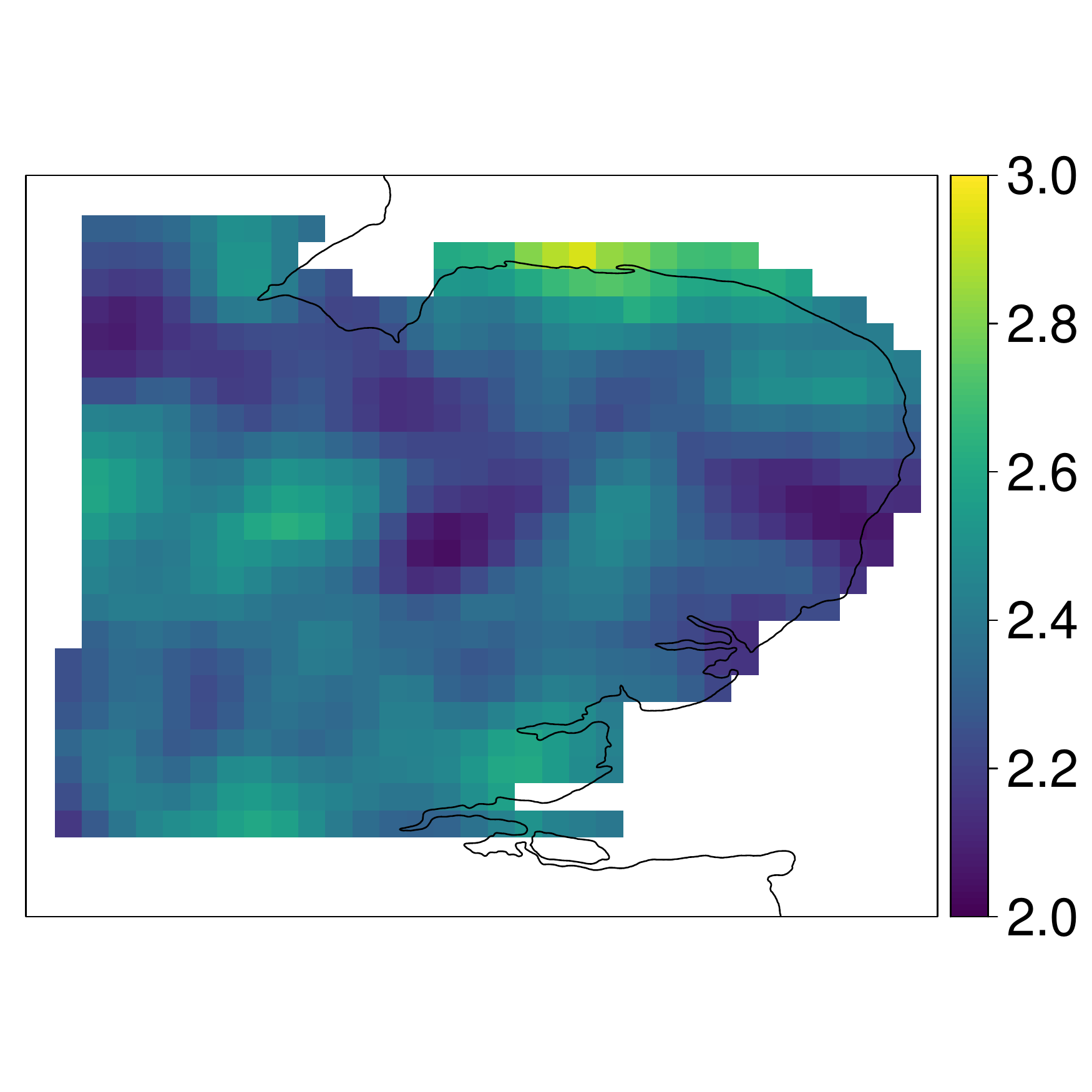} 
\caption{Spatially smoothed estimate of $\upsilon(s)$ for East Anglia.}
\label{GAMscale}
\end{figure}
 \begin{figure}[h]
\centering
\begin{minipage}{0.3\linewidth}
\centering
\includegraphics[width=\linewidth]{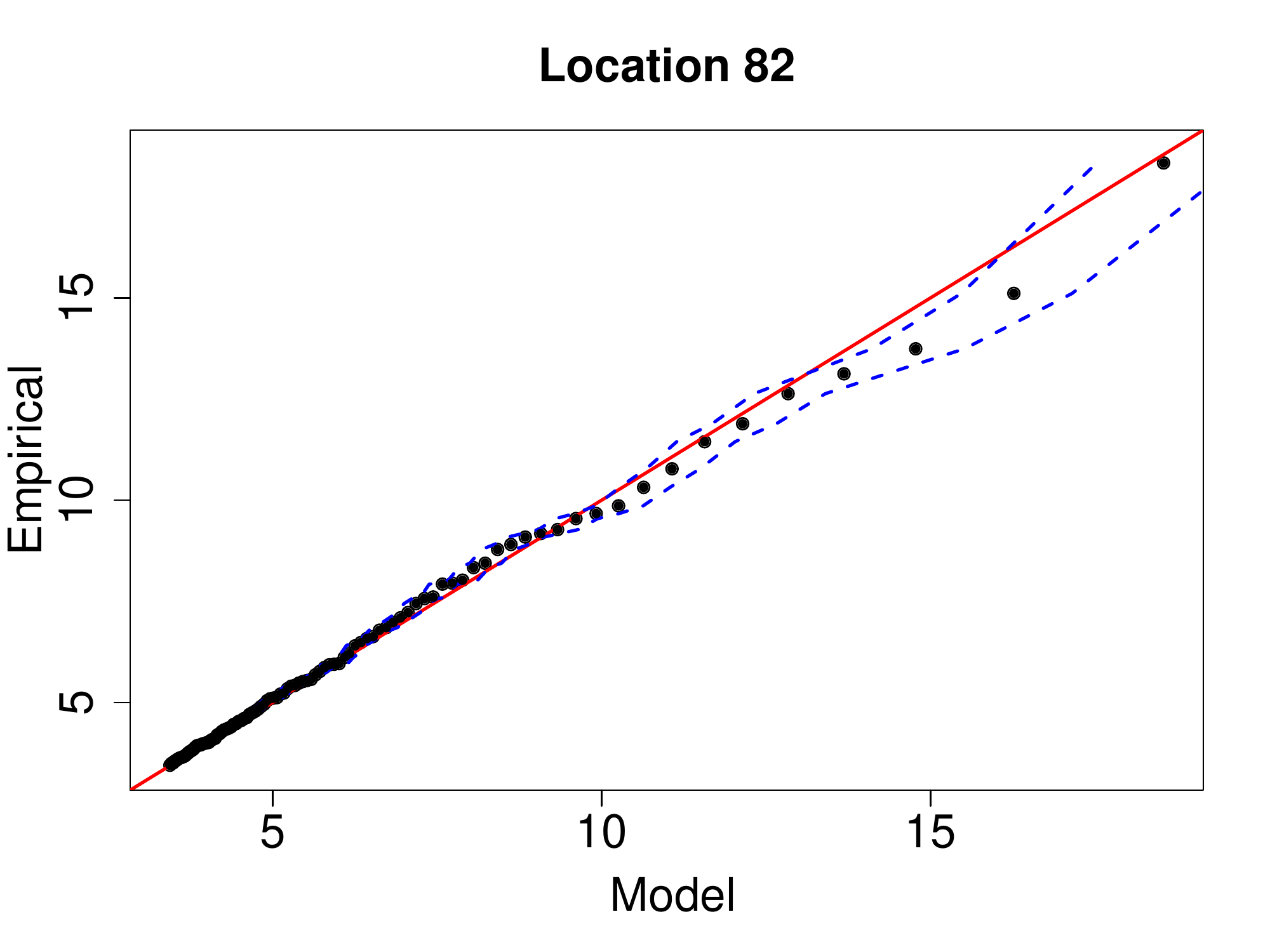} 
\end{minipage}
\begin{minipage}{0.3\linewidth}
\centering
\includegraphics[width=\linewidth]{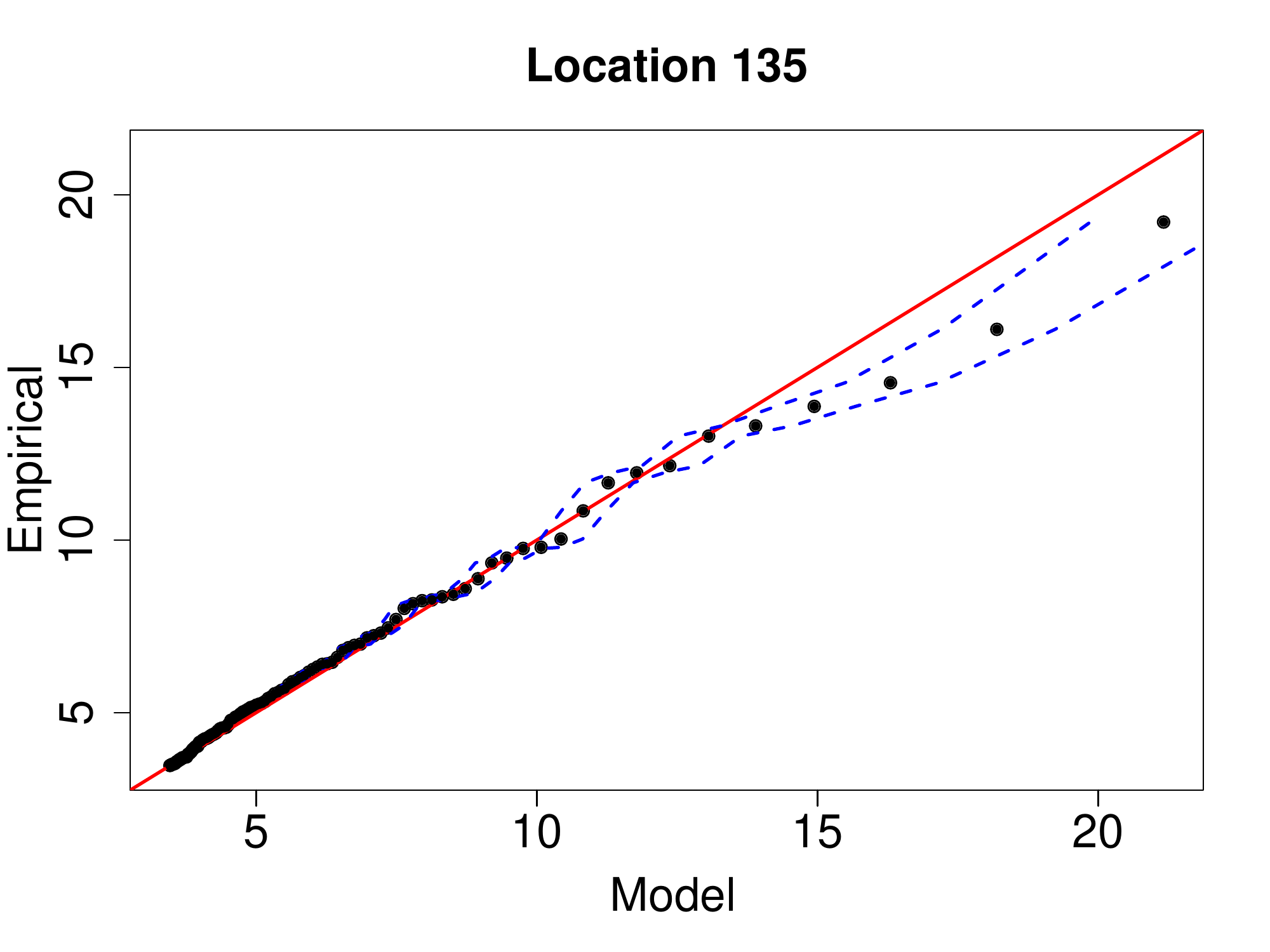} 
\end{minipage}
\begin{minipage}{0.3\linewidth}
\centering
\includegraphics[width=\linewidth]{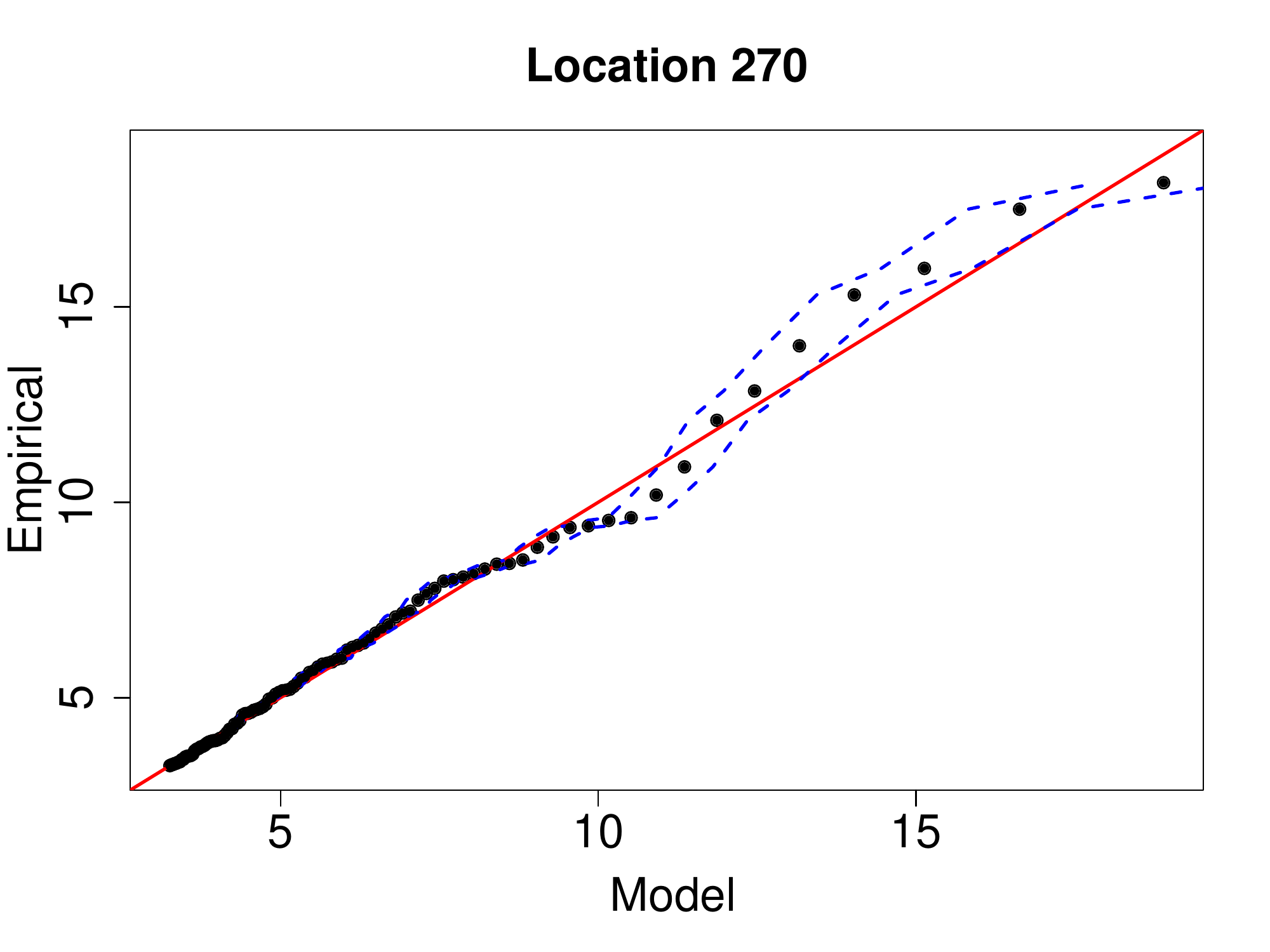} 
\end{minipage}
\begin{minipage}{0.3\linewidth}
\centering
\includegraphics[width=\linewidth]{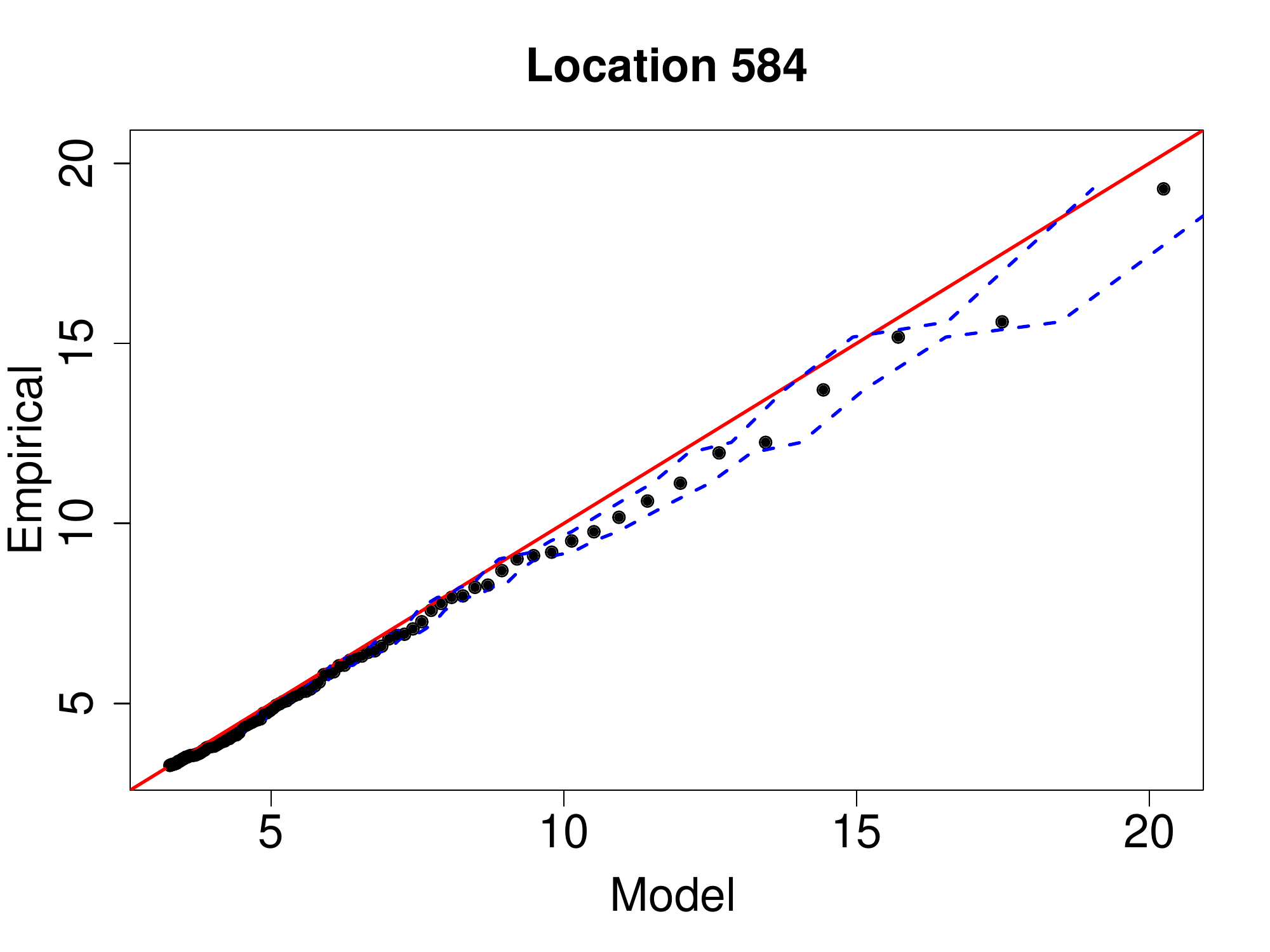} 
\end{minipage}
\begin{minipage}{0.3\linewidth}
\centering
\includegraphics[width=\linewidth]{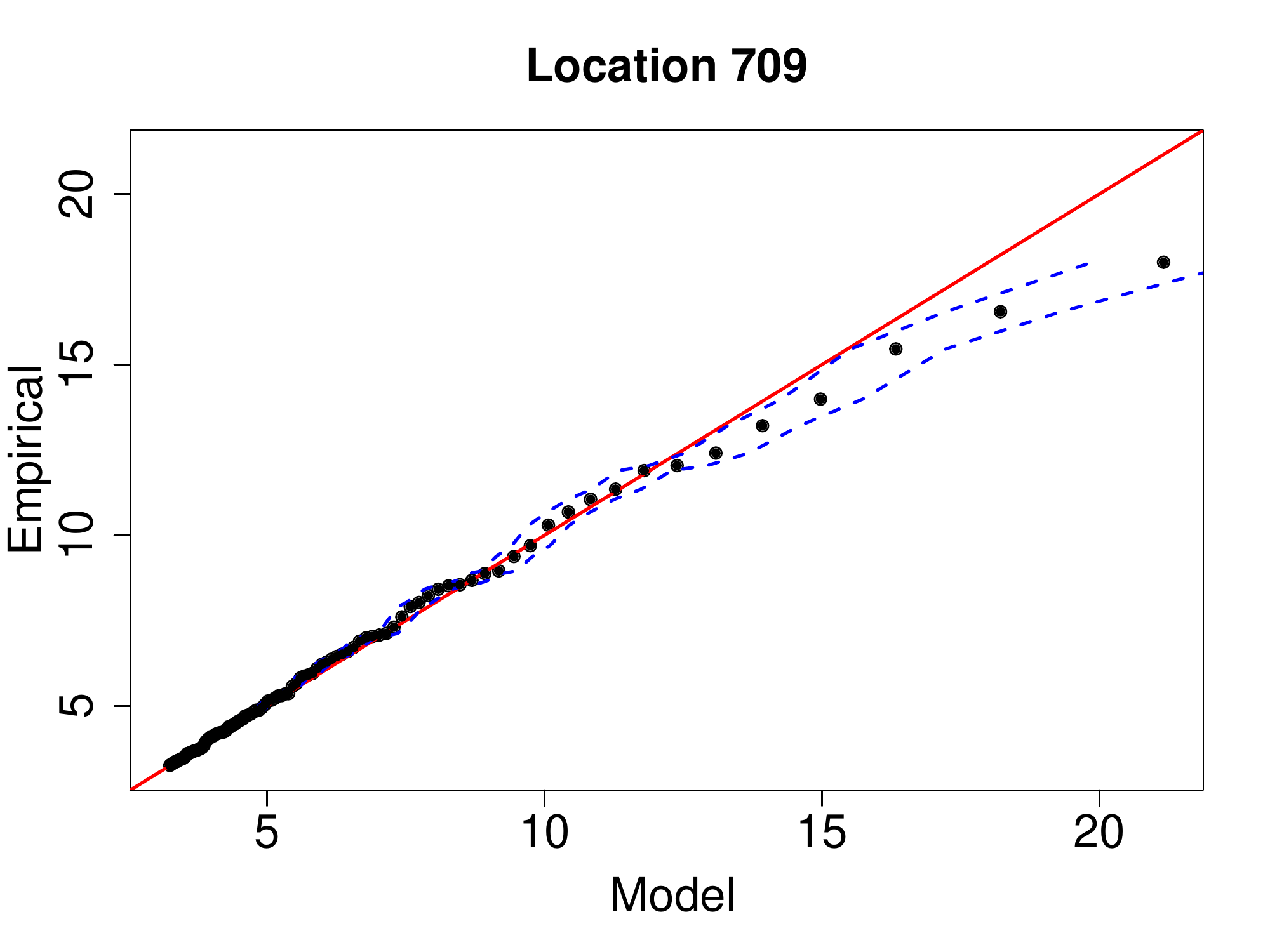} 
\end{minipage}
\begin{minipage}{0.3\linewidth}
\centering
\includegraphics[width=\linewidth]{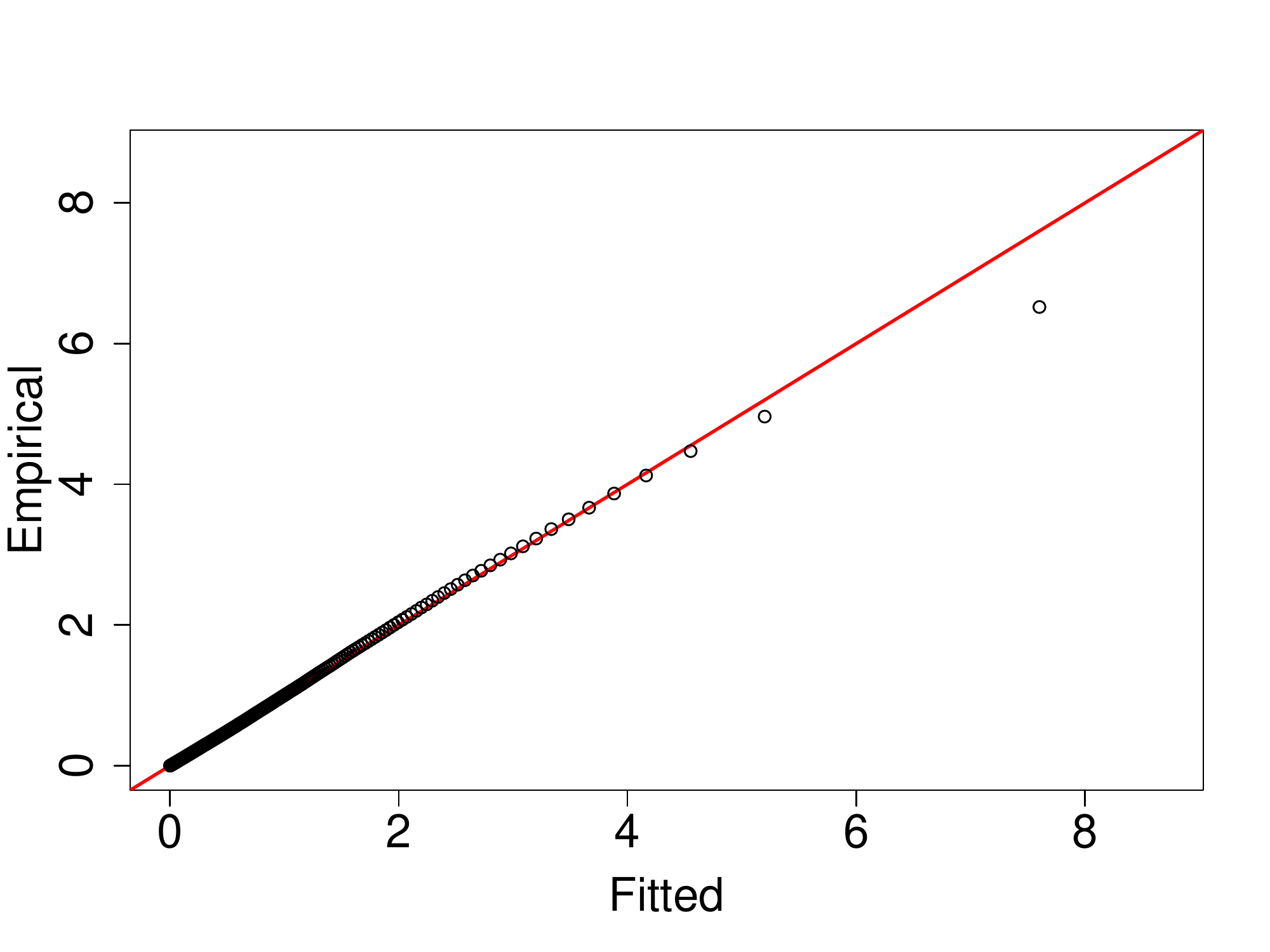} 
\end{minipage}
\caption{ Q-Q plots for the fitted GAM $GPD$ distributions at five randomly sampled sites in $\mathcal{S}$. The $95\%$ confidence intervals are given by the blue dashed lines. Bottom-right: Q-Q plot for pooled marginal transformation over all sites to standard exponential margins. }
\label{Gamdiag}
\end{figure}
\begin{figure}[h]
\centering
\includegraphics[width=0.75\textwidth]{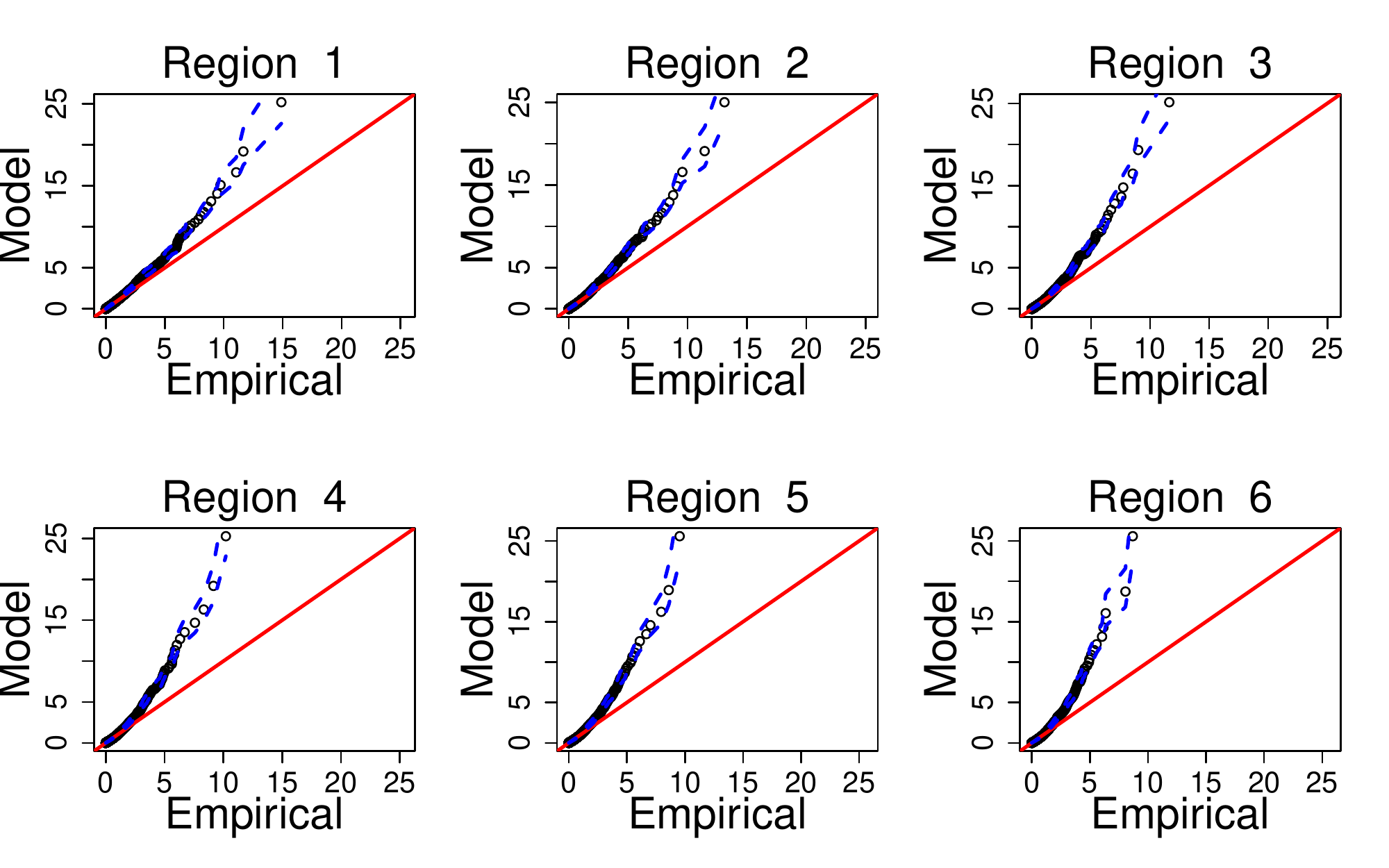} 

\caption{Q-Q plots for AD model, and empirical, $\bar{R}_{\mathcal{A}}$ of regions of increasing size; these are illustrated in Figure~\ref{agg_diag2} of the main text. Probabilities range from $0.7$ to a value corresponding to the 20 year return level, with $95\%$ confidence intervals given by the blue dashed lines.  }
\label{agg_ad}
\end{figure}
\begin{figure}
\begin{minipage}{0.58\textwidth}
\centering
\includegraphics[width=\textwidth]{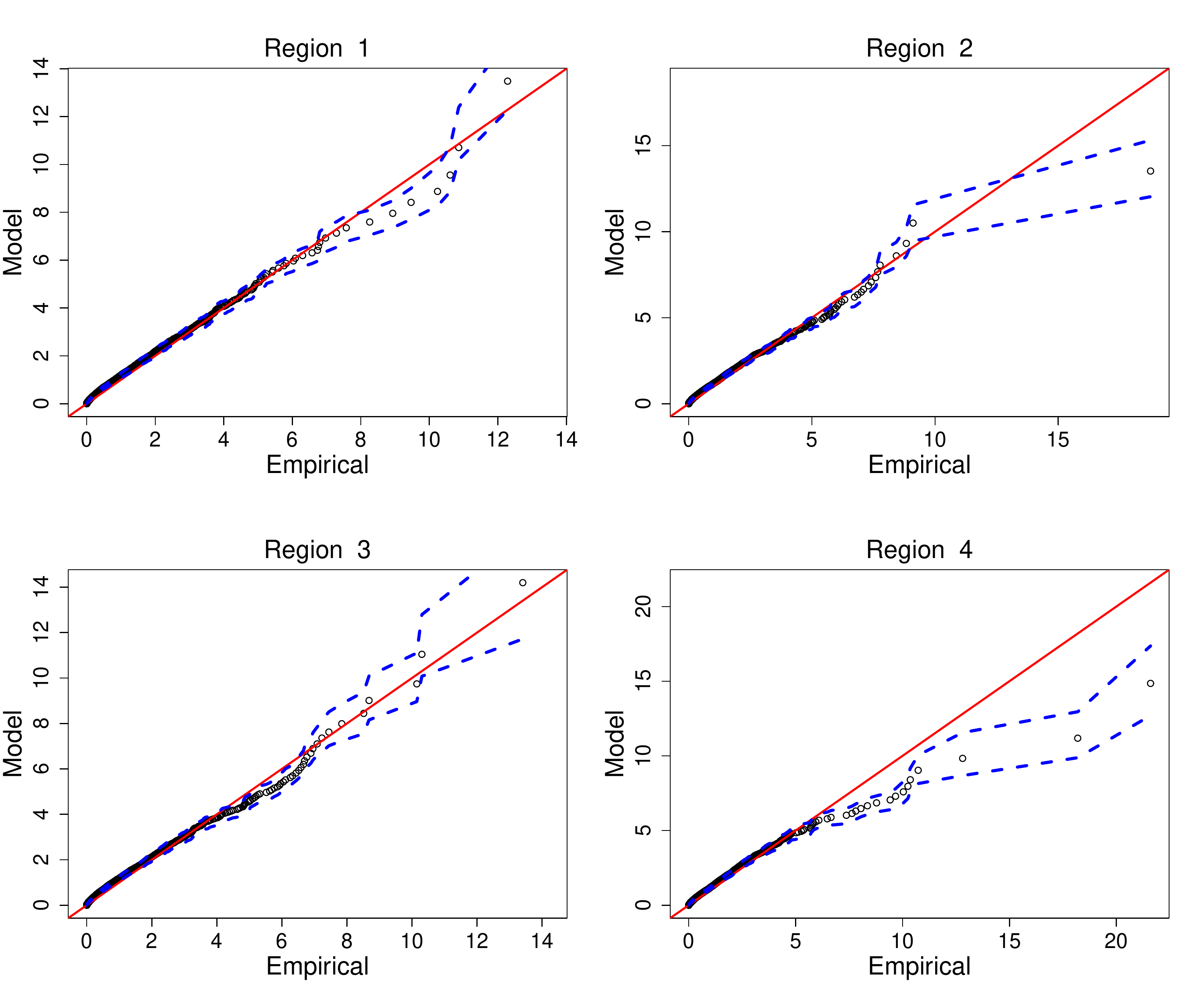} 
\end{minipage}
\begin{minipage}{0.41\textwidth}
\centering
\includegraphics[width=\textwidth]{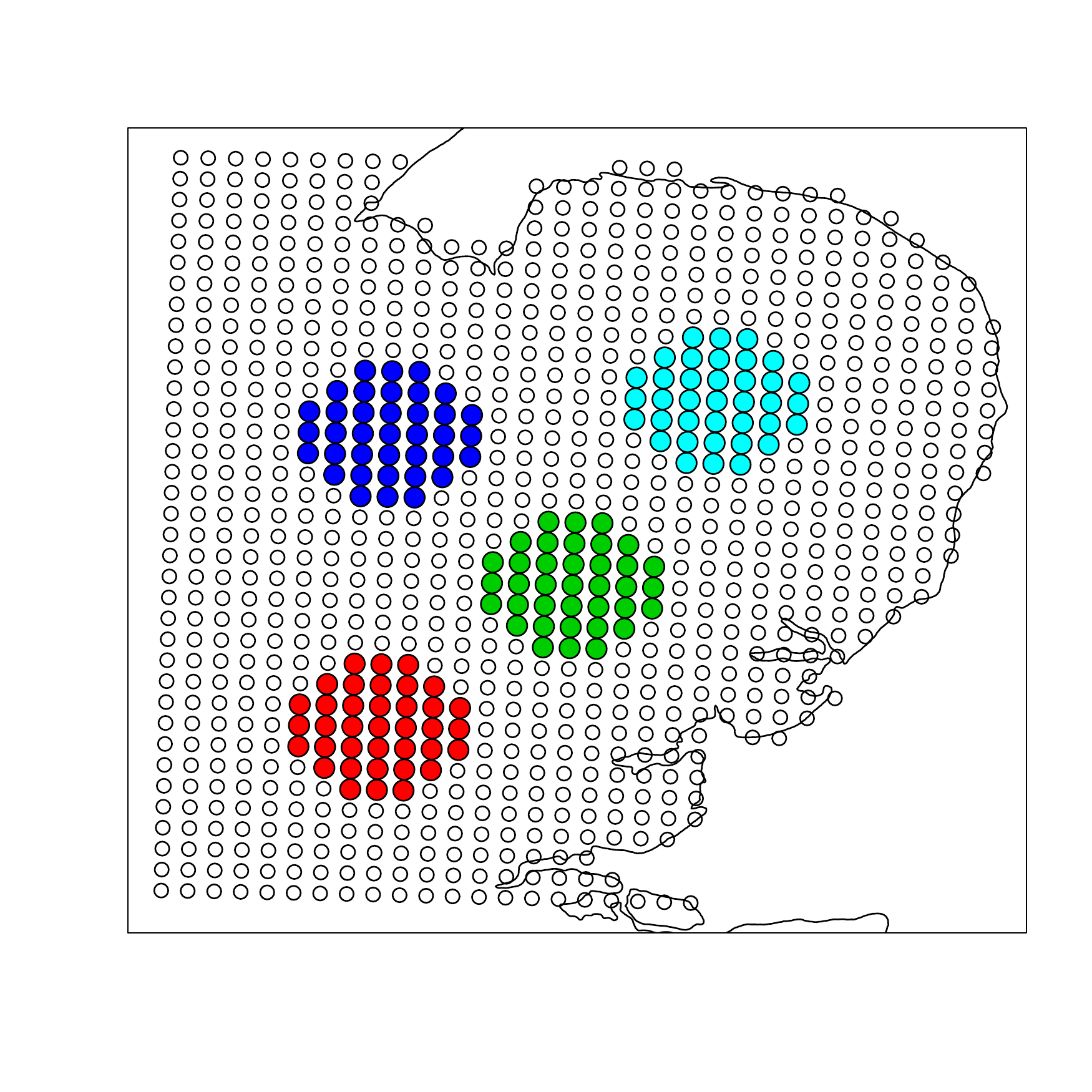} 
\end{minipage}
\caption{Left: Q-Q plots for AI model, and empirical, $\bar{R}_{\mathcal{A}}$ of four regions in $\mathcal{S}$, with $95\%$ confidence intervals given by the dashed lines. Probabilities range from $0.7$ to a value corresponding to the 20 year return level. Right: regions, each with approximate area $925 km^2$, are coloured 1-4 red, green, blue, cyan.}
\label{agg_diag}
\end{figure}

\begin{figure}[h]
\centering
\begin{minipage}{0.32\textwidth}
\centering
\includegraphics[width=\textwidth]{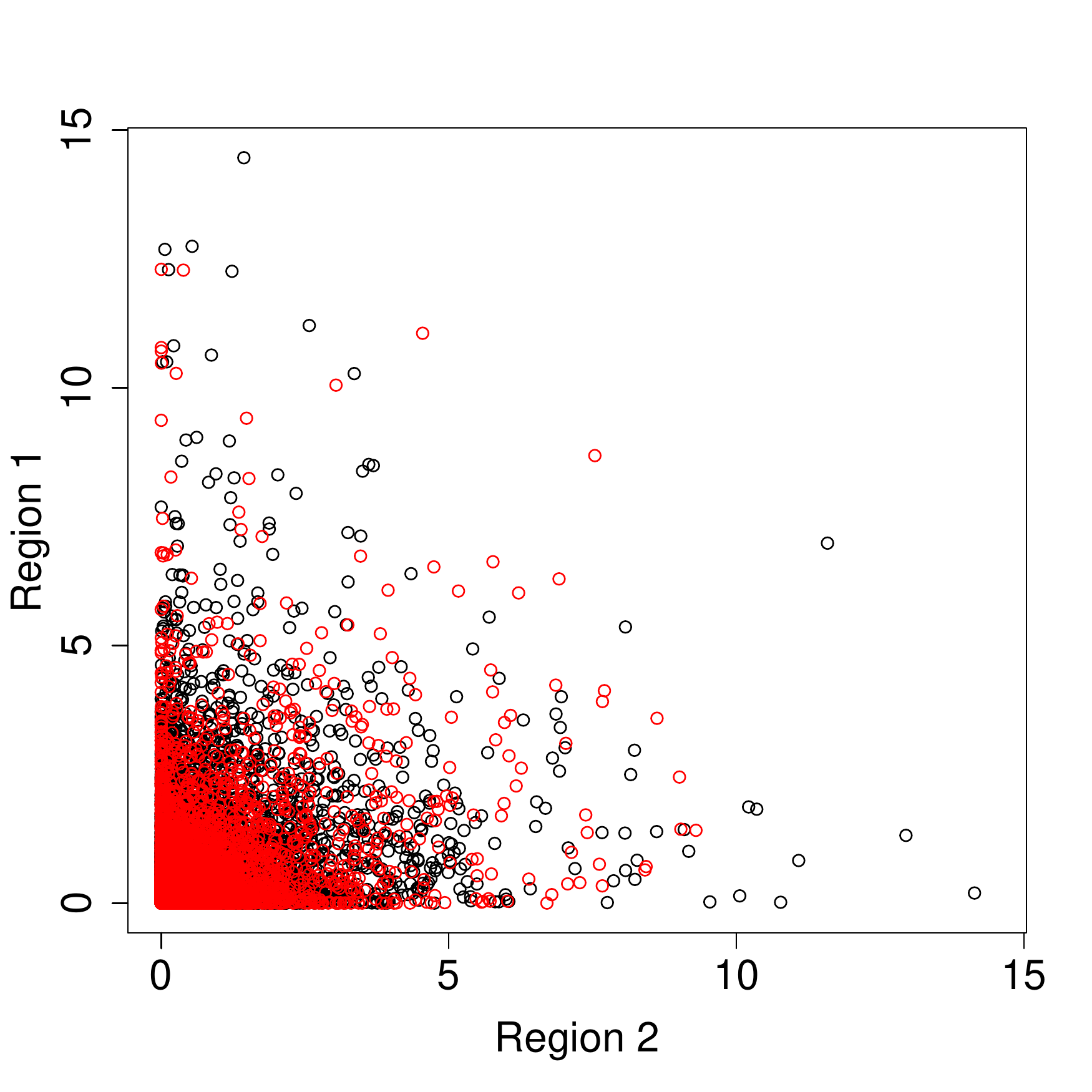} 
\end{minipage}
\begin{minipage}{0.32\textwidth}
\centering
\includegraphics[width=\textwidth]{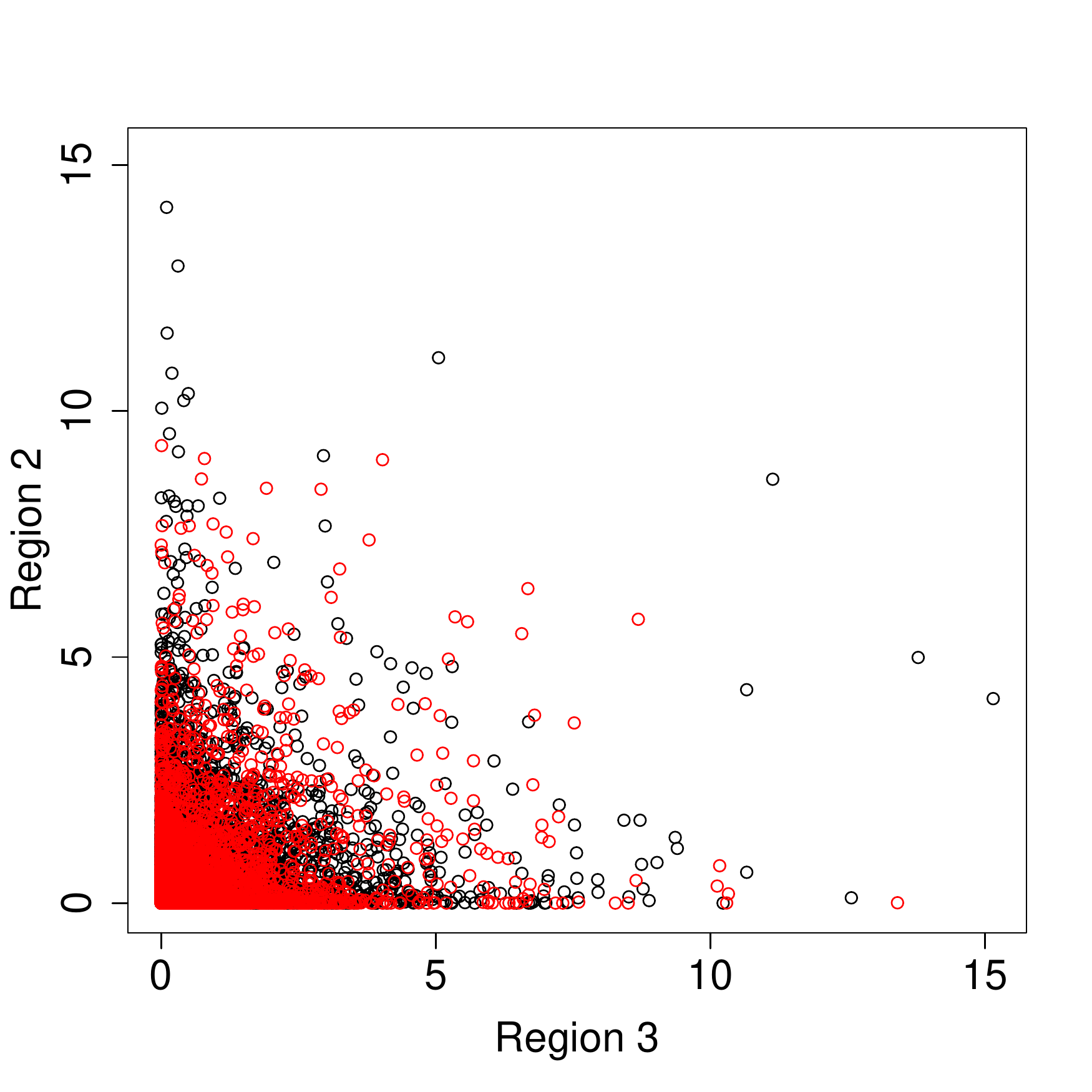} 
\end{minipage}
\begin{minipage}{0.32\textwidth}
\centering
\includegraphics[width=\textwidth]{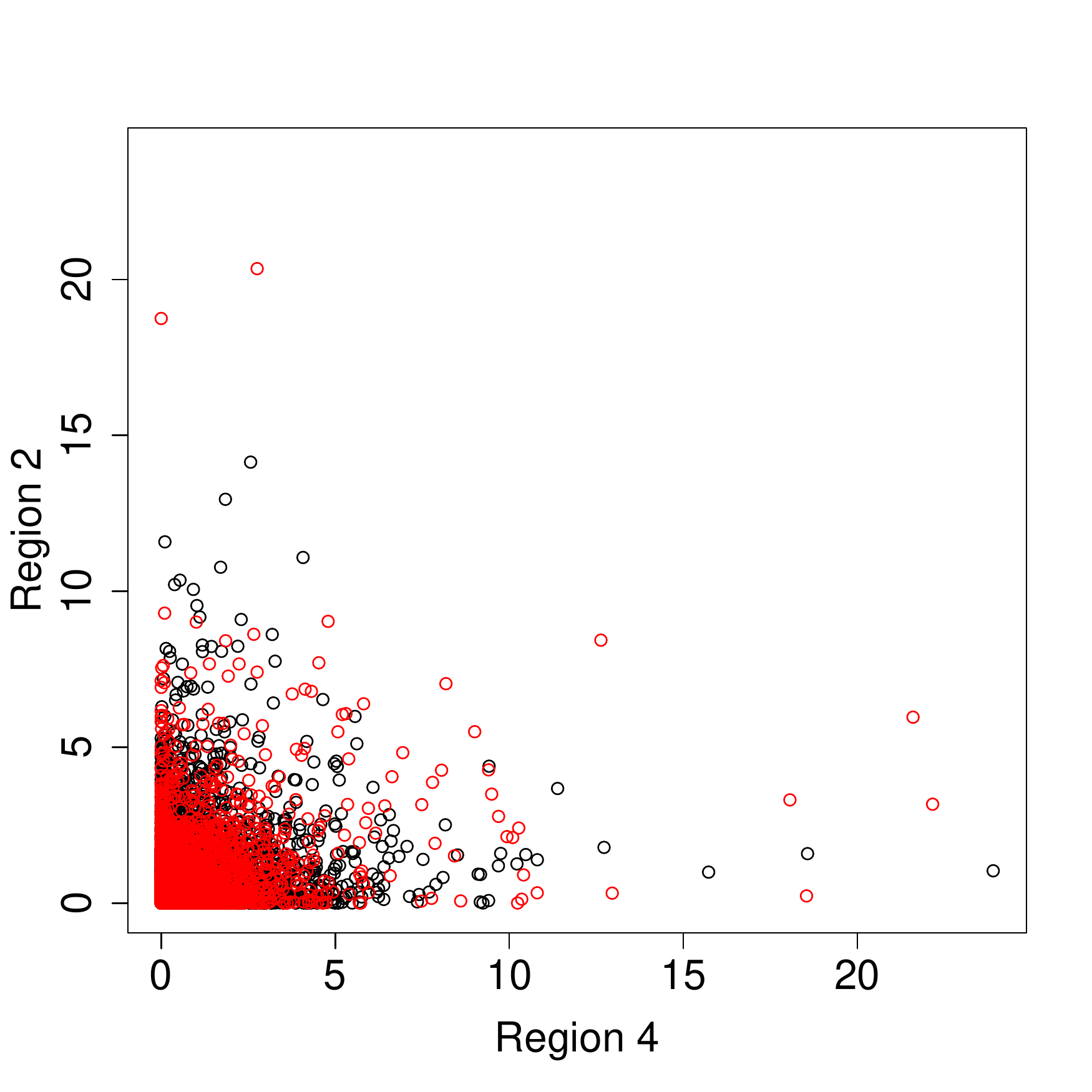} 
\end{minipage}
\caption{Plots of $2\times10^4$ realisations of pairwise $(\bar{R}_{\mathcal{A}},\bar{R}_{\mathcal{B}})$ for non-overlapping regions $\mathcal{A},\mathcal{B}$, illustrated in Figure \ref{agg_diag}. Black points are model estimates, red points are from the data. The regions $\mathcal{A}$ and $\mathcal{B}$ are labelled on the respective panels.}
\label{figjointns}
\end{figure}

\end{appendix}
\end{document}

%% file: header.tex
\usepackage{graphicx}
\usepackage{amsmath}
\usepackage{amsfonts}
\usepackage{amsthm}
\usepackage{hyperref} 
\usepackage{bbm}
\usepackage{amsopn,amssymb,url,mathrsfs,a4wide}
\usepackage{tikz}
\usepackage{booktabs}
\usepackage[margin=2.25cm]{geometry}
\usepackage{parskip}
\usepackage{fancyvrb}
\usepackage{listings}
\usepackage[numbers]{natbib}
\usepackage{float}
\usepackage{verbatim}
\usepackage{caption}
\usepackage{subcaption}
\usepackage{titlesec}
\usepackage{natbib}
\setcitestyle{authoryear,open={(},close={)}}
\newcounter{ctr}

\newcounter{ctr1}

\newcounter{ctr2}

\newcounter{ctr3}

\newenvironment{theorem*}[1]{{\bf Theorem #1} \begin{itshape}}{\end{itshape}}

\newenvironment{corollary*}[1]{{\bf Corollary #1} \begin{itshape}}{\end{itshape}}

\newenvironment{proposition*}[1]{{\bf Proposition #1} \begin{itshape}}{\end{itshape}}

\newcommand{\ud}{\, {\rm d} \kern-.015em }

\newcommand{\modulus}[1]{\left| \kern.05em #1 \kern.05em \right|}
\newcommand{\norm}[1]{\left\| \kern.05em #1 \kern.05em \right\|}
\newcommand{\inner}[1]{\left\langle \kern.05em #1 \kern.05em \right\rangle }

\newcommand{\pick}[2]{\renewcommand{\arraystretch}{0.6}
\left( \kern-.4em \begin{array}{c} #1 \\ #2 \end{array} \kern-.4em \right) }

\newcommand{\var}[1]{\, {\rm var}\left( #1 \right) }

\newsavebox{\FVerbBox}

\titlespacing\chapter{0pt}{3pt plus 4pt minus 2pt}{3pt plus 2pt minus 2pt}
\titlespacing\section{0pt}{3pt plus 4pt minus 2pt}{3pt plus 2pt minus 2pt}
\titlespacing\subsection{0pt}{3pt plus 4pt minus 2pt}{3pt plus 2pt minus 2pt}
\titlespacing\subsubsection{0pt}{3pt plus 4pt minus 2pt}{3pt plus 2pt minus 2pt}
\titlespacing\paragraph{0pt}{3pt plus 2pt minus 2pt}{3pt plus 2pt minus 2pt}